\documentclass[apj]{emulateapj}
\shorttitle{Mass Outflows in the NLR of Mrk 573}
\shortauthors{Revalski et al.}
\slugcomment{Accepted for Publication in ApJ}
\usepackage{graphics}
\usepackage{microtype}
\usepackage{color}
\usepackage{subfigure}
\usepackage{float}
\usepackage{epstopdf}
\usepackage{comment}
\usepackage[colorlinks=true, citecolor=blue, linkcolor=blue, urlcolor=blue]{hyperref}
\usepackage{etoolbox}
\usepackage[normalem]{ulem}

\makeatletter
\patchcmd{\NAT@citex}
  {\@citea\NAT@hyper@{%
     \NAT@nmfmt{\NAT@nm}%
     \hyper@natlinkbreak{\NAT@aysep\NAT@spacechar}{\@citeb\@extra@b@citeb}%
     \NAT@date}}
  {\@citea\NAT@nmfmt{\NAT@nm}%
   \NAT@aysep\NAT@spacechar\NAT@hyper@{\NAT@date}}{}{}
\patchcmd{\NAT@citex}
  {\@citea\NAT@hyper@{%
     \NAT@nmfmt{\NAT@nm}%
     \hyper@natlinkbreak{\NAT@spacechar\NAT@@open\if*#1*\else#1\NAT@spacechar\fi}%
       {\@citeb\@extra@b@citeb}%
     \NAT@date}}
  {\@citea\NAT@nmfmt{\NAT@nm}%
   \NAT@spacechar\NAT@@open\if*#1*\else#1\NAT@spacechar\fi\NAT@hyper@{\NAT@date}}
  {}{}
\makeatother

\begin{document}
\title{\vspace{-25pt} Quantifying Feedback from Narrow Line Region Outflows in Nearby Active Galaxies. I. Spatially Resolved Mass Outflow Rates for the Seyfert 2 galaxy Markarian 573$^{\dagger \star}$}

\author{M. Revalski \altaffilmark{1,6}, D. M. Crenshaw \altaffilmark{1}, S. B. Kraemer \altaffilmark{2}, T. C. Fischer \altaffilmark{3,7}, H. R. Schmitt \altaffilmark{4}, C. Machuca \altaffilmark{5}}
\altaffiltext{1}{Department of Physics and Astronomy, Georgia State University, 25 Park Place, Suite 605, Atlanta, GA 30303, USA; revalski@astro.gsu.edu}
\altaffiltext{2}{Institute for Astrophysics and Computational Sciences, Department of Physics, The Catholic University of America, Washington, DC 20064, USA}
\altaffiltext{3}{Astrophysics Science Division, Goddard Space Flight Center, Code 665, Greenbelt, MD 20771, USA}
\altaffiltext{4}{Naval Research Laboratory, Washington, DC 20375, USA}
\altaffiltext{5}{Department of Astronomy, University of Wisconsin--Madison, 475 North Charter Street, Madison, WI 53707, USA}
\altaffiltext{6}{National Science Foundation Graduate Research Fellow (DGE-1550139)}
\altaffiltext{7}{{\it James Webb Space Telescope} NASA Postdoctoral Program Fellow}
\altaffiltext{$\dagger$}{Based on observations made with the NASA/ESA Hubble Space Telescope, obtained from the Data Archive at the Space Telescope Science Institute, which is operated by the Association of Universities for Research in Astronomy, Inc., under NASA contract NAS 5-26555.}
\altaffiltext{$\star$}{Based in part on observations obtained with the Apache Point Observatory 3.5 m telescope, which is owned and operated by the Astrophysical Research Consortium.}

\begin{abstract}
We present the first spatially resolved mass outflow rate measurements ($\dot M_{out}$) of the optical emission line gas in the narrow line region (NLR) of a Seyfert 2 galaxy, Markarian 573. Using long slit spectra and [O~III] imaging from the {\it Hubble Space Telescope} and {\it Apache Point Observatory} in conjunction with emission line diagnostics and Cloudy photoionization models, we find a peak outflow rate of $\dot M_{out} \approx$ 3.4 $\pm$ 0.5 $M_{\odot}$ yr$^{-1}$ at a distance of 210 pc from the central supermassive black hole (SMBH). The outflow extends to distances of 600 pc from the nucleus with a total mass and kinetic energy of $M \approx 2.2 \times 10^6 M_{\odot}$ and $E \approx 5.1 \times 10^{54}$ erg, revealing the outflows to be more energetic than those in the lower luminosity Seyfert 1 galaxy NGC 4151 \citep{crenshaw2015}. The peak outflow rate is an order of magnitude larger than the mass accretion and nuclear outflow rates, indicating local in-situ acceleration of the circumnuclear NLR gas. We compare these results to global techniques that quantify an average outflow rate across the NLR, and find the latter are subject to larger uncertainties. These results indicate that spatially resolved observations are critical for probing AGN feedback on scales where circumnuclear star formation occurs.
\end{abstract}

\keywords{galaxies: active -- galaxies: individual (Mrk 573) -- galaxies: kinematics and dynamics -- galaxies: Seyfert -- ISM: jets and outflows \vspace{-10pt}}

\section{Introduction}

\subsection{Feedback from Mass Outflows in Active Galaxies}

Observations suggest that active galactic nuclei (AGN) are supermassive black holes (SMBHs) at the centers of galaxies that liberate immense amounts of energy as they accrete matter from the interstellar medium. As this energy propagates into the host galaxy, it can deliver feedback in the form of powerful outflows of ionized and molecular gas that may be key to understanding the observed correlations between SMBHs and their host galaxies. Specifically, these ``AGN winds'' may regulate the SMBH accretion rate and clear the bulge of star forming gas \citep{ciotti2001, hopkins2005, kormendy2013, heckman2014, fiore2017}, thereby establishing the observed relationship between SMBH mass and the stellar velocity dispersion of stars in the bulge ($M_{\bullet} - \sigma_{\star}$; \citealp{ferrarese2000, gebhardt2000, batiste2017}). Outflows may also affect chemical enrichment of the intergalactic medium \citep{hamann1999, khalatyan2008} and the development of large-scale structure in the early universe \citep{scannapieco2004, dimatteo2005}.

To better understand feedback from outflows, especially in unresolved high redshift quasars, it is helpful to perform detailed studies on nearby Seyfert galaxies  (z $\leq$ 0.1, $L_{\mathrm{bol}} \approx 10^{43} - 10^{45}$ erg s$^{-1}$, \citealp{seyfert1943}) because they have bright, spatially resolved ionization structures. Early studies focusing on ultraviolet (UV) and X-ray absorbers found that they are likely driven outward before most of the mass reaches the inner accretion disk, with mass outflow rates 10 $-$ 1000 times larger than the mass accretion rates \citep{crenshaw2003, veilleux2005, crenshaw2012, king2015}.

A logical next step is to quantify feedback from mass outflows on larger scales in the narrow emission line region (NLR), which is composed of ionized gas at distances of $\sim$ 1 $-$ 1000 parsecs (pc) from the central SMBH with densities of $n_\mathrm{H}$ $\approx 10^2 - 10^6$ cm$^{-3}$ \citep{peterson1997}. Kinematic studies of the NLR reveal outflows of ionized gas, often with a biconical geometry \citep{crenshaw2010, mullersanchez2011, fischer2013, fischer2014, bae2016, mullersanchez2016, nevin2016}, that exhibit wide opening angles as compared to narrow relativistic jets that are only present in 5-10\% of AGN \citep{rafter2009}. Following the precedent of others, we define the true NLR to be that with outflow kinematics driven by the AGN, while the extended narrow line region (ENLR) is ionized gas at larger radii exhibiting primarily galactic rotation \citep{unger1987}.

Feedback can be quantified through mass outflow rates ($\dot{M} = Mv/\delta r$) and energetic measures, where all quantities are corrected for observational projection effects \citep{crenshaw2012, crenshaw2015}. Studies during the past decade have produced ``global'' measurements of these quantities that average over the spatial extent of the NLR, and find a range of outflow rates and energetics using geometric \citep{barbosa2009, riffel2009, storchibergmann2010, mullersanchez2011, riffel2011a, riffel2013, schonell2014, schonell2017, schnorrmuller2014, gofford2015, mullersanchez2016, nevin2018, wylezalek2017} and luminosity based techniques \citep{liu2013, forsterschreiber2014, genzel2014, harrison2014, mcelroy2015, karouzos2016, schnorrmuller2016, villarmartin2016, bae2017, bischetti2017, leung2017}.

These global techniques are advantageous because they provide measurements relatively quickly; however, detailed spatial information is lost. We can leverage the power of spatially resolved NLR observations to not only quantify feedback from the outflows, but also reveal in detail where energy is deposited. With this goal, \cite{crenshaw2015} conducted a pilot study to quantify the mass outflow rates at each point along the NLR in the nearby Seyfert 1 galaxy NGC 4151 ($z = 0.00332, D = 13.3$ Mpc). Using their previous kinematic and photoionization models \citep{das2005, kraemer2000a} in conjunction with narrow-band [O~III] imaging \citep{kraemer2008}, they found a peak outflow rate of $\dot M_{out} \approx$ 3.0 $M_{\odot}$ yr$^{-1}$ at a distance of 70 pc from the SMBH, with the outflow extending to $\sim$ 140 pc. This outflow rate exceeds that of the UV/X-ray absorbers and the calculated mass accretion rate based on the AGN bolometric luminosity \citep{crenshaw2012}, indicating that NLR outflows may provide significant feedback to their host galaxies.

This result has motivated us to quantify feedback from NLR outflows in a sample of nearby Seyfert galaxies. In this first paper, we present our results for the Seyfert 2 AGN Markarian 573 (Mrk 573, UGC 01214, UM 363) and give a detailed methodology modeled after \cite{crenshaw2015} that we will apply to additional AGN. This process consists of determining the mass of ionized gas from high spatial resolution spectroscopy (\S2), emission line diagnostics (\S3), [O~III] imaging (\S4), and detailed photoionization models (\S5). We then calculate mass outflow rates and energetics (\S6), presenting our results, discussion, and conclusions in \S7, \S8, and \S9, respectively.

\subsection{Physical Characteristics of Markarian 573}

\begin{figure}
\includegraphics[scale=0.54]{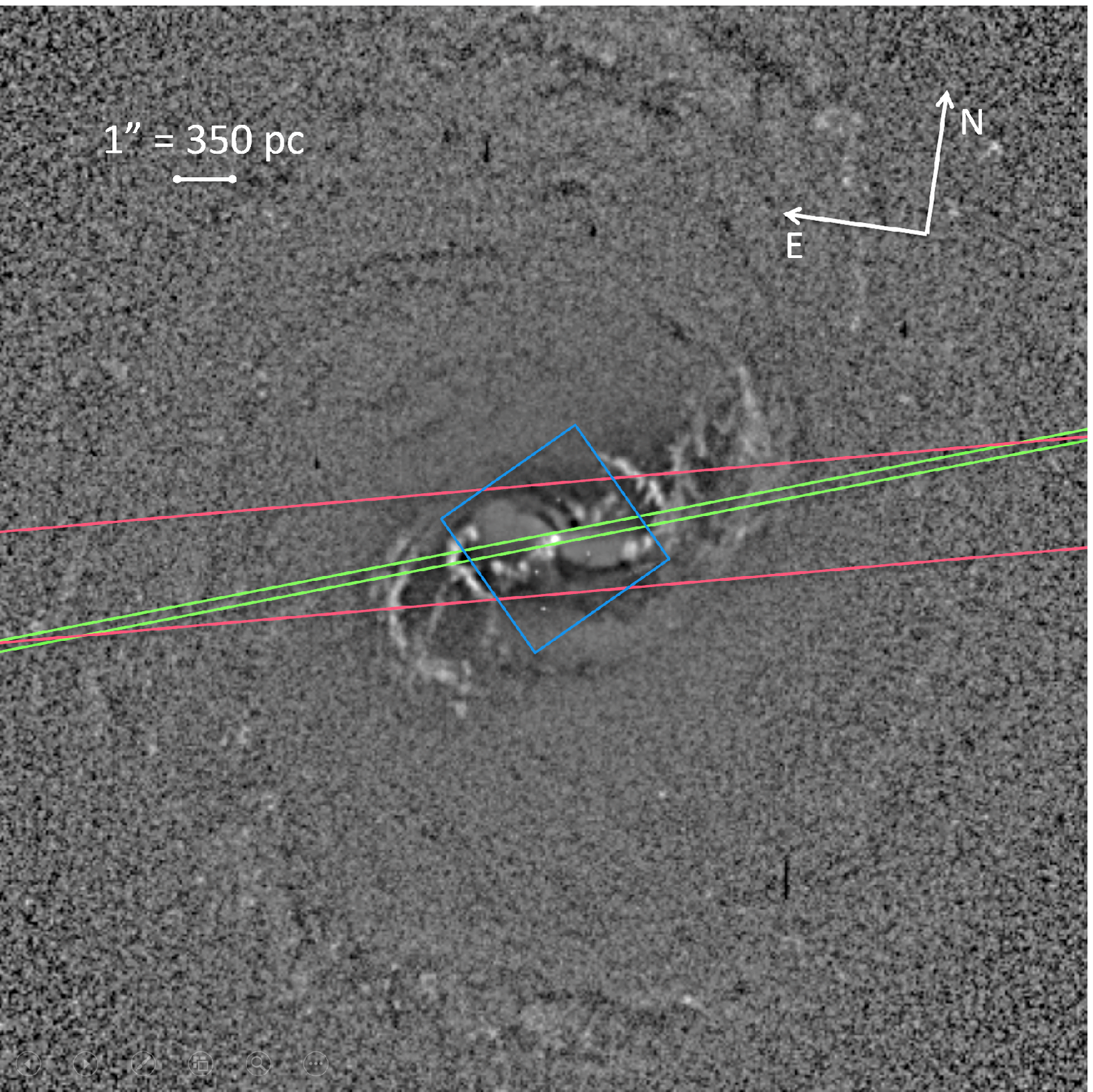}
\caption{A 20$\arcsec$x20$\arcsec$ contrast enhanced structure map from the HST WFPC2 image of Mrk 573 using the F606W filter \citep{pogge2002, fischer2010}. Bright regions correspond to [O~III] and continuum emission, while darker regions are dust lanes. The position of the $0\farcs2$ HST STIS slit is shown in green, the APO DIS $2\farcs0$ slit is shown in red, and the central {\it Gemini} NIFS 3$\arcsec$x3$\arcsec$ field from \cite{fischer2017} is shown in blue.}
\vspace{-5pt}
\label{structure}
\end{figure}

Mrk 573 was chosen as an ideal candidate to refine the techniques developed by \cite{crenshaw2015} because the NLR displays distinct kinematic signatures of outflow and rotation that have been well characterized by previous studies \citep{tsvetanov1989b, tsvetanov1992, afanasiev1996, ruiz2005, schlesinger2009, fischer2010, fischer2017}. Mrk 573 is also a more rapidly accreting and luminous AGN than NGC 4151, allowing us to probe NLR outflows in a more energetic regime \citep{kraemer2009}. Finally, there are high quality spectroscopic and imaging data available from archival resources and our own observing programs.

The central AGN resides in an S0 host galaxy with an (R)SAB(rs) classification \citep{devaucouleurs1995}. There are measured redshifts from stellar kinematics ($z = 0.01718$, \citealp{nelson1995}) and 21 cm observations ($z = 0.01721$, \citealp{springob2005}). The former is most consistent with the NLR gas kinematics and corresponds to $cz \approx 5150$ km s$^{-1}$. We adopt $H_0 = 71$ km s$^{-1}$ Mpc$^{-1}$, corresponding to a Hubble distance of $72$ Mpc and a spatial scale of $\sim 350$ pc/$\arcsec$ on the sky.

As a type 2 AGN under the unified model \citep{khachikian1974, osterbrock1981, antonucci1985, antonucci1993, netzer2015}, Mrk 573 displays narrow permitted and forbidden lines and was suspected to house a hidden BLR \citep{kay1994, tran2001} that was detected with deep spectropolarimetric observations (\citealp{nagao2004a, nagao2004b}; but see also \citealp{ramosalmeida2008, ramosalmeida2009a}). The central SMBH has a mass of Log($M_{\mathrm{BH}}$/$M_{\odot}$) = 7.28  \citep{woo2002, bian2007} and radiates at a bolometric luminosity of Log($L_{\mathrm{bol}}$/erg s$^{-1}$) = 45.5 $\pm$ 0.6 \citep{melendez2008a, melendez2008b, kraemer2009}. This corresponds to an Eddington ratio of $\mathrm{L/L_{Edd} \approx 0.75}$.

Extensive studies of Mrk 573 have revealed a biconical illumination cone of radiation, creating a highly ionized NLR and ENLR spanning several kpc \citep{koski1978, tsvetanov1989a, haniff1991, kinney1991, tsvetanov1992, pogge1993, erkens1997, alonsoherrero1998, storchibergmann1998a, mullaney2008, dopita2015}. The advent of space-based and adaptive optics observations unveiled rich structure, including ionized arcs of emission, spiral dust lanes, and a so dubbed ``linear feature'' of emission line knots extending along a position angle (PA) of $\sim$ 125$\degr$ \citep{pogge1995, malkan1998, ferruit1999, martini2001, pogge2002}. This feature is coincident with a triple-lobed radio source that displays possible interaction with the ENLR \citep{ulvestad1984, unger1987, haniff1988, whittle1988, tsvetanov1989b, capetti1996, falcke1998, ferruit2002}. Several of these features can be seen in our structure map in Figure \ref{structure}.

\section{Observations}

\subsection{Hubble Space Telescope (HST) Spectra \& Imaging}

The archival Space Telescope Imaging Spectrograph (STIS) spectra were retrieved from the Mikulski Archive at the Space Telescope Science Institute and calibrated as part of a detailed study that classified the NLR kinematics in a large sample of Seyfert galaxies \citep{fischer2013, fischer2014}. The observations were obtained on 2001 October 17 as a part of Program ID 9143 (PI: R. Pogge) utilizing the 52$\arcsec$ x $0\farcs2$ slit with G430L and G750M gratings. The G430L observations span $2845 - 5760~\AA$ at a dispersion of 2.744 $\AA$ pixel$^{-1}$, and the G750M observations span $6285 - 6875~\AA$ at a dispersion of 0.555 $\AA$ pixel$^{-1}$. Both data sets have a spatial resolution of $0\farcs05078$ pixel$^{-1}$. Data reduction consisted of aligning the spatially offset exposures using the peaks of the underlying continua to within half a pixel, median combining, and hot pixel removal. We extracted spectra spatially summed over two pixel intervals to match the angular resolution along the slit, and further data reduction details are given in \cite{kraemer2009} and \cite{fischer2010}.

\begin{figure*}
\vspace{-10pt}
\centering
\includegraphics[scale=0.54]{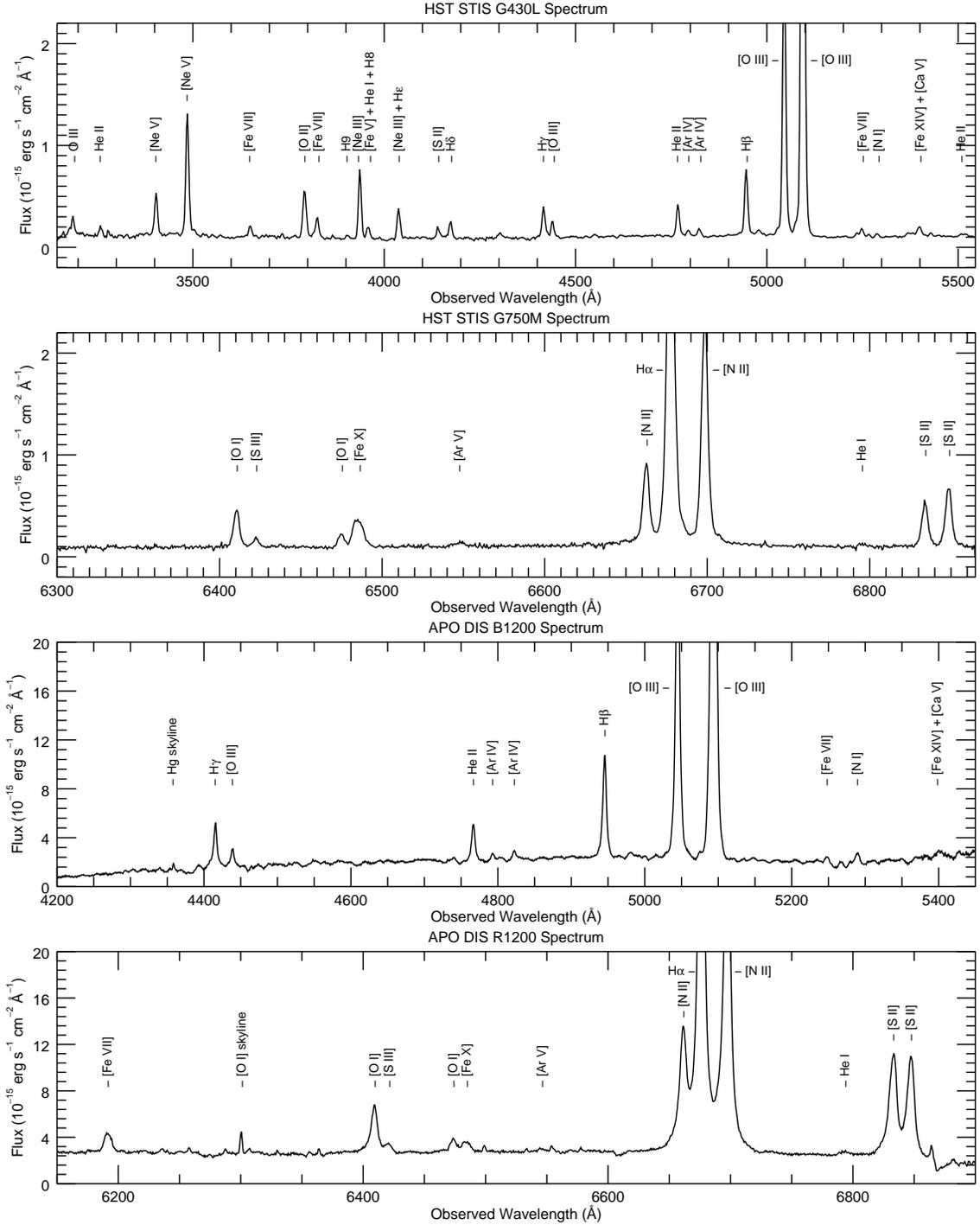}
\vspace{-7pt}
\caption{Spectral traces of the nucleus spatially summed over $0\farcs2$ and $2\farcs0$ for the STIS and DIS data, respectively. From top to bottom are HST STIS G430L, HST STIS G750M, APO DIS B1200, and APO DIS R1200 spectra. The spectra are shown at observed wavelengths, and emission line ratios relative to H$\beta$ are given in Table 1.}
\label{spectra}
\vspace{-2pt}
\end{figure*}

The slit position of $-71\fdg2$ spatially samples the bright nuclear emission and extended arcs, but misses the linear feature of bright emission line knots extending roughly parallel to the radio feature along PA $\sim$ 125${\degr}$. The position of the HST STIS slit relative to the NLR is shown in Figure \ref{structure} and the extracted spectra are shown in Figure \ref{spectra}.

The [O~III] imaging observations were conducted on 1995 November 12 with the HST Wide Field and Planetary Camera 2 (WFPC2) on the WF2 camera with a plate scale of $0\farcs0996$ pixel$^{-1}$ as part of Program ID 6332 (PI: A. Wilson). Two 600~s exposures were taken through the FR533N ramp filter with a wavelength center of 5093 $\AA$, centering on the [O~III] $\lambda$5007 emission line at the redshift of Mrk 537. The images were retrieved and calibrated by \cite{schmitt2003a, schmitt2003b} to study the extended [O~III] morphologies of Seyfert galaxies, and data reduction details are given in those papers.

\subsection{Apache Point Observatory (APO) Spectra}

To characterize NLR emission that falls outside of the narrow STIS slit, we utilized supplementary observations taken with the Dual Imaging Spectrograph (DIS) on the Astrophysical Research Consortium's Apache Point Observatory (ARC's APO) 3.5 meter telescope in Sunspot, New Mexico. The DIS employs a dichroic element that splits light into blue and red channels, allowing for simultaneous data collection in the H$\beta$ and H$\alpha$ portions of the spectrum. Here we focus on our observations along PA = 103$\degr$ with a $2\farcs0$ slit and B/R1200 gratings that were originally presented in our study of the NLR and ENLR kinematics of Mrk 573 \citep{fischer2017}.

These data consist of two 900~s exposures taken on 2013 December 3 at a mean airmass of 1.214 with $1\farcs6$ seeing. The blue channel spans $4190-5455~\AA$ at a dispersion and spatial resolution of 0.6148 $\AA$ pixel$^{-1}$ and $0\farcs42$ pixel$^{-1}$, respectively. The red channel spans $5995-7180~\AA$ at a dispersion and spatial resolution of 0.5796 $\AA$ pixel$^{-1}$ and $0\farcs40$ pixel$^{-1}$, respectively. The corresponding spectral resolutions are approximately 1.23 and 1.16 $\AA$ in the blue and red channels, respectively, yielding resolving powers of $R \approx$ 3400 -- 6200 across the spectra.

We completed a new reduction of the data using the latest calibrations following standard techniques in IRAF \citep{tody1986}.\footnote{IRAF is distributed by the National Optical Astronomy Observatories, which are operated by the Association of Universities for Research in Astronomy, Inc., under cooperative agreement with the National Science Foundation.} This consisted of bias subtraction, image trimming, bad pixel replacement, flat fielding, Laplacian edge cosmic ray removal \citep{vandokkum2001}, and image combining. Wavelength calibration was completed using comparison lamp images taken immediately before the science exposures, and flux calibration with the airmass at mid exposure using the standard stars Feige 110 and BD+28 \citep{oke1990}. The DIS dispersion and spatial directions are not precisely perpendicular, so we fit a line to the underlying galaxy continuum and resampled the data to a new grid to ensure that measurements of different emission lines from the same pixel row sample the same spatial location. We extracted spectra over single pixel spatial intervals to probe large-scale NLR gradients.

\section{Spectral Analysis}

\subsection{Gaussian Template Fitting}

We identified emission lines in the spectra using previous studies and photoionization model predictions. Care was taken to measure weak lines that serve as temperature or density diagnostics to constrain the physical conditions in the NLR gas. We fit all emission lines with S/N $> 2$, and tentatively identified lines that fell below this threshold at all positions, including He~I+Fe~II $\lambda$4923, [Fe~VII] $\lambda$5158, [Fe~VI] $\lambda$5176, [N~I] $\lambda$5199, [Fe~XIV] $\lambda$5303, [Ca~V] $\lambda$5310, and [Ar~X] $\lambda$5536 \citep{storchibergmann1996}.

Photoionization modeling requires accurate emission line flux ratios for comparison. To accomplish this we fit gaussian profiles to all emission lines at each spatial extraction using a template method that calculates the centroids and widths of all lines based on the best fit to the strong, velocity resolved H$\alpha$ emission line. We found that H$\alpha$ produced more accurate centroid predictions than the stronger [O~III] emission line due to the higher spectral resolution of the G750M data, and we corrected the line widths to maintain the same intrinsic velocity width as the template and account for the different line-spread functions of the gratings. The height was free to vary to encompass the total line flux, with errors calculated from the residuals between the data and fits.

This method has the flexibility to fit multiple kinematic components and heavily blended lines; however, information is lost about intrinsic differences in the radial velocity and full width at half maximum (FWHM) of individual lines. While studies have shown that the radial velocity or FWHM can correlate with the critical density or ionization potential of an emission line \citep{whittle1985, derobertis1986, kraemer2000b, rodriguezardila2006}, our goal was to obtain accurate emission line fluxes. We conducted tests and found the integrated fluxes from free versus template fitting methods differed by $\sim$ 2 -- 8\%, and that the template method produced more accurate fits to weak emission lines in noisy regions. A single exception is the [Fe~X] $\lambda 6375$ line that is blue shifted relative to all other lines, as noted by \cite{kraemer2009} and \cite{schlesinger2009}. In this case the fluxes agree to within $15\%$.

\subsection{Ionized Gas Kinematics}

Calculating mass outflow rates requires knowing the NLR mass and its radial velocity at each location. The kinematics of Mrk 573 have been studied extensively, exhibiting a linearly increasing velocity profile with distance from the nucleus, followed by a turnover and linear decrease \citep{tsvetanov1992, afanasiev1996, schlesinger2009, fischer2010, fischer2017}.

\begin{figure}
\centering
\subfigure{
\includegraphics[scale=0.98]{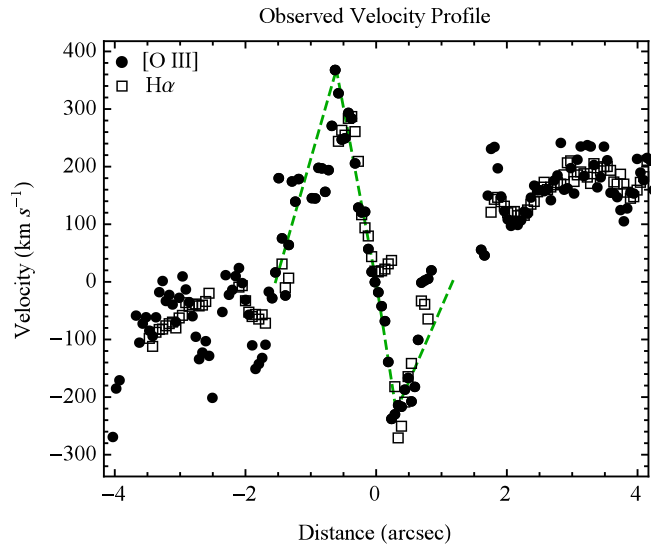}}
\vspace{-10pt}
\caption{Observed HST STIS [O III] and H$\alpha$ kinematics shown with filled circles and empty squares, respectively. The best fit linear velocity laws are shown as dashed green lines. SE is to the left and NW is to the right.}
\label{kinematics}
\end{figure}

To determine accurate outflow velocity laws, we fit a series of linear functions to single row extractions of the [O~III] emission line velocity centroids. We used these fits to derive the observed velocity of the gas at each projected radial distance, as shown in Figure \ref{kinematics}. The outflow terminates at $\sim$ $1\farcs7$ (600 pc), beyond which galactic rotation dominates \citep{fischer2017}. The spectra display a single kinematic component at all locations, except for a weak secondary line at $+0\farcs25$ that was not fit.

From our results presented in \cite{fischer2017} we adopt a host disk inclination of $i = 38\degr$ with the northeast (NE) side of the disk closest to us, a revision from our earlier model \citep{fischer2010}, and in agreement with previous findings \citep{pogge1995}. We assume the NLR gas is approximately coplanar with the stellar disk \citep{keel1980, afanasiev1996, xanthopoulos1996, afanasiev1998, schmitt2000} and is flowing radially along the disk \citep{fischer2017}. The PAs of the inner disk and NLR are approximately 97$\degr$ and 128$\degr$, respectively. This yields an offset of $\varphi = 31\degr$ between the disk and NLR major axes. While the HST spectral slit is not exactly parallel to the NLR major axis, all material out to $\sim$ $0\farcs6$ can travel along the NLR major axis and remain within the spectral slit, so we adopt this angle ($\varphi = 31\degr$) for deprojecting the velocities. With this model of radial outflow, the observed Doppler velocities were deprojected to their intrinsic values via
\begin{equation}
V_{\mathrm{int}}(i,\varphi) = \left(\frac{V_{\mathrm{obs}}}{sin(i)sin(\varphi)}\right),
\end{equation}
\noindent
where $V_{\mathrm{int}}$ and $V_{\mathrm{obs}}$ are the intrinsic and observed velocities for an inclination $i$ and phase angle $\varphi$ from the major axis. This is similar to the expression for rotational velocities, with the final term sine instead of cosine. The true distance from the SMBH is calculated from the observed distance and inverting the equation for distance from the center of an ellipse,\footnote{Our terminology assumes that the distances from the central SMBH and the ``nucleus'', as defined by the peak [O~III] emission, are the same; however, there is evidence that AGN nuclei and their SMBHs can be offset by tens of pc \citep{kraemer2010, comerford2017}.}
\begin{equation}
R_{\mathrm{int}}(i,\varphi) = \sqrt{R^2_{\mathrm{obs}} \left(cos^2(\varphi) + \frac{sin^2(\varphi)}{cos^2(i)} \right)},
\end{equation}
\noindent
where $R_{\mathrm{int}}$ and $R_{\mathrm{obs}}$ are the intrinsic and observed distances for an inclination $i$ and phase angle $\varphi$ from the major axis. We adopt the convention that $i = 0\degr$ is face on and $i = 90\degr$ is edge on. The phase angle $\varphi$ is defined such that $\varphi = 0\degr$ is along the major axis and $\varphi = 90\degr$ is along the minor axis.

With our adopted angles ($i = 38\degr$, $\varphi = 31\degr$) the intrinsic velocities are $\sim 3.15$ times larger than the observed values. Thus the peak observed outflow velocity of $\sim$ 350 km s$^{-1}$ corresponds to an intrinsic velocity of 1100 km s$^{-1}$, typical of Seyferts \citep{fischer2013, fischer2014}.

\setlength{\tabcolsep}{0.01in}
\tabletypesize{\scriptsize}
\begin{deluxetable*}{l|c|c|c|c|c|c|c|c|c|c|}
\vspace{-26pt}
\tablenum{1}
\tablecaption{Observed HST STIS Emission Line Ratios \vspace{-5pt}}
\tablehead{
\colhead{Emission Line} & \colhead{+$0\farcs46$} & \colhead{+$0\farcs36$} & \colhead{+$0\farcs25$} & \colhead{+$0\farcs15$} & \colhead{+$0\farcs05$} & \colhead{--$0\farcs05$} & \colhead{--$0\farcs15$} & \colhead{--$0\farcs25$} & \colhead{--$0\farcs36$} & \colhead{--$0\farcs46$} \vspace{-3pt}}
\startdata
 \vspace{-1pt} O III $\lambda$3133	& ... $\pm$ ...	& ... $\pm$ ...	& ... $\pm$ ...	& ... $\pm$ ...	& 0.25 $\pm$ 0.06	& 0.25 $\pm$ 0.03	& 0.28 $\pm$ 0.02	& 0.20 $\pm$ 0.05	& ... $\pm$ ...	& ... $\pm$ ...	\\  \relax
 \vspace{-1pt} He II $\lambda$3203	& ... $\pm$ ...	& ... $\pm$ ...	& ... $\pm$ ...	& ... $\pm$ ...	& 0.12 $\pm$ 0.02	& 0.15 $\pm$ 0.02	& 0.22 $\pm$ 0.03	& 0.10 $\pm$ 0.05	& ... $\pm$ ...	& ... $\pm$ ...	\\ \relax
 \vspace{-1pt} [Ne V] $\lambda$3346	& ... $\pm$ ...	& ... $\pm$ ...	& ... $\pm$ ...	& ... $\pm$ ...	& 0.65 $\pm$ 0.05	& 0.64 $\pm$ 0.04	& 0.81 $\pm$ 0.05	& 0.40 $\pm$ 0.05	& ... $\pm$ ...	& ... $\pm$ ...	\\ \relax
 \vspace{-1pt} [Ne V] $\lambda$3426	& 1.92 $\pm$ 0.30	& 2.44 $\pm$ 0.78	& ... $\pm$ ...	& 1.78 $\pm$ 0.47	& 1.88 $\pm$ 0.15	& 1.79 $\pm$ 0.12	& 1.96 $\pm$ 0.19	& 1.22 $\pm$ 0.08	& 0.87 $\pm$ 0.07	& ... $\pm$ ...	\\ \relax
 \vspace{-1pt} [Fe VII] $\lambda$3586	& ... $\pm$ ...	& ... $\pm$ ...	& ... $\pm$ ...	& ... $\pm$ ...	& 0.16 $\pm$ 0.02	& 0.17 $\pm$ 0.03	& 0.15 $\pm$ 0.02	& ... $\pm$ ...	& ... $\pm$ ...	& ... $\pm$ ...	\\ \relax
 \vspace{-1pt} [O II] $\lambda$3727	& 1.75 $\pm$ 0.28	& 1.55 $\pm$ 0.20	& 2.24 $\pm$ 1.01	& 1.67 $\pm$ 0.28	& 0.86 $\pm$ 0.05	& 0.68 $\pm$ 0.06	& 0.87 $\pm$ 0.04	& 1.11 $\pm$ 0.07	& 1.53 $\pm$ 0.10	& 1.85 $\pm$ 0.36	\\ \relax
 \vspace{-1pt} [Fe VII] $\lambda$3759	& ... $\pm$ ...	& ... $\pm$ ...	& ... $\pm$ ...	& 0.33 $\pm$ 0.05	& 0.36 $\pm$ 0.03	& 0.28 $\pm$ 0.02	& 0.18 $\pm$ 0.01	& 0.06 $\pm$ 0.02	& ... $\pm$ ...	& ... $\pm$ ...	\\ \relax
 \vspace{-1pt} [Ne III] $\lambda$3869	& 1.44 $\pm$ 0.30	& 0.90 $\pm$ 0.20	& ... $\pm$ ...	& 1.26 $\pm$ 0.17	& 1.09 $\pm$ 0.10	& 0.85 $\pm$ 0.06	& 1.06 $\pm$ 0.10	& 1.02 $\pm$ 0.07	& 0.94 $\pm$ 0.06	& 1.50 $\pm$ 0.18	\\ \relax
 \vspace{-1pt} He I $\lambda$3889$^a$	& ... $\pm$ ...	& ... $\pm$ ...	& ... $\pm$ ...	& 0.24 $\pm$ 0.07	& 0.22 $\pm$ 0.02	& 0.16 $\pm$ 0.02	& 0.18 $\pm$ 0.02	& 0.10 $\pm$ 0.01	& 0.24 $\pm$ 0.03	& ... $\pm$ ...	\\ \relax
 \vspace{-1pt} [Ne III] $\lambda$3969$^b$ & ... $\pm$ ...	& ... $\pm$ ...	& ... $\pm$ ...	& 0.35 $\pm$ 0.07	& 0.52 $\pm$ 0.04	& 0.37 $\pm$ 0.03	& 0.44 $\pm$ 0.06	& 0.38 $\pm$ 0.04	& 0.43 $\pm$ 0.07	& ... $\pm$ ...	\\ \relax
 \vspace{-1pt} [S II] $\lambda$4074	& ... $\pm$ ...	& ... $\pm$ ...	& ... $\pm$ ...	& ... $\pm$ ...	& 0.18 $\pm$ 0.05	& 0.10 $\pm$ 0.02	& 0.08 $\pm$ 0.03	& ... $\pm$ ...	& ... $\pm$ ...	& ... $\pm$ ...	\\ \relax
 \vspace{-1pt} H$\delta$ $\lambda$4102	& ... $\pm$ ...	& ... $\pm$ ...	& ... $\pm$ ...	& ... $\pm$ ...	& 0.26 $\pm$ 0.03	& 0.25 $\pm$ 0.02	& 0.25 $\pm$ 0.02	& 0.22 $\pm$ 0.03	& ... $\pm$ ...	& ... $\pm$ ...	\\ \relax
 \vspace{-1pt} H$\gamma$ $\lambda$4340	& ... $\pm$ ...	& ... $\pm$ ...	& 0.40 $\pm$ 0.30	& 0.59 $\pm$ 0.11	& 0.52 $\pm$ 0.03	& 0.45 $\pm$ 0.03	& 0.48 $\pm$ 0.05	& 0.42 $\pm$ 0.03	& 0.42 $\pm$ 0.05	& 0.64 $\pm$ 0.09	\\ \relax
 \vspace{-1pt} [O III] $\lambda$4363	& ... $\pm$ ...	& ... $\pm$ ...	& ... $\pm$ ...	& 0.28 $\pm$ 0.08	& 0.27 $\pm$ 0.02	& 0.22 $\pm$ 0.02	& 0.22 $\pm$ 0.03	& 0.19 $\pm$ 0.03	& ... $\pm$ ...	& ... $\pm$ ...	\\ \relax
 \vspace{-1pt} He II $\lambda$4686	& ... $\pm$ ...	& ... $\pm$ ...	& ... $\pm$ ...	& 0.39 $\pm$ 0.07	& 0.46 $\pm$ 0.03	& 0.50 $\pm$ 0.03	& 0.51 $\pm$ 0.04	& 0.46 $\pm$ 0.03	& 0.40 $\pm$ 0.03	& ... $\pm$ ...	\\ \relax
 \vspace{-1pt} [Ar IV] $\lambda$4711	& ... $\pm$ ...	& ... $\pm$ ...	& ... $\pm$ ...	& ... $\pm$ ...	& 0.07 $\pm$ 0.01	& 0.11 $\pm$ 0.01	& 0.12 $\pm$ 0.02	& 0.10 $\pm$ 0.01	& ... $\pm$ ...	& ... $\pm$ ...	\\ \relax
 \vspace{-1pt} [Ar IV] $\lambda$4740	& ... $\pm$ ...	& ... $\pm$ ...	& ... $\pm$ ...	& ... $\pm$ ...	& 0.12 $\pm$ 0.01	& 0.11 $\pm$ 0.01	& 0.12 $\pm$ 0.02	& 0.07 $\pm$ 0.01	& ... $\pm$ ...	& ... $\pm$ ...	\\ \relax
 \vspace{-1pt} H$\beta$ $\lambda$4861	& 1.00 $\pm$ 0.13	& 1.00 $\pm$ 0.11	& 1.00 $\pm$ 0.38	& 1.00 $\pm$ 0.13	& 1.00 $\pm$ 0.05	& 1.00 $\pm$ 0.05	& 1.00 $\pm$ 0.04	& 1.00 $\pm$ 0.05	& 1.00 $\pm$ 0.04	& 1.00 $\pm$ 0.08	\\ \relax
 \vspace{-1pt} [O III] $\lambda$4959	& 5.25 $\pm$ 0.69	& 4.00 $\pm$ 0.52	& 5.09 $\pm$ 2.10	& 4.87 $\pm$ 0.73	& 4.45 $\pm$ 0.27	& 4.40 $\pm$ 0.28	& 4.72 $\pm$ 0.29	& 4.52 $\pm$ 0.38	& 4.53 $\pm$ 0.22	& 5.36 $\pm$ 0.59	\\ \relax
 \vspace{-1pt} [O III] $\lambda$5007	& 15.55 $\pm$ 2.01	& 11.44 $\pm$ 1.58	& 13.05 $\pm$ 5.58	& 14.06 $\pm$ 2.14	& 13.49 $\pm$ 0.81	& 13.53 $\pm$ 0.86	& 13.93 $\pm$ 0.79	& 13.53 $\pm$ 1.11	& 13.17 $\pm$ 0.65	& 16.07 $\pm$ 1.58	\\ \relax
 \vspace{-1pt} [O I] $\lambda$6300	& ... $\pm$ ...	& ... $\pm$ ...	& ... $\pm$ ...	& 1.48 $\pm$ 0.21	& 0.38 $\pm$ 0.02	& 0.17 $\pm$ 0.01	& 0.18 $\pm$ 0.01	& ... $\pm$ ...	& ... $\pm$ ...	& ... $\pm$ ...	\\ \relax
 \vspace{-1pt} [S III] $\lambda$6312	& ... $\pm$ ...	& ... $\pm$ ...	& ... $\pm$ ...	& 0.15 $\pm$ 0.03	& 0.08 $\pm$ 0.01	& 0.05 $\pm$ 0.01	& 0.06 $\pm$ 0.01	& ... $\pm$ ...	& ... $\pm$ ...	& ... $\pm$ ...	\\ \relax
 \vspace{-1pt} [O I] $\lambda$6363	& ... $\pm$ ...	& ... $\pm$ ...	& ... $\pm$ ...	& 0.47 $\pm$ 0.07	& 0.14 $\pm$ 0.01	& 0.07 $\pm$ 0.00	& 0.05 $\pm$ 0.00	& ... $\pm$ ...	& ... $\pm$ ...	& ... $\pm$ ...	\\ \relax
 \vspace{-1pt} [Fe X] $\lambda$6375	& ... $\pm$ ...	& ... $\pm$ ...	& ... $\pm$ ...	& 1.20 $\pm$ 0.17	& 0.39 $\pm$ 0.03	& 0.16 $\pm$ 0.01	& 0.06 $\pm$ 0.00	& ... $\pm$ ...	& ... $\pm$ ...	& ... $\pm$ ...	\\ \relax
 \vspace{-1pt} [N II] $\lambda$6548	& 1.31 $\pm$ 0.20	& 0.57 $\pm$ 0.09	& 2.16 $\pm$ 0.84	& 3.22 $\pm$ 0.42	& 0.91 $\pm$ 0.05	& 0.42 $\pm$ 0.02	& 0.47 $\pm$ 0.02	& 0.57 $\pm$ 0.05	& 0.63 $\pm$ 0.05	& 0.21 $\pm$ 0.17	\\ \relax
 \vspace{-1pt} H$\alpha$ $\lambda$6563	& 5.69 $\pm$ 0.75	& 2.85 $\pm$ 0.31	& 7.09 $\pm$ 2.70	& 13.70 $\pm$ 1.81	& 4.51 $\pm$ 0.23	& 2.53 $\pm$ 0.14	& 2.69 $\pm$ 0.14	& 2.32 $\pm$ 0.15	& 2.26 $\pm$ 0.12	& 2.57 $\pm$ 0.21	\\ \relax
 \vspace{-1pt} [N II] $\lambda$6584	& 3.79 $\pm$ 0.49	& 2.35 $\pm$ 0.26	& 5.06 $\pm$ 1.94	& 8.71 $\pm$ 1.17	& 2.35 $\pm$ 0.15	& 1.25 $\pm$ 0.08	& 1.27 $\pm$ 0.06	& 1.49 $\pm$ 0.11	& 1.56 $\pm$ 0.11	& 1.71 $\pm$ 0.15	\\ \relax
 \vspace{-1pt} [S II] $\lambda$6716	& ... $\pm$ ...	& ... $\pm$ ...	& 1.25 $\pm$ 0.50	& 1.70 $\pm$ 0.24	& 0.45 $\pm$ 0.03	& 0.22 $\pm$ 0.01	& 0.27 $\pm$ 0.02	& 0.33 $\pm$ 0.03	& 0.49 $\pm$ 0.03	& 0.53 $\pm$ 0.06	\\ \relax
 \vspace{-1pt} [S II] $\lambda$6731	& ... $\pm$ ...	& ... $\pm$ ...	& 1.21 $\pm$ 0.48	& 2.18 $\pm$ 0.30	& 0.59 $\pm$ 0.04	& 0.32 $\pm$ 0.02	& 0.35 $\pm$ 0.02	& 0.40 $\pm$ 0.04	& 0.59 $\pm$ 0.05	& 0.53 $\pm$ 0.12	\\ \hline
H$\beta$ Flux$\times$10$^{-15}$ &	0.119$\pm$0.015 & 	0.234$\pm$0.025 & 	0.113$\pm$0.043 & 	0.447$\pm$0.058 & 	3.172$\pm$0.152 & 	3.350$\pm$0.166 & 	2.101$\pm$0.078 & 	1.205$\pm$0.062 & 	0.534$\pm$0.020 & 	0.154$\pm$0.012
\enddata
\tablecomments{Observed emission line ratios relative to H$\beta$ at each spatial distance from the nucleus in arcseconds, with positive numbers corresponding to the NW direction and negative to the SE. Emission lines were fit using widths and centroids calculated from fits to H$\alpha$ and error bars are the quadrature sum of the fractional flux uncertainty in H$\beta$ and each respective line. Rows marked with ``...	$\pm$	...'' represent nondetections. Wavelengths are approximate vacuum values used as markers in Cloudy. $^a$ blend of He I 3889 + [Fe V] 3892 + H8. $^b$ blend of [Ne~III] 3968 + H$\epsilon$ 3970.}
\vspace{-2pt}
\end{deluxetable*}

\setlength{\tabcolsep}{0.01in} 
\tabletypesize{\scriptsize}
\begin{deluxetable*}{l|c|c|c|c|c|c|c|c|c|c|}
\vspace{-26pt}
\tablenum{2}
\tablecaption{Reddening-Corrected HST STIS Emission Line Ratios \vspace{-5pt}}
\tablehead{
\colhead{Emission Line} & \colhead{+$0\farcs46$} & \colhead{+$0\farcs36$} & \colhead{+$0\farcs25$$^\dagger$} & \colhead{+$0\farcs15$$^\dagger$} & \colhead{+$0\farcs05$} & \colhead{--$0\farcs05$} & \colhead{--$0\farcs15$} & \colhead{--$0\farcs25$} & \colhead{--$0\farcs36$} & \colhead{--$0\farcs46$} \vspace{-3pt}}
\startdata
 \vspace{-1pt} O III $\lambda$3133	& ... $\pm$ ...	& ... $\pm$ ...	& ... $\pm$ ...	& ... $\pm$ ...	& 0.46 $\pm$ 0.07	& 0.25 $\pm$ 0.06	& 0.28 $\pm$ 0.04	& 0.20 $\pm$ 0.08	& ... $\pm$ ...	& ... $\pm$ ...	\\ \relax
 \vspace{-1pt} He II $\lambda$3203	& ... $\pm$ ...	& ... $\pm$ ...	& ... $\pm$ ...	& ... $\pm$ ...	& 0.21 $\pm$ 0.02	& 0.15 $\pm$ 0.03	& 0.22 $\pm$ 0.04	& 0.10 $\pm$ 0.06	& ... $\pm$ ...	& ... $\pm$ ...	\\ \relax
 \vspace{-1pt} [Ne V] $\lambda$3346	& ... $\pm$ ...	& ... $\pm$ ...	& ... $\pm$ ...	& ... $\pm$ ...	& 1.07 $\pm$ 0.08	& 0.64 $\pm$ 0.11	& 0.81 $\pm$ 0.09	& 0.40 $\pm$ 0.11	& ... $\pm$ ...	& ... $\pm$ ...	\\ \relax
 \vspace{-1pt} [Ne V] $\lambda$3426	& 3.93 $\pm$ 0.63	& 2.44 $\pm$ 0.83	& ... $\pm$ ...	& 4.96 $\pm$ 1.00	& 3.01 $\pm$ 0.22	& 1.79 $\pm$ 0.29	& 1.96 $\pm$ 0.25	& 1.22 $\pm$ 0.31	& 0.87 $\pm$ 0.24	& ... $\pm$ ...	\\ \relax
 \vspace{-1pt} [Fe VII] $\lambda$3586	& ... $\pm$ ...	& ... $\pm$ ...	& ... $\pm$ ...	& ... $\pm$ ...	& 0.24 $\pm$ 0.02	& 0.17 $\pm$ 0.04	& 0.15 $\pm$ 0.02	& ... $\pm$ ...	& ... $\pm$ ...	& ... $\pm$ ...	\\ \relax
 \vspace{-1pt} [O II] $\lambda$3727	& 2.97 $\pm$ 0.42	& 1.55 $\pm$ 0.24	& 4.09 $\pm$ 3.20	& 3.55 $\pm$ 0.54	& 1.21 $\pm$ 0.07	& 0.68 $\pm$ 0.09	& 0.87 $\pm$ 0.07	& 1.11 $\pm$ 0.21	& 1.53 $\pm$ 0.32	& 1.85 $\pm$ 0.40	\\ \relax
 \vspace{-1pt} [Fe VII] $\lambda$3759	& ... $\pm$ ...	& ... $\pm$ ...	& ... $\pm$ ...	& 0.69 $\pm$ 0.10	& 0.50 $\pm$ 0.04	& 0.28 $\pm$ 0.04	& 0.18 $\pm$ 0.01	& 0.06 $\pm$ 0.02	& ... $\pm$ ...	& ... $\pm$ ...	\\ \relax
 \vspace{-1pt} [Ne III] $\lambda$3869	& 2.28 $\pm$ 0.36	& 0.90 $\pm$ 0.21	& ... $\pm$ ...	& 2.44 $\pm$ 0.33	& 1.47 $\pm$ 0.11	& 0.85 $\pm$ 0.10	& 1.06 $\pm$ 0.11	& 1.02 $\pm$ 0.17	& 0.94 $\pm$ 0.17	& 1.50 $\pm$ 0.22	\\ \relax
 \vspace{-1pt} He I $\lambda$3889$^a$	& ... $\pm$ ...	& ... $\pm$ ...	& ... $\pm$ ...	& 0.46 $\pm$ 0.09	& 0.30 $\pm$ 0.02	& 0.16 $\pm$ 0.02	& 0.18 $\pm$ 0.02	& 0.10 $\pm$ 0.02	& 0.24 $\pm$ 0.05	& ... $\pm$ ...	\\ \relax
 \vspace{-1pt} [Ne III] $\lambda$3969$^b$ & ... $\pm$ ...	& ... $\pm$ ...	& ... $\pm$ ...	& 0.64 $\pm$ 0.10	& 0.68 $\pm$ 0.05	& 0.37 $\pm$ 0.04	& 0.44 $\pm$ 0.06	& 0.38 $\pm$ 0.07	& 0.43 $\pm$ 0.10	& ... $\pm$ ...	\\ \relax
 \vspace{-1pt} [S II] $\lambda$4074	& ... $\pm$ ...	& ... $\pm$ ...	& ... $\pm$ ...	& ... $\pm$ ...	& 0.23 $\pm$ 0.05	& 0.10 $\pm$ 0.02	& 0.08 $\pm$ 0.03	& ... $\pm$ ...	& ... $\pm$ ...	& ... $\pm$ ...	\\ \relax
 \vspace{-1pt} H$\delta$ $\lambda$4102	& ... $\pm$ ...	& ... $\pm$ ...	& ... $\pm$ ...	& ... $\pm$ ...	& 0.33 $\pm$ 0.03	& 0.25 $\pm$ 0.03	& 0.25 $\pm$ 0.02	& 0.22 $\pm$ 0.04	& ... $\pm$ ...	& ... $\pm$ ...	\\ \relax
 \vspace{-1pt} H$\gamma$ $\lambda$4340	& ... $\pm$ ...	& ... $\pm$ ...	& 0.54 $\pm$ 0.36	& 0.86 $\pm$ 0.12	& 0.62 $\pm$ 0.03	& 0.45 $\pm$ 0.04	& 0.48 $\pm$ 0.05	& 0.42 $\pm$ 0.05	& 0.42 $\pm$ 0.07	& 0.64 $\pm$ 0.10	\\ \relax
 \vspace{-1pt} [O III] $\lambda$4363	& ... $\pm$ ...	& ... $\pm$ ...	& ... $\pm$ ...	& 0.40 $\pm$ 0.08	& 0.32 $\pm$ 0.02	& 0.22 $\pm$ 0.02	& 0.22 $\pm$ 0.03	& 0.19 $\pm$ 0.03	& ... $\pm$ ...	& ... $\pm$ ...	\\ \relax
 \vspace{-1pt} He II $\lambda$4686	& ... $\pm$ ...	& ... $\pm$ ...	& ... $\pm$ ...	& 0.44 $\pm$ 0.07	& 0.49 $\pm$ 0.03	& 0.50 $\pm$ 0.03	& 0.51 $\pm$ 0.04	& 0.46 $\pm$ 0.03	& 0.40 $\pm$ 0.03	& ... $\pm$ ...	\\ \relax
 \vspace{-1pt} [Ar IV] $\lambda$4711	& ... $\pm$ ...	& ... $\pm$ ...	& ... $\pm$ ...	& ... $\pm$ ...	& 0.07 $\pm$ 0.01	& 0.11 $\pm$ 0.01	& 0.12 $\pm$ 0.02	& 0.10 $\pm$ 0.01	& ... $\pm$ ...	& ... $\pm$ ...	\\ \relax
 \vspace{-1pt} [Ar IV] $\lambda$4740	& ... $\pm$ ...	& ... $\pm$ ...	& ... $\pm$ ...	& ... $\pm$ ...	& 0.12 $\pm$ 0.01	& 0.11 $\pm$ 0.01	& 0.12 $\pm$ 0.02	& 0.07 $\pm$ 0.01	& ... $\pm$ ...	& ... $\pm$ ...	\\ \relax
 \vspace{-1pt} H$\beta$ $\lambda$4861	& 1.00 $\pm$ 0.13	& 1.00 $\pm$ 0.11	& 1.00 $\pm$ 0.38	& 1.00 $\pm$ 0.13	& 1.00 $\pm$ 0.05	& 1.00 $\pm$ 0.05	& 1.00 $\pm$ 0.04	& 1.00 $\pm$ 0.05	& 1.00 $\pm$ 0.04	& 1.00 $\pm$ 0.08	\\ \relax
 \vspace{-1pt} [O III] $\lambda$4959	& 5.02 $\pm$ 0.69	& 4.00 $\pm$ 0.52	& 4.84 $\pm$ 2.12	& 4.57 $\pm$ 0.73	& 4.32 $\pm$ 0.27	& 4.40 $\pm$ 0.28	& 4.72 $\pm$ 0.29	& 4.52 $\pm$ 0.39	& 4.53 $\pm$ 0.23	& 5.36 $\pm$ 0.59	\\ \relax
 \vspace{-1pt} [O III] $\lambda$5007	& 14.53 $\pm$ 2.02	& 11.44 $\pm$ 1.59	& 11.93 $\pm$ 5.60	& 12.02 $\pm$ 2.15	& 12.90 $\pm$ 0.81	& 13.53 $\pm$ 0.88	& 13.93 $\pm$ 0.80	& 13.53 $\pm$ 1.15	& 13.17 $\pm$ 0.73	& 16.07 $\pm$ 1.59	\\ \relax
 \vspace{-1pt} [O I] $\lambda$6300	& ... $\pm$ ...	& ... $\pm$ ...	& ... $\pm$ ...	& 0.38 $\pm$ 0.21	& 0.26 $\pm$ 0.02	& 0.17 $\pm$ 0.02	& 0.18 $\pm$ 0.02	& ... $\pm$ ...	& ... $\pm$ ...	& ... $\pm$ ...	\\ \relax
 \vspace{-1pt} [S III] $\lambda$6312	& ... $\pm$ ...	& ... $\pm$ ...	& ... $\pm$ ...	& 0.04 $\pm$ 0.03	& 0.05 $\pm$ 0.01	& 0.05 $\pm$ 0.01	& 0.06 $\pm$ 0.01	& ... $\pm$ ...	& ... $\pm$ ...	& ... $\pm$ ...	\\ \relax
 \vspace{-1pt} [O I] $\lambda$6363	& ... $\pm$ ...	& ... $\pm$ ...	& ... $\pm$ ...	& 0.12 $\pm$ 0.07	& 0.09 $\pm$ 0.01	& 0.07 $\pm$ 0.01	& 0.05 $\pm$ 0.00	& ... $\pm$ ...	& ... $\pm$ ...	& ... $\pm$ ...	\\ \relax
 \vspace{-1pt} [Fe X] $\lambda$6375	& ... $\pm$ ...	& ... $\pm$ ...	& ... $\pm$ ...	& 0.29 $\pm$ 0.17	& 0.26 $\pm$ 0.03	& 0.16 $\pm$ 0.02	& 0.06 $\pm$ 0.00	& ... $\pm$ ...	& ... $\pm$ ...	& ... $\pm$ ...	\\ \relax
 \vspace{-1pt} [N II] $\lambda$6548	& 0.67 $\pm$ 0.22	& 0.57 $\pm$ 0.11	& 0.89 $\pm$ 0.92	& 0.69 $\pm$ 0.43	& 0.59 $\pm$ 0.06	& 0.42 $\pm$ 0.06	& 0.47 $\pm$ 0.04	& 0.57 $\pm$ 0.14	& 0.63 $\pm$ 0.17	& ... $\pm$ ...	\\ \relax
 \vspace{-1pt} H$\alpha$ $\lambda$6563	& 2.90 $\pm$ 0.84	& 2.85 $\pm$ 0.44	& 2.91 $\pm$ 2.95	& 2.91 $\pm$ 1.85	& 2.90 $\pm$ 0.27	& 2.53 $\pm$ 0.38	& 2.69 $\pm$ 0.25	& 2.32 $\pm$ 0.55	& 2.26 $\pm$ 0.58	& 2.57 $\pm$ 0.38	\\ \relax
 \vspace{-1pt} [N II] $\lambda$6584	& 1.92 $\pm$ 0.55	& 2.35 $\pm$ 0.37	& 2.06 $\pm$ 2.12	& 1.82 $\pm$ 1.20	& 1.51 $\pm$ 0.17	& 1.25 $\pm$ 0.19	& 1.27 $\pm$ 0.11	& 1.49 $\pm$ 0.36	& 1.56 $\pm$ 0.41	& 1.71 $\pm$ 0.26	\\ \relax
 \vspace{-1pt} [S II] $\lambda$6716	& ... $\pm$ ...	& ... $\pm$ ...	& 0.48 $\pm$ 0.54	& 0.33 $\pm$ 0.24	& 0.28 $\pm$ 0.03	& 0.22 $\pm$ 0.03	& 0.27 $\pm$ 0.03	& 0.33 $\pm$ 0.09	& 0.49 $\pm$ 0.14	& 0.53 $\pm$ 0.09	\\ \relax
 \vspace{-1pt} [S II] $\lambda$6731	& ... $\pm$ ...	& ... $\pm$ ...	& 0.47 $\pm$ 0.52	& 0.41 $\pm$ 0.31	& 0.37 $\pm$ 0.04	& 0.32 $\pm$ 0.05	& 0.35 $\pm$ 0.03	& 0.40 $\pm$ 0.11	& 0.59 $\pm$ 0.17	& 0.53 $\pm$ 0.14	\\ \hline \relax
E(B-V) & 0.61 $\pm$ 0.12 & 	0.00 $\pm$ 0.10 & 	0.70 $\pm$ 0.80 & 	0.88 $\pm$ 0.15 & 	0.40 $\pm$ 0.05 & 	0.00 $\pm$ 0.13 & 	0.00 $\pm$ 0.07 & 	0.00 $\pm$ 0.21 & 	0.00 $\pm$ 0.23 & 	0.00 $\pm$ 0.11
\enddata
\tablecomments{Same as in Table 1 with line ratios corrected for galactic extinction using a galactic reddening curve \citep{savage1979}. The H$\alpha$/H$\beta$ ratios were fixed at 2.90 and negative E(B-V) values were set to zero. Error bars are the quadrature sum of the fractional flux uncertainty in H$\beta$ and each respective line along with the reddening uncertainty. ``$\dagger$'' indicates a row with secondary correction to the reddening (\S 3.3); original values for +$0\farcs15$ and +$0\farcs25$ were E(B-V) = 1.42 $\pm$ 0.12 and 0.82 $\pm$ 0.37, respectively.}
\vspace{-2pt}
\end{deluxetable*}

\subsection{Emission Line Ratios}

We used the integrated line fluxes to calculate emission line ratios relative to the standard hydrogen Balmer (H$\beta$) emission line at each position along the NLR (Table 1). Near the nucleus, lines with a wide range of ionization potentials (IPs) are seen, from neutral [O I] to [Fe X] with an IP of 234 eV. The number and fluxes of detectable lines decreases steadily with increasing distance from the nucleus until the bright arcs of emission are encountered at $\sim$ $1\farcs8$. Fits were made to all visible emission lines across the NLR, but only emission within $\sim$ $1\farcs7$ of the nucleus displays kinematic signatures of outflow and are included in our tables and models presented here. All measurements are available by request to M.R.

Before interpreting the line ratios, we applied a reddening correction using a standard galactic reddening curve \citep{savage1979} and color excesses calculated from the observed H$\alpha$/H$\beta$ ratios, assuming an intrinsic recombination value of 2.90 \citep{osterbrock2006}. The extinction was calculated from
\begin{equation}
E(B-V)\equiv-\frac{2.5log\left(\frac{F_{o}}{F_{i}}\right)}{R_\lambda}=\frac{2.5log\left(\frac{(\mathrm{H}\alpha/\mathrm{H}\beta)_{i}}{(\mathrm{H}\alpha/\mathrm{H}\beta)_{o}}\right)}{R_{\mathrm{H}\alpha}-R_{\mathrm{H}\beta}},
\end{equation}
\noindent
where $E(B-V)$ is the color excess, $A_\lambda$ is the extinction in magnitudes, $R_\lambda$ is the reddening value at a particular wavelength, and $F_o$ and $F_i$ are the observed and intrinsic fluxes, respectively. The flux ratios can be expanded to the intrinsic and observed H$\alpha$/H$\beta$ ratios, and the galactic reddening values are $R_{\mathrm{H}\alpha} \approx$ 2.497 and $R_{\mathrm{H}\beta} \approx$ 3.687. Corrected line ratios relative to H$\beta$ are then given by
\begin{equation}
H_{int} = H_{obs} \cdot 10~exp[0.4 \cdot E(B-V)\cdot (R_j - R_{\mathrm{H}\beta})],
\end{equation}
\noindent
where $H_{\mathrm{int}}$ and $H_{\mathrm{obs}}$ are the intrinsic and observed emission line ratios and $R_j$ is the reddening value at the wavelength of the emission line being corrected. Uncertainties were propagated from the H$\alpha$ and H$\beta$ fit errors.

\begin{figure*}
\vspace{-6pt}
\centering
\subfigure{
\includegraphics[scale=0.46]{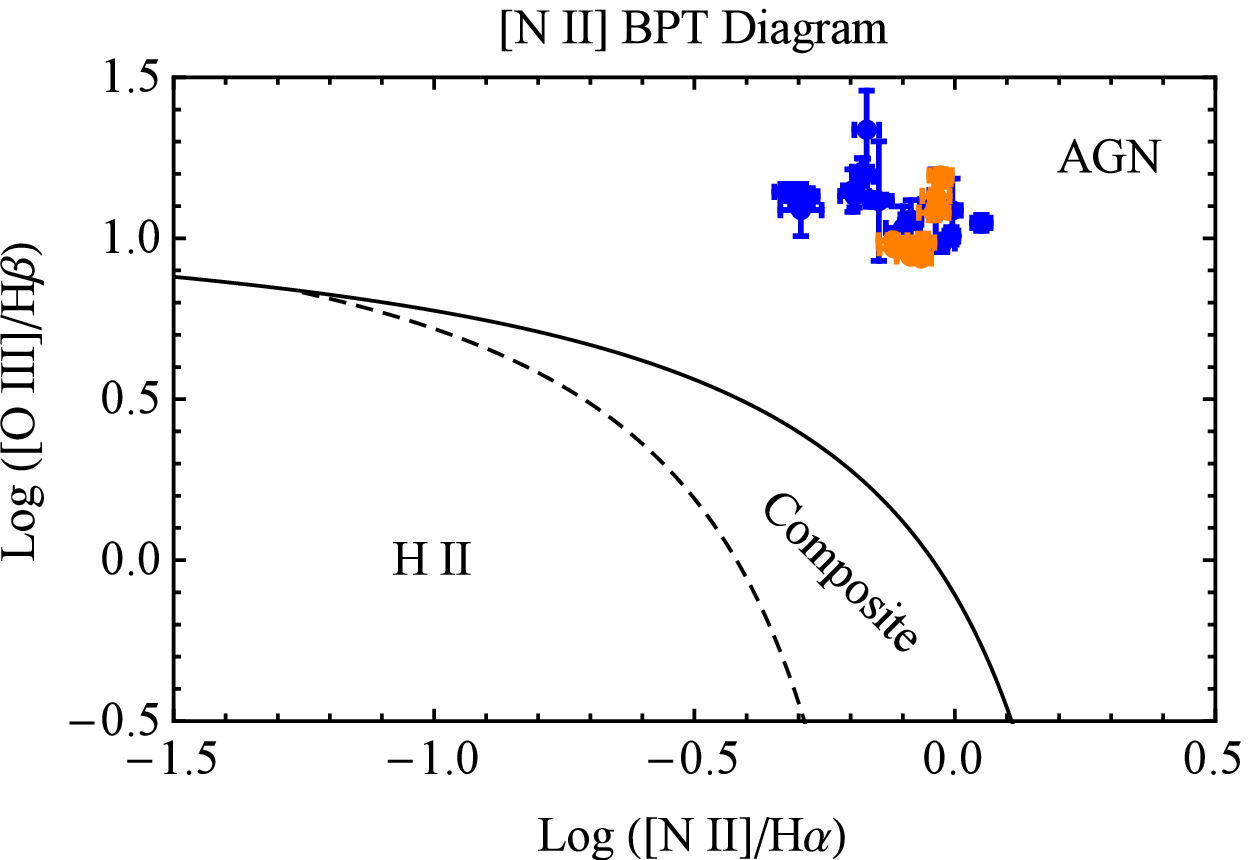}}
\subfigure{
\includegraphics[scale=0.46]{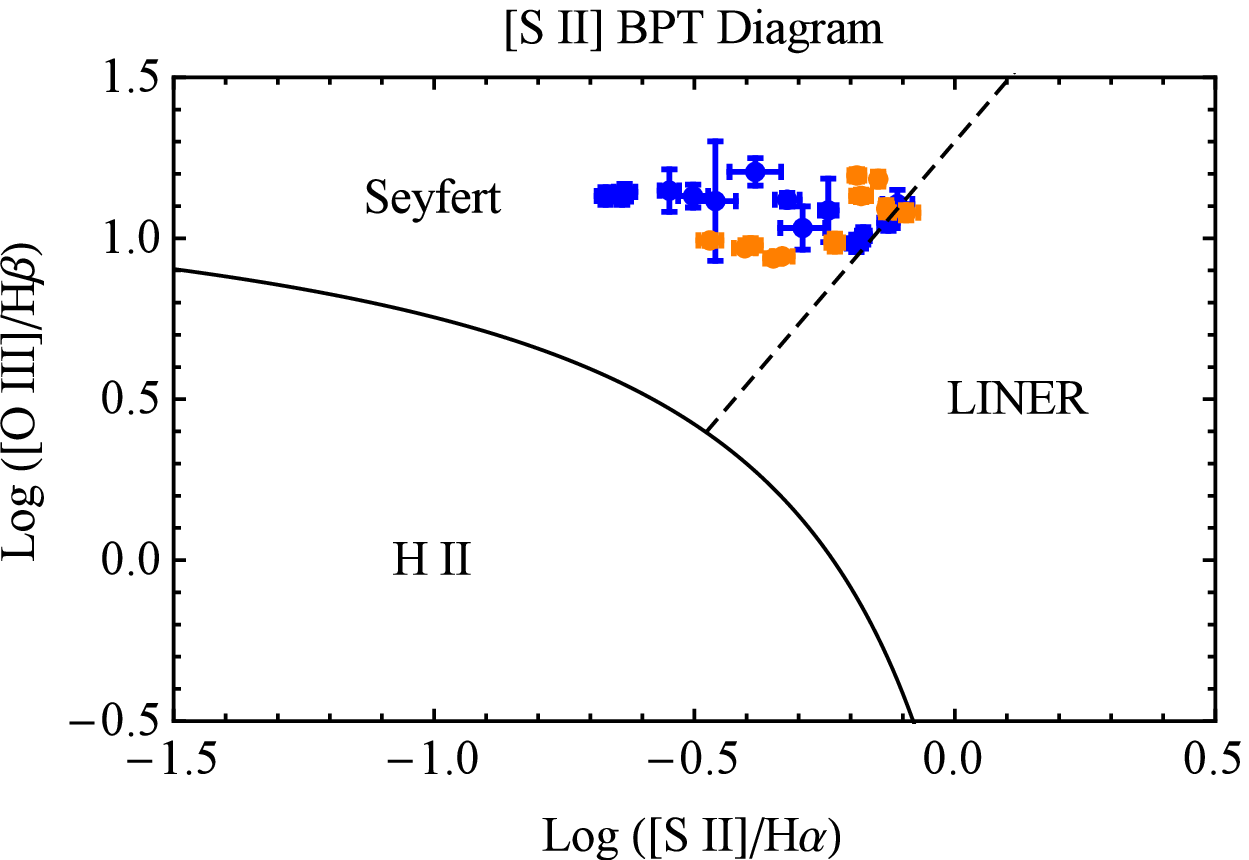}}
\subfigure{
\includegraphics[scale=0.46]{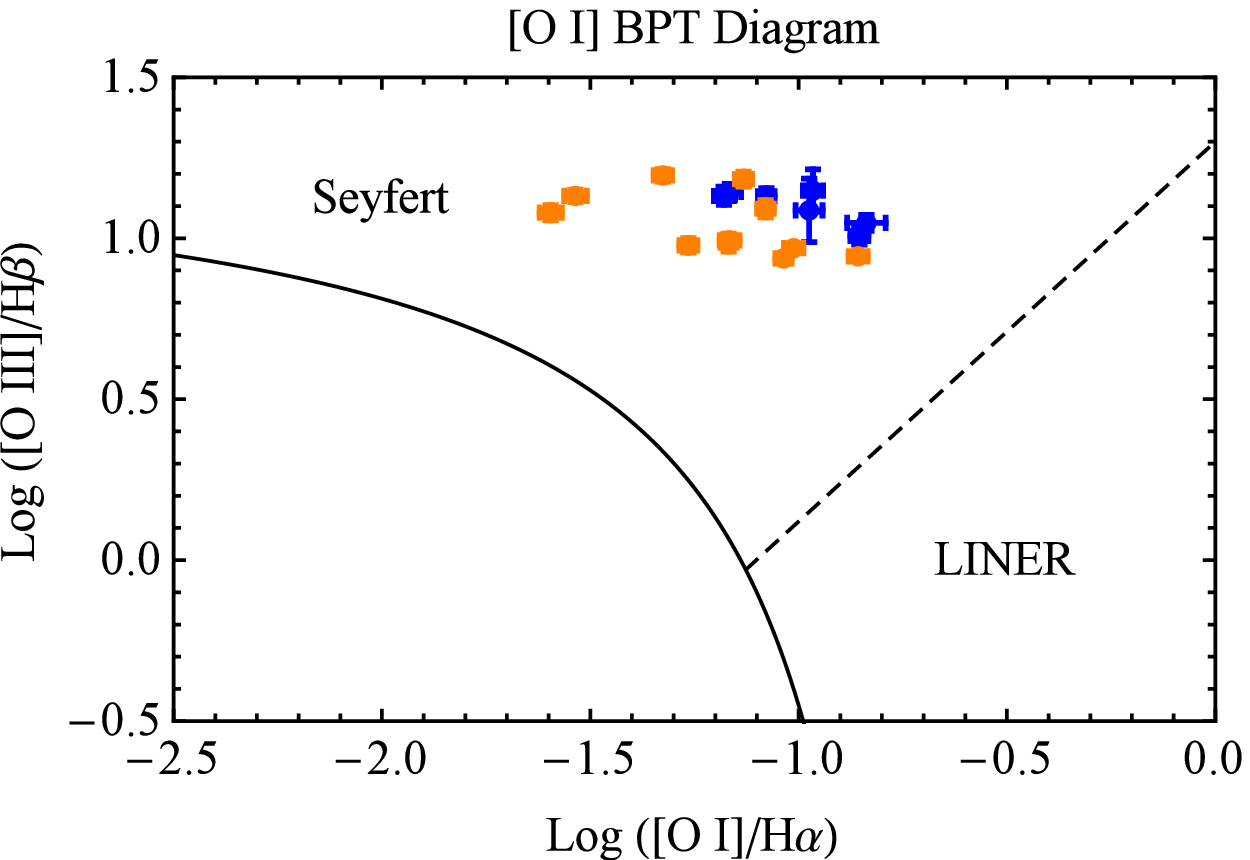}}
\vspace{-8pt}
\caption{BPT ionization diagrams for [N II], [S II], and [O I] for all measurements within $2\arcsec$ of the nucleus. HST STIS and APO DIS points are in blue and orange circles, respectively. Further extended emission also shows AGN ionization \citep{fischer2017}. The demarcation lines for distinguishing ionization mechanisms are from \cite{kewley2001, kewley2006, kauffmann2003}.}
\label{bpts}
\vspace{-2pt}
\end{figure*}

Several extractions had H$\alpha$/H$\beta$ $\leq$ 2.90 within errors, and for these locations we assumed no reddening. At two locations, $0\farcs15$ and $0\farcs25$ NW of the nucleus, this procedure derived large extinctions, and the resulting corrected line ratios were unphysically large in the blue portion of the spectrum. Our investigations concluded that residual sub-pixel offset in the spatial direction between the G430L and G750M spectra, combined with an extremely steep flux gradient at those locations, was the most plausible explanation. Another possibility is preferential absorption of H$\beta$, which can be enhanced by a young-intermediate age ($10 - 100$ Myr) stellar population. Stellar population modeling of Mrk 573 is consistent with an old nuclear stellar population \citep{alonsoherrero1998, gonzalezdelgado2001, raimann2003, ramosalmeida2009a}, although some studies have found evidence of recent star formation within the inner few hundred pc \citep{riffel2006, riffel2007, diniz2017}. To quantify any absorption, we measured the equivalent width (EW) of H$\beta$ by comparing the integrated line flux to the surrounding continuum level. For most nuclear extractions ($r<0\farcs5$) we measured EWs $>30~\AA$. For positions $+0\farcs15$ through $+0\farcs46$ we found EWs $\approx 6-12~\AA$, indicating possible absorption at the $\leq$ 10\% level. These studies and measurements, together with a lack of visible absorption in the higher order Balmer lines, suggests that stellar absorption is a minor secondary affect. For positions $+0\farcs15$ and $+0\farcs25$ we used the H$\alpha$/H$\gamma$ ratios to calculate the extinction values for the blue spectrum to obtain more physically realistic line ratios to compare with our models. The shortest wavelength lines such as [O~II] and [Ne~V] at these two positions retain a minor residual over correction.

\subsection{Emission Line Diagnostics}

The dereddened line ratios allow us to limit our model parameter ranges by constraining the abundances, ionization, temperature, and density of the gas at each location. We ultimately used these diagnostics to extrapolate a density law to larger distances where there are not sufficient emission lines to create detailed photoionization models.

\subsubsection{Ionization}

To properly model the NLR, we confirmed that our HST and APO spectra, which have vastly different spatial resolutions and slit widths, both sample completely AGN ionized gas using Baldwin--Phillips--Terlevich (BPT) diagrams shown in Figure \ref{bpts} \citep{baldwin1981, veilleux1987}. These diagrams exploit ratios of emission lines with small wavelength separations to avoid the effects of extinction. Specifically, the ratio of [O~III] $\lambda$5007/H$\beta$ $\lambda$4861 compared to [N~II] $\lambda$6584, [S~II] $\lambda$$\lambda$6716, 6731, and [O~I] $\lambda$6300, relative to H$\alpha$ $\lambda$6563. 

These tests indicate that all of the NLR gas participating in the outflow is ionized by the central AGN, in agreement with previous findings \citep{unger1987, kraemer2009, schlesinger2009, fischer2017}. All of the line ratios fall in the AGN/Seyfert regions of the diagrams using the explored separation criteria (\citealp{kewley2001, kewley2006, kauffmann2003}, see also \citealp{stasinska2006, schawinski2007, kewley2013a, kewley2013b, melendez2014, bar2017}). [S~II] arises from low ionization gas, and we interpret the points near the AGN/LINER border as emission from gas that is subject to a partially absorbed, or ``filtered'', ionizing spectrum, as discussed in \S5.1.

In addition, He II $\lambda$4686/H$\beta$ $\lambda$4861 is useful for constraining the column density of the gas ($N_\mathrm{H}$ cm$^{-2}$). As radiation propagates through the NLR gas, both emission lines strengthen until He II ionizing photons ($E >$ 54.4 eV) are exhausted. As a result, He II/H$\beta$ is $\sim 1$ in optically thin (matter-bounded) gas and reduces to $\sim 0.2$ in optically thick (radiation-bounded) gas, with the exact values dependent on the SED. Intermediate values indicate a mixture of these cases and are representative of our ratios in Table 2.

\subsubsection{Abundances}

Elemental abundances play an important role in determining the heating and cooling balance within the gas and are determined by the true abundances and the fraction of each element trapped in dust grains. NLR abundances are typically solar or greater, but are observed to vary between objects over the range $Z \approx 0.6 - 1.8~Z_\odot$ \citep{storchibergmann1998b, hamann1999, nagao2006, dors2014, dors2015, castro2017}.

\begin{figure}
\centering
\subfigure{
\includegraphics[scale=0.71]{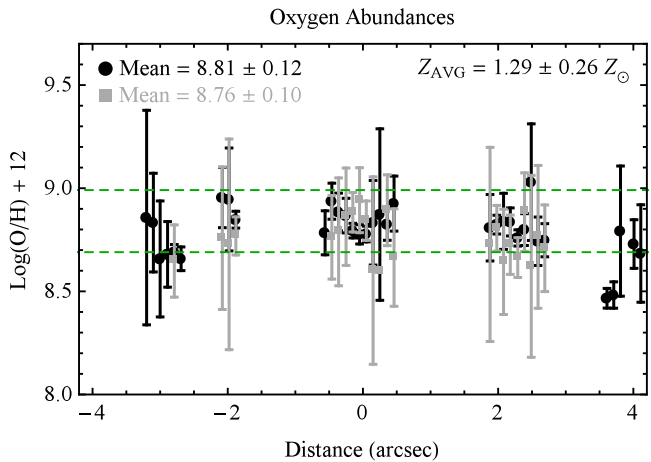}}
\caption{Oxygen abundance as a function of distance from the nucleus using the methods of \cite{storchibergmann1998b} and \cite{castro2017} in black circles and gray squares, respectively, for the HST STIS data. The lower and upper dotted green lines represent one and two times solar abundance values, respectively. SE is to the left and NW is to the right.}
\label{abundances}
\end{figure}

We determined the abundances of elements across the NLR of Mrk 573 by first finding the abundance of oxygen and then scaled other elements by that factor. We adopt solar abundances ($Z_\odot$) from \cite{asplund2009}, which lists oxygen as Log(O/H)+12 = 8.69. For Mrk 573 we determined the oxygen abundance using equations 2 and 4 from \cite{storchibergmann1998b} and \cite{castro2017}, respectively, which compare the ratios of strong emission lines in the spectra. These yield a mean oxygen abundance of Log(O/H)+12 = 8.78, or $Z = 1.29~\pm~0.26~Z_\odot$. The distributions are shown in Figure \ref{abundances}, with errors propagated from the individual uncertainties in the equations. This result neglects a minor density-dependent correction that would decrease points near the nucleus and increase outer points by $\leq$ 0.04 dex.

This abundance is in excellent agreement with the global NLR value found by \cite{dors2015} and \cite{castro2017} for Mrk 573. Other relationships in \cite{storchibergmann1998b} and \cite{dors2015} yielded higher and lower abundances, respectively, and were not included due to their sensitivities to the ionizing spectrum and temperature, as discussed in those papers.

From this result we adopted a NLR metallicity of $Z \approx 1.3~Z_\odot$ for all elements, with nitrogen scaled by $Z^2$ as it is enhanced by secondary nucleosynthesis processes \citep{hamann2002, nagao2006}. 

\subsubsection{Temperature}

Density sensitive line ratios that are useful for determining masses also exhibit a weak temperature dependence that must be accounted for to derive accurate densities \citep{osterbrock2006, draine2011}. The strongest indicator of the electron density ($n_e$ cm$^{-3}$) in our data is the [S~II] $\lambda \lambda$6716/6731 ratio. We employed the available temperature dependent ratio of [O~III] $\lambda \lambda$4363/5007 that may come from hotter gas and scaled it to derive a temperature of the [S~II] gas \citep{wilson1997}. The measured [O~III] ratios and photoionization model predictions are shown in Figure \ref{tempdensity}.

The temperature sensitive [O~III] line ratio only changes appreciably due to density affects for $n_e > 10^4$ cm$^{-3}$ as [O~III] $\lambda$5007 begins to be collisionally suppressed. For the APO DIS data the observed [S~II] $\lambda \lambda$6716/6731 ratios do not drop below 0.5, indicating $n_e < 10^4$ cm$^{-3}$ over any range of temperatures typically seen in NLRs, so we derived temperatures from [O~III] in the low density limit.

We calculated Cloudy \citep{ferland2013} photoionization models over a wide range of parameters and found the mean [S~II] temperature to be $\sim 0.18 - 0.25$ dex cooler than the mean [O~III] emitting gas. We adopted the upper end of this range for scaling because the [S~II] emissivity peaks deeper within a cloud as the gas begins to become neutral. The uncertainty in this scaling from model to model variation is $\sim$ 0.1 dex.

Using this procedure we determined the mean temperature of $\mathrm{T_{[O~III]}} = 13,500 \pm 650 \mathrm{~K}$ for our APO data (corresponding to an H$\gamma$ reddening-corrected flux ratio of $\lambda 4363/ \lambda 5007 = 0.0175\pm0.0017$) and $\mathrm{T_{[O~III]}} = 13,450 \pm 750 \mathrm{~K}$ for the HST data. These values are in excellent agreement with previous studies \citep{tsvetanov1992, spinoglio2000, schlesinger2009}. Adopting $\mathrm{T_{[S~II]}} = \mathrm{T_{[O~III]}} - 0.25 \mathrm{~dex}$ we find $\mathrm{T_{[S~II]}} = 7590 \pm 375 \mathrm{~K}$, with uncertainties calculated from the variation in the derived temperatures.

\subsubsection{Density}

\begin{figure*}
\vspace{-10pt}
\centering
\subfigure{
\includegraphics[scale=0.965]{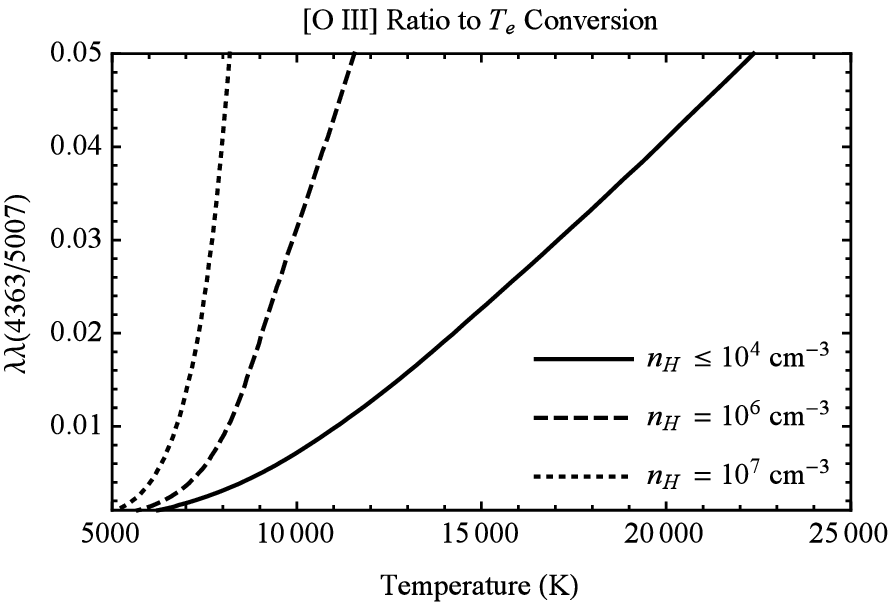}}
\vspace{-4pt}
\subfigure{
\includegraphics[scale=0.965]{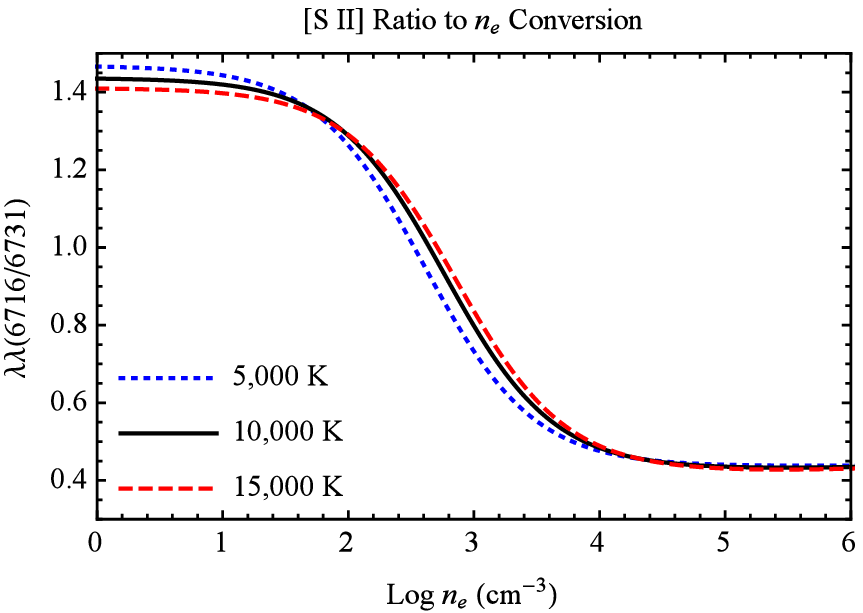}}
\subfigure{
\includegraphics[scale=0.965]{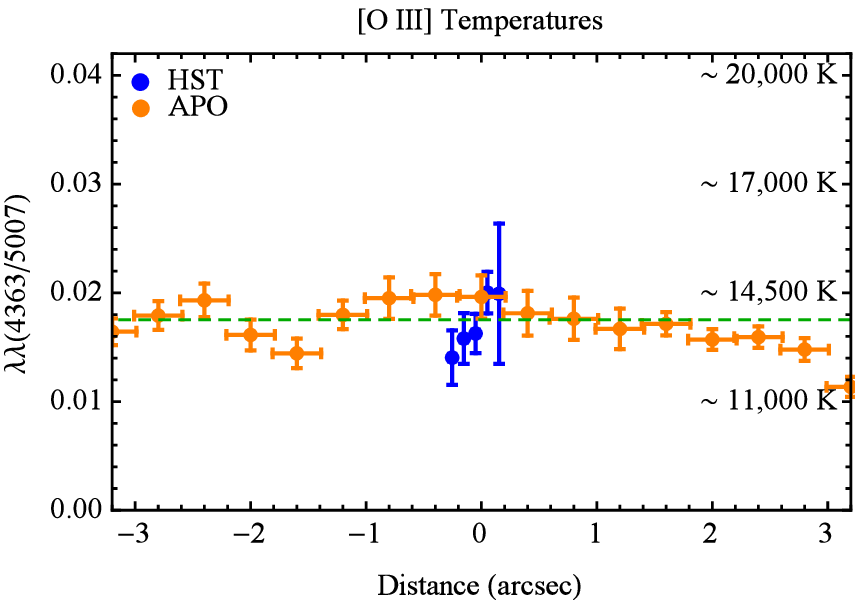}}
\subfigure{
\includegraphics[scale=0.965]{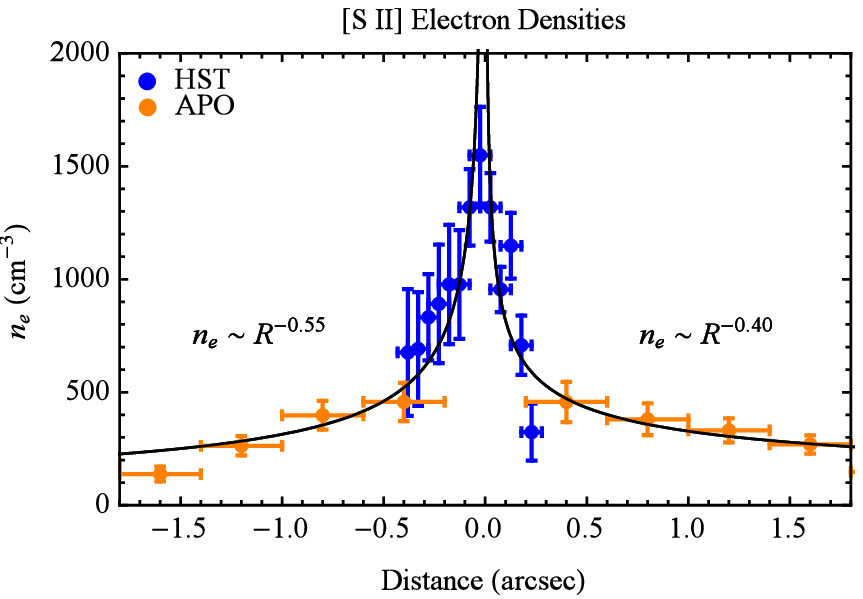}}
\vspace{-10pt}
\caption{Top left: The theoretical behavior of the [O~III] $\lambda \lambda$4363/5007 line ratio as a function of temperature for several densities calculated using Cloudy. Bottom left: The derived temperature of the [O~III] emitting gas for the HST STIS and APO DIS data sets in blue and orange circles, respectively. The mean of $\mathrm{T_{[O~III]}} = 13,500 \pm 650 \mathrm{~K}$ was calculated for all points within $\pm 1\farcs7$ of the nucleus. Top right: The theoretical conversion of the [S~II] $\lambda \lambda$6716/6731 ratios to electron density for several temperatures calculated using Cloudy. Bottom right: The derived [S~II] densities and power law fits with SE to the left and NW to the right.}
\label{tempdensity}
\end{figure*}

We derived the electron density ($n_e$ cm$^{-3}$) at each location using the observed [S~II] $\lambda \lambda$6716/6731 line ratios, aforementioned temperature, and photoionization model grids. Each ratio was matched to the closest grid value and corresponding density, and the results are shown in Figure \ref{tempdensity}. Errors were propagated from the original fit residuals, temperature uncertainties, and grid step size.

We fit independent power laws to the density profiles in either direction from the nucleus, with points beyond the outflow ($r > 1\farcs7$) not included in the fit. Interestingly, the density profiles have shallow power law indexes with $n_e \propto r^{-0.4} - r^{-0.6}$. This overall decreasing profile is consistent with the previous study by \cite{tsvetanov1992}; however, we derive a peak nuclear density that is 2--3$\times$ higher due to the lack of atmospheric smearing in the high spatial resolution HST spectra. This is a clear demonstration of the powerful selection effects that arise in blended ground based observations.

Finally, within the NLR some elements heavier than hydrogen will contribute more than one electron per nuclei, and the electron density will be higher than the hydrogen density. From models we adopt the conversion $n_\mathrm{H} = 0.85\times n_e$ \citep{crenshaw2015}.

\begin{figure}[H]
\centering
\subfigure{
\includegraphics[scale=0.56]{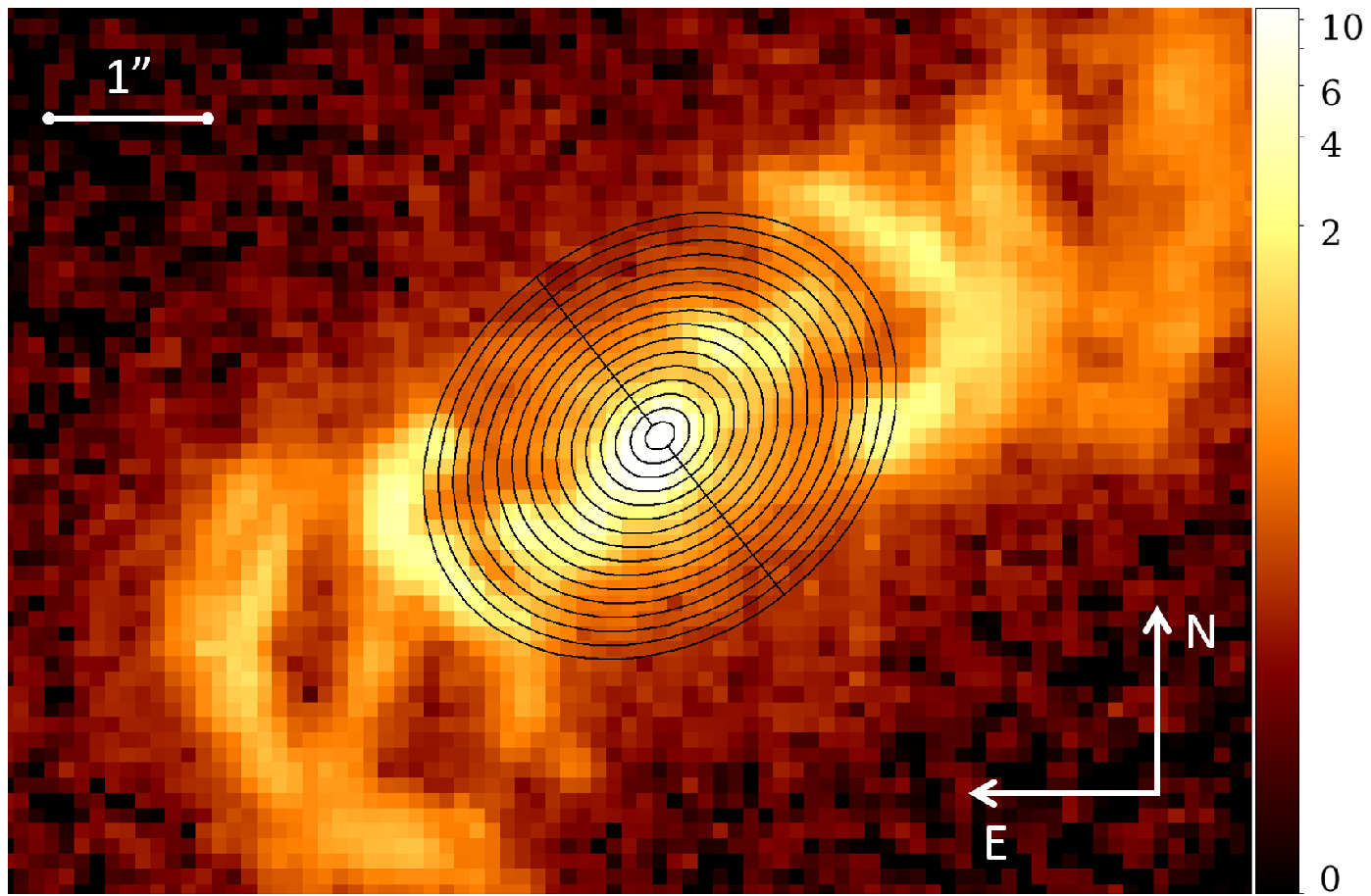}}
\subfigure{
\includegraphics[scale=0.71]{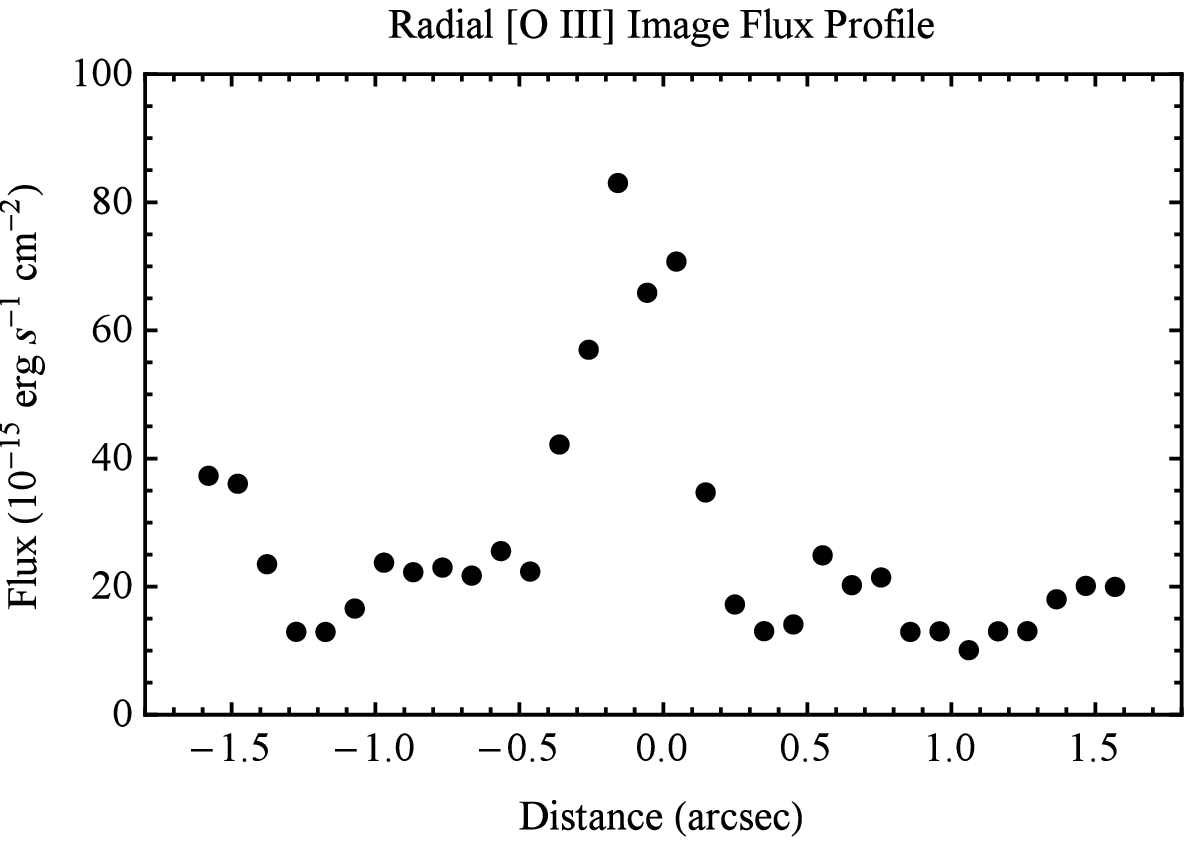}}
\caption{Top: A portion of the HST WFPC2 [O~III] image with overlaid elliptical semi-annuli representing rings of constant distance from the nucleus in black. The color bar gives fluxes in units of $10^{-15}$ erg s$^{-1}$ cm$^{-2}$. Bottom: The azimuthally summed [O~III] semi-annuli fluxes; SE is to the left and NW is to the right. Typical errors are smaller than the size of the points.}
\label{imaging}
\end{figure}

\section{Image Analysis}

To account for the NLR mass outside of our spectral slit observations, we employ [O~III] imaging and use the physical quantities derived from the spectra and models to convert [O~III] fluxes to mass at each point along the NLR. Here we improve on the methodology of \cite{crenshaw2015} by dividing the NLR in half, which is necessary due to the asymmetry of the velocity laws and NLR flux distribution on either side of the nucleus.

We determined the total [O~III] flux as a function of distance from the nucleus by analyzing an HST WFPC2 [O~III] image of Mrk 573 using the Elliptical Panda routine within the SAOImage DS9 image software \citep{joye2003}. We constructed a series of concentric semi-ellipses centered on the nucleus with spacings equal to the spatial sampling of our extracted HST spectra (2 pixels, or $0\farcs10156$). The ellipticity for each ring was calculated based on our adopted inclination of 38$\degr$ via $b/a = cos(i)$, where $a$ and $b$ are the major and minor axis lengths, respectively. A portion of the [O~III] image with overlaid semi-annuli and the resulting azimuthally summed flux profile are shown in Figure \ref{imaging}. Errors were calculated from line-free regions in the image, and we derived $\sigma \approx 2 \cdot 10^{-17}$ erg s$^{-1}$ cm$^{-2}$ pixel$^{-1}$, in agreement with that found by \cite{schmitt2003a}. The error in any given annulus was calculated as $\sqrt{N_{\mathrm{pix}}}\cdot \sigma$, where $N_{\mathrm{pix}}$ is the number of pixels in the annulus. Typical fractional errors are $< 1\%$.

\section{Photoionization Models}

To accurately convert the [O~III] image fluxes (Figure \ref{imaging}) to the amount of mass at each position in the NLR, we created photoionization models that match the physical conditions of the emitting clouds in our high spatial resolution HST STIS spectra. This is critical because the emissivity of the gas will depend on the physical conditions at each location, and detailed models are needed to derive a scale factor between [O~III] flux and mass. To model our dereddened line ratios (Table 2), we employed the Cloudy photoionization code version 13.04 and all hotfixes \citep{ferland2013}.

\subsection{Input Parameters}

To create a physically consistent model, Cloudy must be able to determine the number and energy distribution of photons striking the face of a cloud of known composition and geometry. The first of these is described by the ionization parameter ($U$), which is a ratio of the number of ionizing photons to the number of atoms at the face of the cloud, and is given by \citep[\S 13.6]{osterbrock2006}
\begin{equation}
U = \frac{1}{4 \pi r^2 n_H c} \int_{\nu_0}^{\infty} \frac{L_{\nu}}{h\nu} d\nu,
\end{equation}
where $r$ is the radial distance of the emitting cloud from the AGN, $n_\mathrm{H}$ is the hydrogen number density cm$^{-3}$, and $c$ is the speed of light. The integral represents the number of ionizing photons s$^{-1}$ emitted by the AGN, $Q(H)$, across the spectral energy distribution (SED). Specifically, $Q(H) = \int_{\nu_0}^{\infty} (L_{\nu}/h\nu) d\nu$, where $L_{\nu}$ is the luminosity of the AGN as a function of frequency, $h$ is Planck's constant, and $\nu_0 = 13.6 eV/h$ is the ionization potential of hydrogen \citep[\S 14.3]{osterbrock2006}. \footnote{Within the X-ray community the ionization parameter is frequently described by $\xi = L_i / n_\mathrm{H} r^2$, where $L_i$ is the radiation energy density between 1 and 1000 Rydbergs (13.6 eV--13.6 keV). A close approximation for typical Seyfert power law SEDs is Log($U$) = Log($\xi$) -- 1.5 \citep{crenshaw2012}.}

For our SED we adopted the broken power law of \cite{kraemer2009}, with $L_{\nu} \propto \nu^{\alpha}$ and $\alpha = -1.0$ for energies $< 13.6$ eV, $\alpha = -1.5$ for $13.6~eV \leq h\nu < 0.5$ keV, and $\alpha = -0.8$ for energies above $0.5$ keV, with low and high energy cutoffs at 1 eV and 100 keV, respectively. \cite{kraemer2009} determined $Q(H) \approx 6 \cdot 10^{54}$ photons s$^{-1}$, similar to earlier studies \citep{wilson1988}.

With this ionizing photon luminosity it is difficult to explain the presence of strong low ionization lines such as [S~II] at small distances from the nucleus that form at low ionization parameters $\sim$ log($U$) = --3. In order to maintain physical continuity in Equation 5 the resulting densities would be much higher than the critical density of [S~II], and these lines would be collisionally suppressed. Two possible explanations are that either the gas is high out of the NLR plane and only close to the nucleus in projection, or that some of the NLR gas is exposed to a heavily filtered continuum where much of the ionizing flux has been removed by a closer in absorber \citep{ferland1982, binette1996, alexander1999, collins2009, kraemer2009}. Our previous studies concluded that the NLR material is approximately coplanar with the host disk \citep{fischer2017}, and we adopt the later explanation that was successfully modeled by \cite{kraemer2009}. Using a filter with Log($U$) = --1.50 the best fitting column densities for the absorber were Log(N$_\mathrm{H}$) = 21.50--21.60 cm$^{-2}$, similar to \cite{kraemer2009}.

\begin{figure}
\centering
\subfigure
{\includegraphics[scale=0.99]{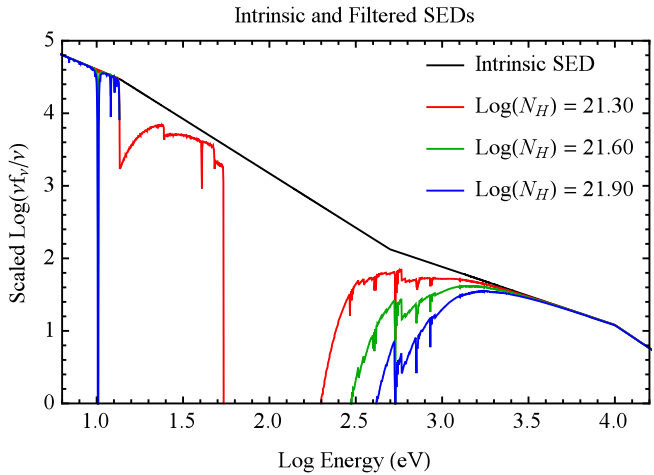}}
\vspace{-10pt}
\caption{A portion of the intrinsic and transmitted SEDs for different absorber column densities, with Log(N$_\mathrm{H}$) = $21.50 - 21.60$ best matching the observations.}
\label{filters}
\vspace{-10pt}
\end{figure}

To fully model the gas, we use multiple components with different ionization states, which we refer to as ``HIGH'', ``MED'' and ``LOW'' ION ionization components. The HIGH and MED ION components were directly ionized by the AGN SED, while the LOW ION component was calculated using the filtered SED. The intrinsic SED and filtered continua for several absorber column densities are shown in Figure \ref{filters}.

The composition of the gas is specified by the abundances, dust content, and corresponding depletions of certain heavy elements out of gas phase into dust grains. We adopt abundances of $\sim$ 1.3 Z$_{\odot}$, as determined in \S3.4.2. The inclusion of dust is important, as it removes coolants from the gas and contributes to photoelectric heating \citep{vanhoof2001, vanhoof2004, weingartner2001, draine2003, draine2011, weingartner2006, krugel2008}. In more highly ionized gas, strong iron emission indicates the gas is primarily dust free \citep{nagao2003}. In addition, we examined the {\it International Ultraviolet Explorer} (IUE) spectrum of Mrk 573 \citep{macalpine1988}. The IUE aperture encompasses a large portion of the NLR such that the spectrum is heavily weighted toward the denser gas that is emitting most efficiently. Nonetheless, \cite{macalpine1988} reports Ly$\alpha$/C~IV $\lambda$1549 = 8.0. Our dusty models predict this ratio to be $\sim$ 0.6 for a typical HIGH ION component, while for a dust free and optically thin model, it is $\sim$ 7. This indicates dust free gas, as there is little to no suppression of Ly$\alpha$, and was further motivation to adopt a dust free HIGH ION component. From previous studies a dust content of approximately half the levels seen in the ISM reproduce the observed medium and low ionization lines seen in the spectra \citep{collins2009, kraemer2009}, and we adopt that for MED and LOW ION.

The exact logarithmic abundances relative to hydrogen by number for dust free models are: He = -1,  C = -3.47, N = -3.92, O = -3.17, Ne = -3.96, Na = -5.76, Mg = -4.48 Al = -5.55, Si = -4.51, P = -6.59,  S = -4.82, Ar = -5.60, Ca = -5.66, Fe = -4.40, and Ni = -5.78. For models with a dust level of 0.50 relative to the ISM, we accounted for the depletion of certain heavy elements onto graphite and silicate grains. Nitrogen is not depleted, as it is deposited onto ice mantles in grains that dissociate in the NLR  \citep{seab1983, snow1996, collins2009}. The logarithmic abundances relative to hydrogen by number for these dusty models are: He = -1, C = -3.59, N = -3.92, O = -3.21, Ne = -3.96, Na = -5.76, Mg = -4.74, Al = -5.81, Si = -4.76, P = -6.59, S = -4.82, Ar = -5.60, Ca = -5.92, Fe = -4.66, and Ni = -6.04. We consider only the effects of the default atomic data within Cloudy on our predictions (see, e.g. \citealp{juandedios2017, laha2017} for discussions).

\subsection{Model Selection}

With these input parameters, we ran grids of models over a range of ionization parameters for each location along the slit located a distance $r$ from the nucleus. From Equation 5 the only unknown quantities are $U$ and $n_\mathrm{H}$, so for each $U$ we solved for the corresponding density to maintain physical continuity. If $U$ and $n_\mathrm{H}$ are allowed to vary independently, then the corresponding distance $r$ would be incorrect. The number density of the LOW ION component was constrained to a small range encompassing typical errors around the power law fits in Figure \ref{tempdensity}, while a range of densities were explored for the HIGH and MED ION components, as they are less constrained. Using the limits from our diagnostics in \S3, we explored a range of parameters for number and column density, column density of the filter, and number of components. We then add fractional combinations of the multiple components to create a final composite model that matches the total observed H$\beta$ luminosity.

\setlength{\tabcolsep}{0.035in}
\tabletypesize{\footnotesize}
\begin{deluxetable}{c|c|c|c|c|c|c|c|}
\vspace{-30pt}
\tablenum{3}
\tablecaption{Cloudy Model Input Parameters \vspace{-3pt}}
\tablehead{
\colhead{Distance} & \colhead{Comp} & \colhead{Ionization} & \colhead{Column} & \colhead{Number} & \colhead{Dust} & \colhead{Input}\\
\colhead{from} & \colhead{ION} & \colhead{Parameter} & \colhead{Density} & \colhead{Density} & \colhead{Content} & \colhead{SED}\\
\colhead{Nucleus} & \colhead{Name} & \colhead{Log($U$)} & \colhead{Log(N$_\mathrm{H}$)} & \colhead{Log(n$_\mathrm{H}$)} & \colhead{Relative} & \colhead{Type}\\
\colhead{(arcsec)}& \colhead{} & \colhead{(unitless)} & \colhead{(cm$^{-2}$)} & \colhead{(cm$^{-3}$)} & \colhead{to ISM} & \colhead{(I/F)}\\
\colhead{(1)} & \colhead{(2)} & \colhead{(3)} & \colhead{(4)} & \colhead{(5)} & \colhead{(6)} & \colhead{(7)}
}
\startdata
0.46	 & 	High		 & 	-0.70	 & 	21.90	 & 	2.44	 & 	0.0	 & 	I	 \\ 
0.46	 & 	Med		 & 	-1.40	 & 	21.60	 & 	3.14	 & 	0.5	 & 	I	 \\ 
0.46	 & 	Low		 & 	-2.83	 & 	22.10	 & 	2.20	 & 	0.5	 & 	F	 \\ 
0.36	 & 	High		 & 	-1.20	 & 	21.30	 & 	3.16	 & 	0.0	 & 	I	 \\ 
0.36	 & 	Med		 & 	-1.50	 & 	21.00	 & 	3.46	 & 	0.5	 & 	I	 \\ 
0.36	 & 	Low		 & 	-2.82	 & 	21.80	 & 	2.40	 & 	0.5	 & 	F	 \\ 
0.25	 & 	High		 & 	-1.10	 & 	19.70	 & 	3.37	 & 	0.0	 & 	I	 \\ 
0.25	 & 	Med		 & 	-1.50	 & 	21.30	 & 	3.77	 & 	0.5	 & 	I	 \\ 
0.25	 & 	Low		 & 	-2.70	 & 	22.00	 & 	2.60	 & 	0.5	 & 	F	 \\ 
0.15	 & 	High		 & 	-0.60	 & 	22.00	 & 	3.32	 & 	0.0	 & 	I	 \\ 
0.15	 & 	Med		 & 	-1.40	 & 	20.20	 & 	4.12	 & 	0.5	 & 	I	 \\ 
0.15	 & 	Low		 & 	-1.82	 & 	21.80	 & 	3.70	 & 	0.5	 & 	F	 \\ 
0.05	 & 	High		 & 	-0.70	 & 	19.80	 & 	4.37	 & 	0.0	 & 	I	 \\ 
0.05	 & 	Med		 & 	-1.40	 & 	21.40	 & 	5.07	 & 	0.5	 & 	I	 \\ 
0.05	 & 	Low		 & 	-2.61	 & 	22.60	 & 	3.90	 & 	0.5	 & 	F	 \\ 
-0.05	 & 	High		 & 	-1.10	 & 	22.20	 & 	4.77	 & 	0.0	 & 	I	 \\ 
-0.05	 & 	Med		 & 	-1.60	 & 	20.20	 & 	5.27	 & 	0.5	 & 	I	 \\ 
-0.05	 & 	Low		 & 	-0.96	 & 	22.70	 & 	3.80	 & 	0.5	 & 	F	 \\ 
-0.15	 & 	High		 & 	-0.70	 & 	20.10	 & 	3.42	 & 	0.0	 & 	I	 \\ 
-0.15	 & 	Med		 & 	-1.70	 & 	20.20	 & 	4.42	 & 	0.5	 & 	I	 \\ 
-0.15	 & 	Low		 & 	-1.92	 & 	21.70	 & 	3.80	 & 	0.5	 & 	F	 \\ 
-0.25	 & 	High		 & 	...	 & 	...	 & 	...	 & 	...	 & 	I	 \\ 
-0.25	 & 	Med		 & 	-1.80	 & 	20.20	 & 	4.07	 & 	0.5	 & 	I	 \\ 
-0.25	 & 	Low		 & 	-2.16	 & 	21.60	 & 	3.60	 & 	0.5	 & 	F	 \\ 
-0.36	 & 	High		 & 	-0.60	 & 	19.80	 & 	2.56	 & 	0.0	 & 	I	 \\ 
-0.36	 & 	Med		 & 	-1.70	 & 	21.90	 & 	3.66	 & 	0.5	 & 	I	 \\ 
-0.36	 & 	Low		 & 	-2.48	 & 	21.72	 & 	2.80	 & 	0.5	 & 	F	 \\ 
-0.46	 & 	High		 & 	-0.70	 & 	21.90	 & 	2.44	 & 	0.0	 & 	I	 \\ 
-0.46	 & 	Med		 & 	-1.40	 & 	21.50	 & 	3.14	 & 	0.5	 & 	I	 \\ 
-0.46	 & 	Low		 & 	-2.83	 & 	22.10	 & 	2.20	 & 	0.5	 & 	F	 
\enddata
\tablecomments{The Cloudy photoionization model input parameters. The columns are: (1) position, (2) component name, (3) log$_{10}$ ionization parameter, (4) log$_{10}$ column density, (5) log$_{10}$ number density, (6) dust fraction relative to ISM, and (7) the implemented SED (Intrinsic/Filtered, see \S5.1). The ionization parameters for LOW ION models are computed by Cloudy using the filtered SED.}
\end{deluxetable}

\setlength{\tabcolsep}{0.036in}
\tabletypesize{\footnotesize}
\begin{deluxetable}{c|c|c|c|c|c|c|}
\vspace{-30pt}
\tablenum{4}
\tablecaption{Cloudy Model Output Parameters \vspace{-3pt}}
\tablehead{
\colhead{Distance} & \colhead{Comp} & \colhead{Fraction} & \colhead{Log(F$_{\mathrm{H}\beta}$)} & \colhead{Cloud} & \colhead{Cloud} & \colhead{Cloud}\\
\colhead{from} & \colhead{ION} & \colhead{of Total} & \colhead{Model} & \colhead{Surface} & \colhead{Model} & \colhead{Model}\\
\colhead{Nucleus} & \colhead{Name} & \colhead{Model} & \colhead{Flux} & \colhead{Area} & \colhead{Thickness} & \colhead{Depth}\\
\colhead{(arcsec)} & \colhead{} & \colhead{} & \colhead{(cgs)} & \colhead{(pc$^2$)} & \colhead{(pc)} & \colhead{(pc)}\\
\colhead{(1)} & \colhead{(2)} & \colhead{(3)} & \colhead{(4)} & \colhead{(5)} & \colhead{(6)} & \colhead{(7)}
}
\startdata
0.46	 & 	High	 & 	0.10	 & 	-0.80	 & 	39.	 & 	9.3	 & 	0.6	 \\ 
0.46	 & 	Med	 & 	0.65	 & 	-0.47	 & 	120.	 & 	0.9	 & 	1.7	 \\ 
0.46	 & 	Low	 & 	0.25	 & 	-1.51	 & 	502.	 & 	25.7	 & 	7.2	 \\ 
0.36	 & 	High	 & 	0.30	 & 	-0.60	 & 	18.	 & 	0.4	 & 	0.3	 \\ 
0.36	 & 	Med	 & 	0.45	 & 	-0.58	 & 	26.	 & 	0.1	 & 	0.4	 \\ 
0.36	 & 	Low	 & 	0.25	 & 	-1.39	 & 	93.	 & 	8.1	 & 	1.3	 \\ 
0.25	 & 	High	 & 	0.20	 & 	-2.00	 & 	1618.	 & 	$<$0.1	 & 	23.1	 \\ 
0.25	 & 	Med	 & 	0.50	 & 	0.08	 & 	33.	 & 	0.1	 & 	0.5	 \\ 
0.25	 & 	Low	 & 	0.30	 & 	-1.02	 & 	252.	 & 	8.1	 & 	3.6	 \\ 
0.15	 & 	High	 & 	0.25	 & 	0.15	 & 	103.	 & 	1.6	 & 	1.5	 \\ 
0.15	 & 	Med	 & 	0.15	 & 	-0.76	 & 	498.	 & 	$<$0.1	 & 	7.1	 \\ 
0.15	 & 	Low	 & 	0.60	 & 	-0.03	 & 	374.	 & 	0.4	 & 	5.3	 \\ 
0.05	 & 	High	 & 	0.25	 & 	-1.01	 & 	2057.	 & 	$<$0.1	 & 	29.4	 \\ 
0.05	 & 	Med	 & 	0.60	 & 	1.43	 & 	18.	 & 	$<$0.1	 & 	0.3	 \\ 
0.05	 & 	Low	 & 	0.15	 & 	0.57	 & 	33.	 & 	1.6	 & 	0.5	 \\ 
-0.05	 & 	High	 & 	0.55	 & 	1.82	 & 	2.	 & 	0.1	 & 	$<$0.1	 \\ 
-0.05	 & 	Med	 & 	0.30	 & 	0.40	 & 	26.	 & 	$<$0.1	 & 	0.4	 \\ 
-0.05	 & 	Low	 & 	0.15	 & 	0.70	 & 	7.	 & 	2.6	 & 	0.1	 \\ 
-0.15	 & 	High	 & 	0.05	 & 	-1.66	 & 	311.	 & 	$<$0.1	 & 	4.4	 \\ 
-0.15	 & 	Med	 & 	0.45	 & 	-0.42	 & 	164.	 & 	$<$0.1	 & 	2.3	 \\ 
-0.15	 & 	Low	 & 	0.50	 & 	-0.02	 & 	71.	 & 	0.3	 & 	1.0	 \\ 
-0.25	 & 	High	 & 	0.00	 & 	...	 & 	...	 & 	...	 & 	...		 \\ 
-0.25	 & 	Med	 & 	0.40	 & 	-0.75	 & 	177.	 & 	$<$0.1	 & 	2.5	 \\ 
-0.25	 & 	Low	 & 	0.60	 & 	-0.42	 & 	124.	 & 	0.3	 & 	1.8	 \\ 
-0.36	 & 	High	 & 	0.25	 & 	-2.85	 & 	6173.	 & 	0.1	 & 	88.2	 \\ 
-0.36	 & 	Med	 & 	0.60	 & 	-0.10	 & 	26.	 & 	0.6	 & 	0.4	 \\ 
-0.36	 & 	Low	 & 	0.15	 & 	-1.35	 & 	117.	 & 	2.7	 & 	1.7	 \\ 
-0.46	 & 	High	 & 	0.05	 & 	-0.80	 & 	3.	 & 	9.3	 & 	$<$0.1	 \\ 
-0.46	 & 	Med	 & 	0.70	 & 	-0.48	 & 	21.	 & 	0.7	 & 	0.3	 \\ 
-0.46	 & 	Low	 & 	0.25	 & 	-1.51	 & 	81.	 & 	25.7	 & 	1.2	 
\enddata
\tablecomments{The best fitting Cloudy model output parameters. The columns are: (1) position, (2) component name, (3) fraction of model contributing to the final H$\beta$ luminosity, (4) log$_{10}$ H$\beta$ model flux (erg s$^{-1}$ cm$^{-2}$), (5) surface area of the emitting gas, (6) gas cloud thickness (N$_\mathrm{H}$/n$_\mathrm{H}$), and (7) depth into the plane of the sky.}
\end{deluxetable}

To determine the best fit model at each location, we employed a simple numerical scheme that computed the model line ratios for all possible fractional combinations of our HIGH, MED, and LOW ION components across our range of parameters. No assumptions were made about the number of components, and the technique could find that one, two, or three components was a best fit to the observations. Model/dereddened line ratios were calculated for each composite model, with an ideal match having a ratio of one. 

Our criteria for a successful fit were the following. The sensitive diagnostics lines of [O~III], [N~II], and [S~II] must match the observations to within 30\%. In addition, the column density sensitive He~II $\lambda$4686 line ratios must match within 10\%. Furthermore, we imposed the limits that all remaining lines must match their predicted value within a factor of five, and the global model/dereddened ratios must center on a mean of unity to within 20\%.

In cases where a position was well matched by two or more similar models, we chose the composite model that best matched the strong emission lines and those sensitive to column density, as these should provide the most realistic mass determination. For all but one position ($-0\farcs25$) three components produced the best fit. The best fit absorber column density was Log(N$_\mathrm{H}$) = 21.50 for positions +$0\farcs15$, $-0\farcs05$, $-0\farcs15$, and $-0\farcs25$, and Log(N$_\mathrm{H}$) = 21.60 for all other positions. The input and output parameters for these best fit models are given in Tables 3 and 4, respectively. The final predicted emission line ratios from all components weighted by their fractional contributions are given in Table 5.

\begin{figure*}
\vspace{-12pt}
\centering
\subfigure
{\includegraphics[scale=0.63]{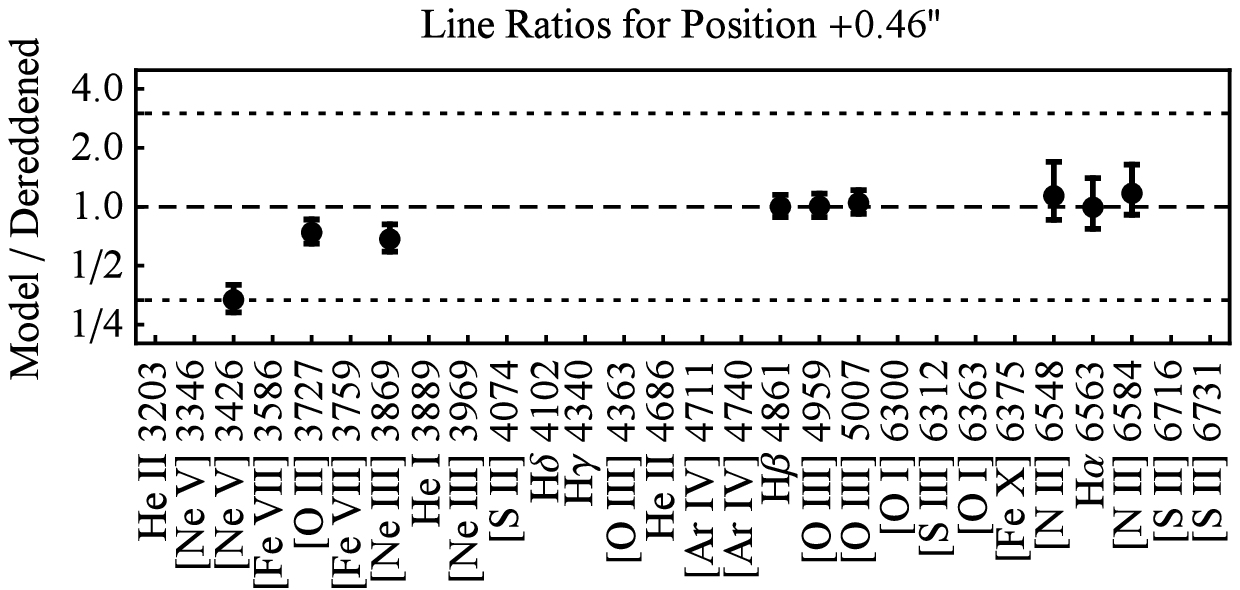}}
\vspace{-10pt}
\subfigure
{\includegraphics[scale=0.63]{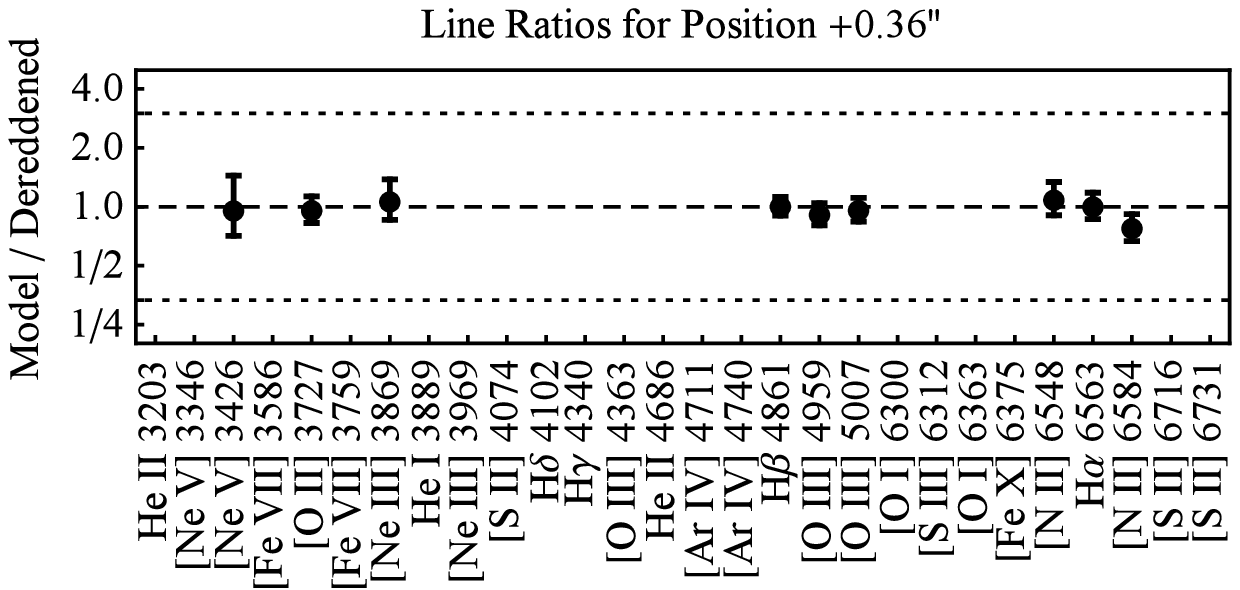}}
\vspace{-10pt}
\subfigure
{\includegraphics[scale=0.63]{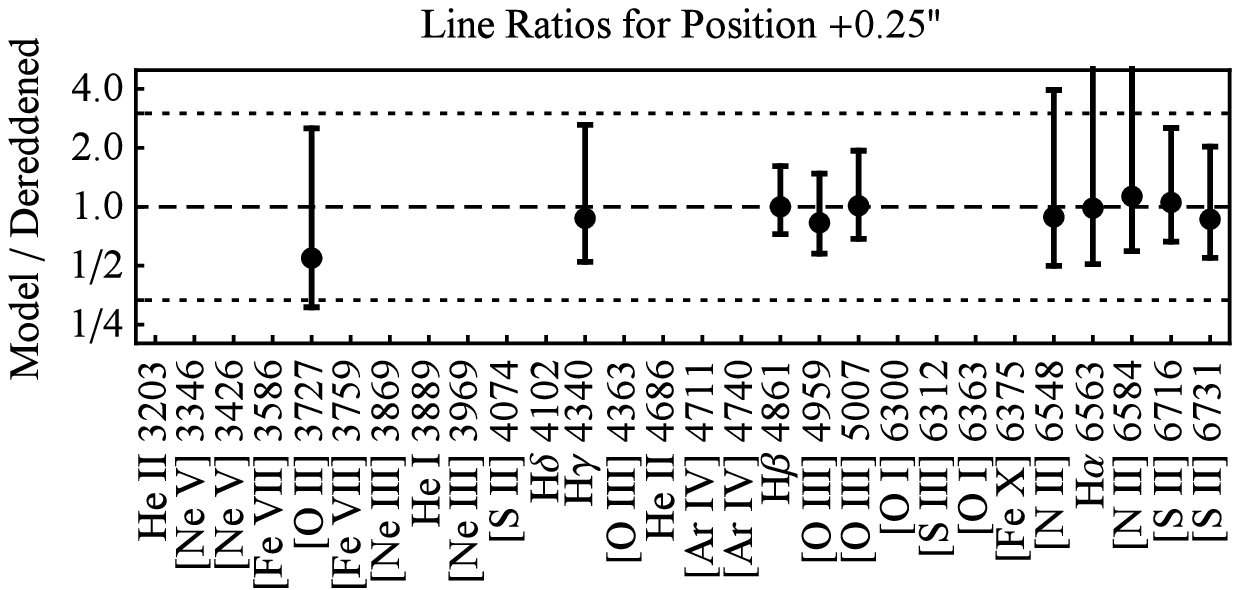}}
\subfigure
{\includegraphics[scale=0.63]{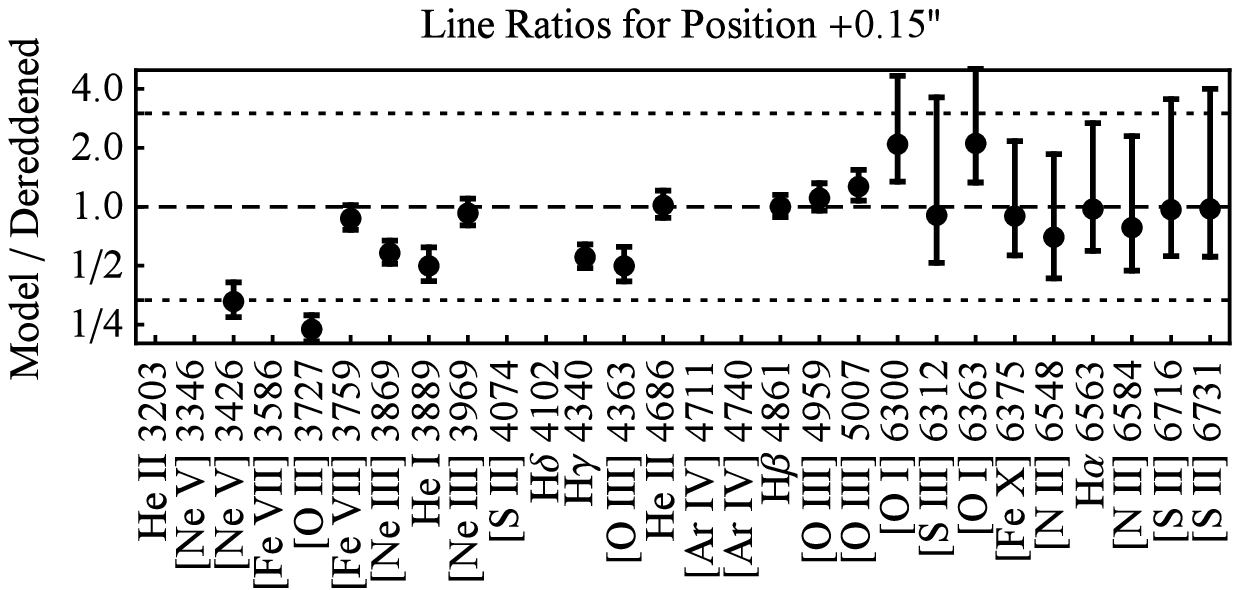}}
\vspace{-10pt}
\subfigure
{\includegraphics[scale=0.63]{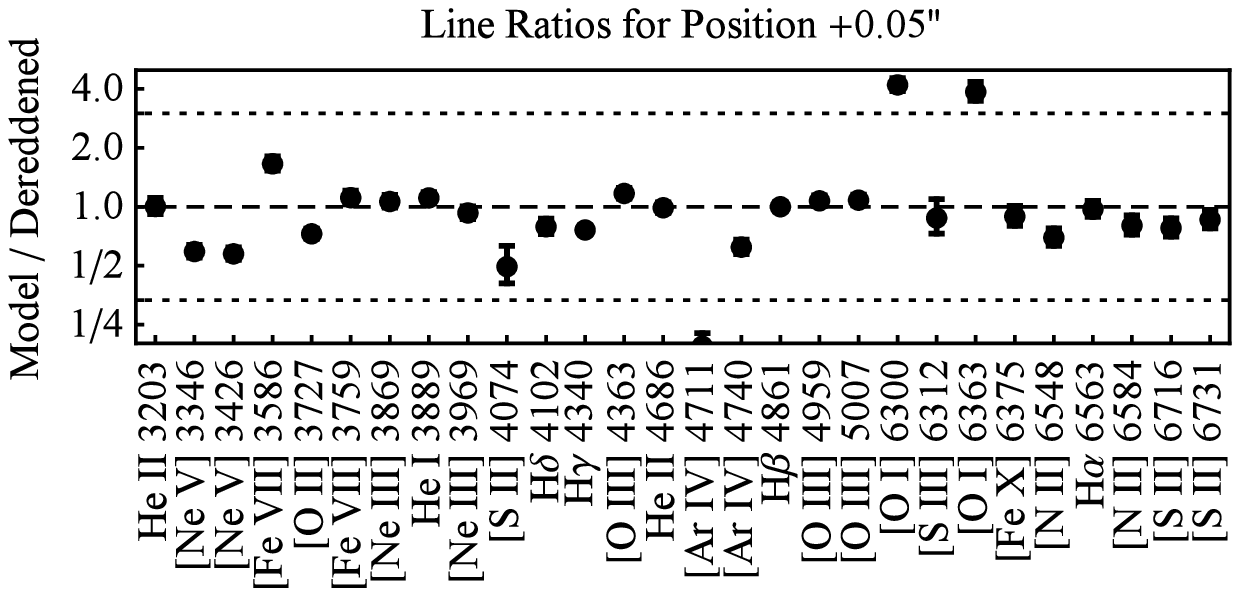}}
\subfigure
{\includegraphics[scale=0.63]{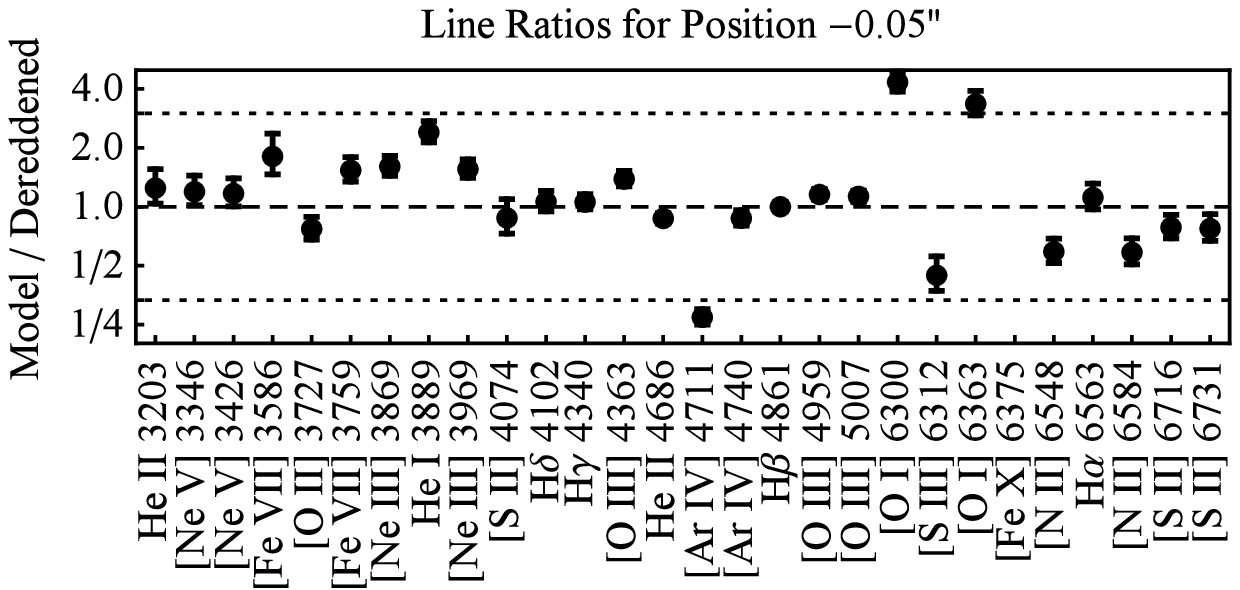}}
\vspace{-10pt}
\subfigure
{\includegraphics[scale=0.63]{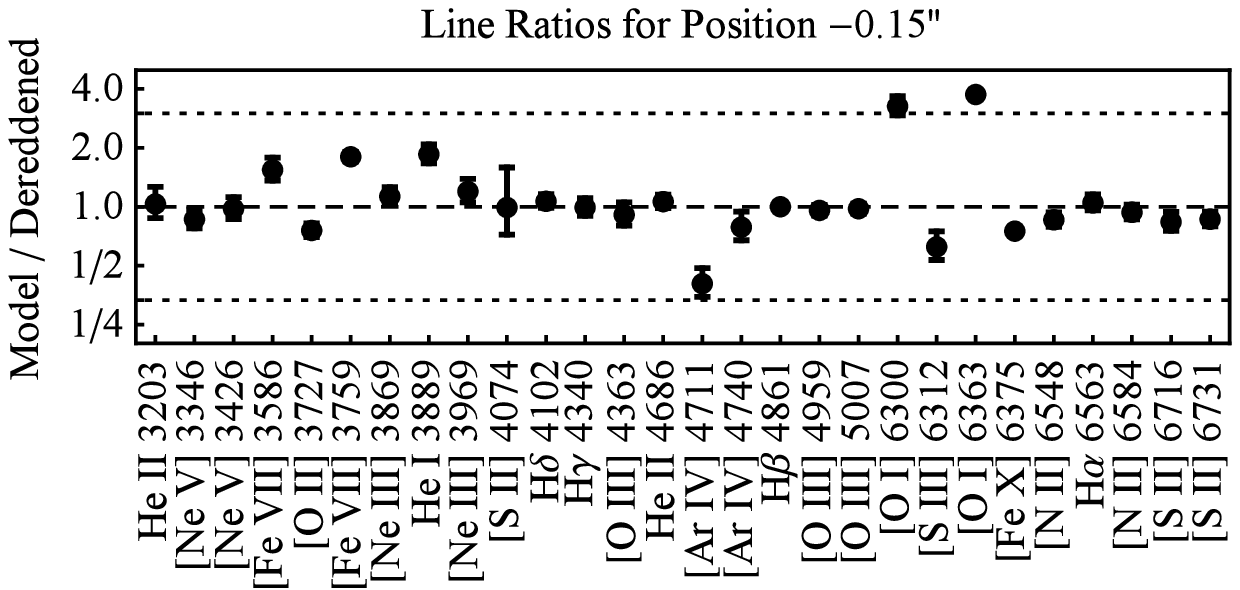}}
\subfigure
{\includegraphics[scale=0.63]{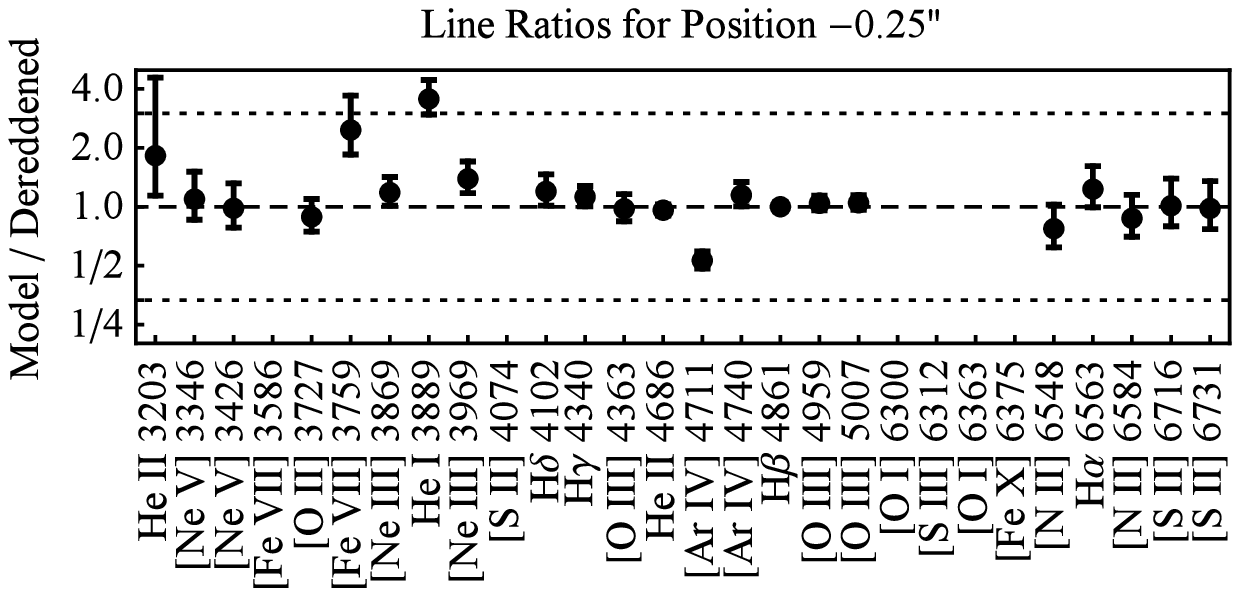}}
\subfigure
{\includegraphics[scale=0.63]{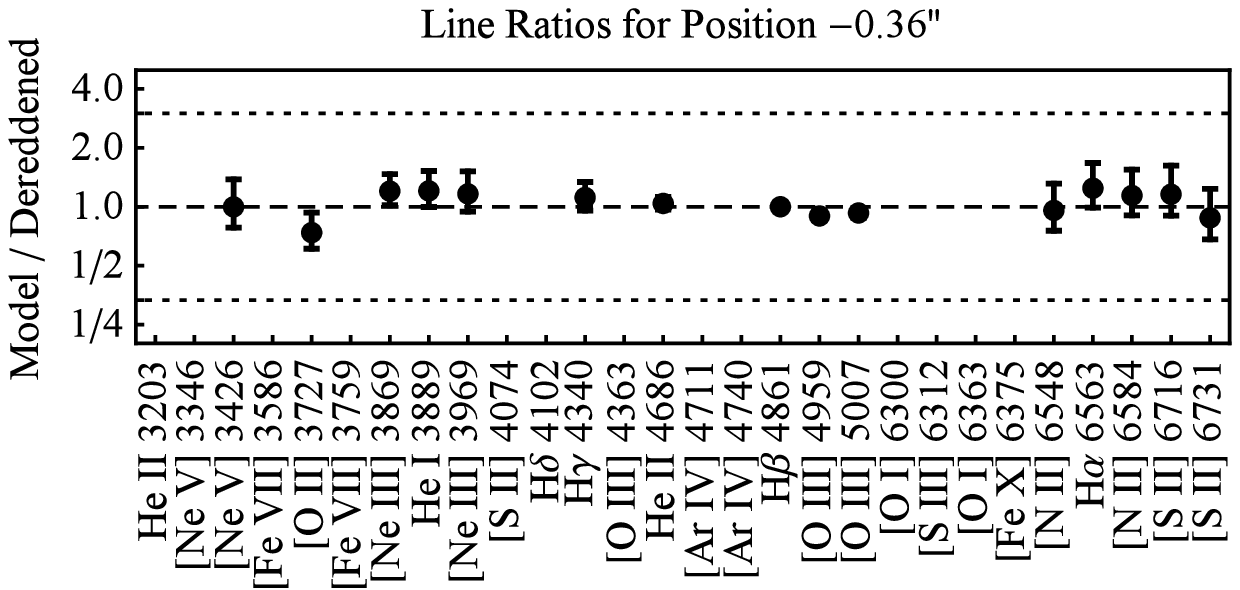}}
\subfigure
{\includegraphics[scale=0.63]{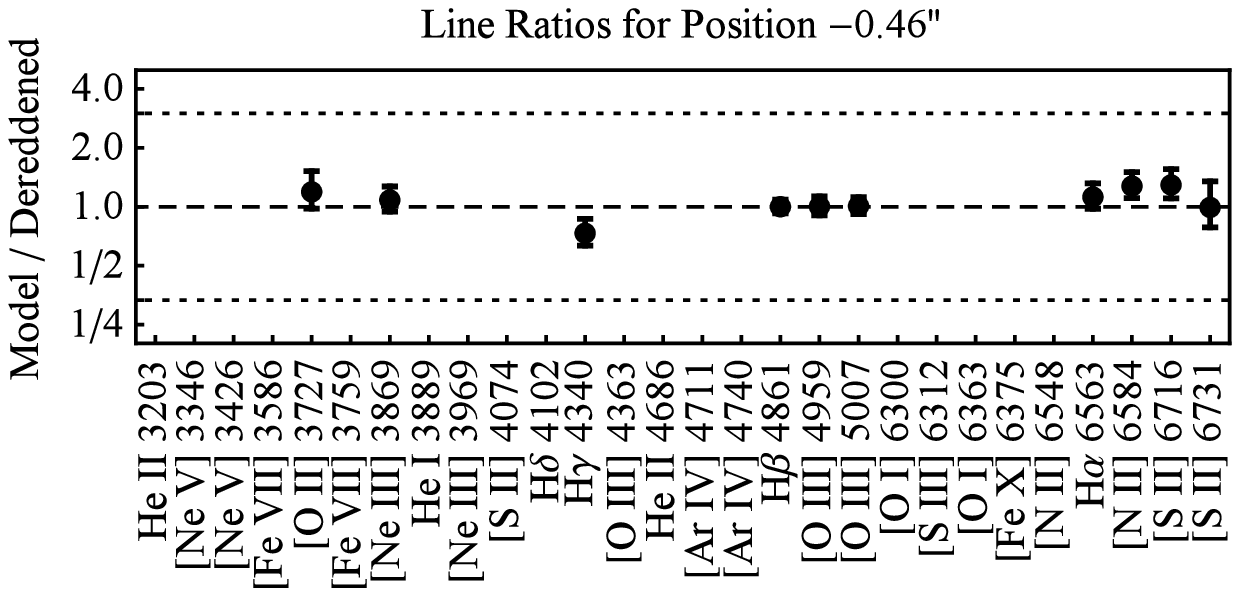}}
\vspace{-5pt}
\caption{The composite Cloudy model line ratios divided by the dereddened values for each position. The dashed unity lines indicate an exact match, while the dotted lines are factor of three difference intervals (tick marks are logarithmically spaced for even ratio distributions above and below the unity line). Points above the unity line are over-predicted by the models, while points below the unity line are under-predicted. Lines with no data point were too weak to measure at that location.}
\label{models}
\vspace{-5pt}
\end{figure*}

\subsection{Model Comparison to the Observations}

The comparison between our dereddened and model emission line ratios is shown in Figure \ref{models}. The dashed unity line indicates an exact match, and all points between the dotted lines represent agreement to within better than a factor of three. A variety of factors contribute to the deviations of each line from an exact match, including a poor gaussian fit, low S/N, accuracy of atomic data, and the accuracy of our multi-component models. Here we discuss deviations greater than a factor of three within errors for the important diagnostic lines at each position.

The positions at larger radial distances from the nucleus ($\pm0\farcs46$, $\pm0\farcs36$, $\pm0\farcs25$) all show excellent agreement between the observations and models. The apparent under-prediction of lines in the blue at $+0\farcs46$, $+0\farcs25$, and $+0\farcs15$ is indicative of the residual over-extinction correction discussed in \S3.3. As the ratios are over-corrected by a factor of $\sim 1.5-2$, their apparent under-prediction by a similar factor indicates a good fit. The large errors in the red at $+0\farcs25$ and $+0\farcs15$ are due to the large reddening uncertainties (\S3.3).

For positions at small radial distances from the nucleus ($\pm0\farcs15$ and $\pm0\farcs05$) with up to 28 emission lines, we also find excellent agreement between the observations and models. The sole exception is the over-prediction of the neutral oxygen doublet [O~I] $\lambda \lambda$6300,6363. For these positions, we explored using less stringent criteria on [O~III], [N~II], [S~II], and He~II, which decreased the over-prediction from $\sim 4.5$ times to $\sim 3.8$ times, at the expense of other line fits.

The collisionally excited [O~I] emission line is strongest in the neutral zone of a cloud, and is sensitive to the temperature \citep{kraemer2000a} and gas turbulence \citep{moy2002}. This may indicate an excess of X-rays transmitted by the absorbing filter, increasing the temperature in the neutral zone.  Because the extraction bin is $\sim 0\farcs1$x$0\farcs2$ (2x4 pixels, $\sim$ 36x72 pc) some of the lower ionization material could be slightly offset from the peak [O III] emission, but still within the spectral extraction bin, as seen in Mrk 3 \citep{collins2009}. As the centers of these two extractions are just $\sim$18 pc from the nucleus, the X-ray flux could be artificially high if the material is located toward the extremes of the bins. Given the excellent agreement of all other diagnostic lines at these locations, we opted against further fine tuning.

\begin{figure*}
\vspace{-5pt}
\centering
\subfigure{
\includegraphics[scale=0.46]{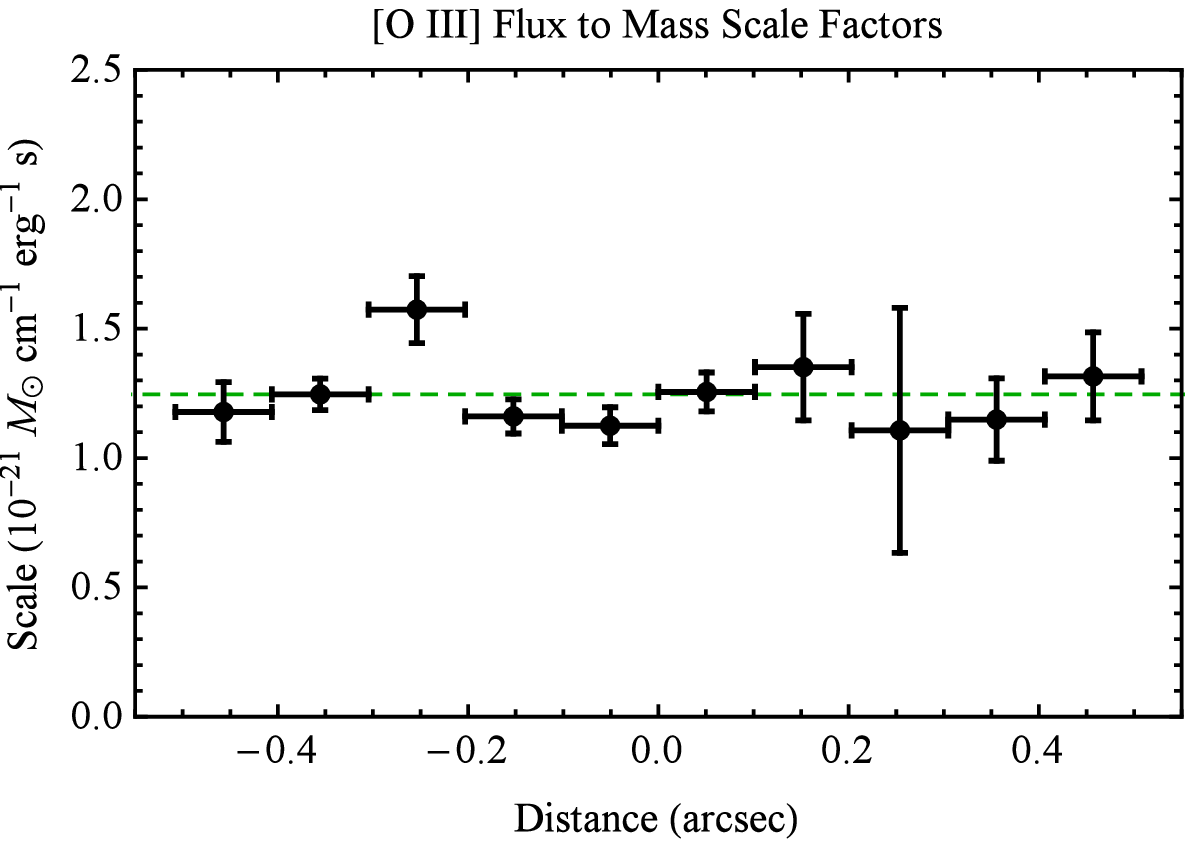}}
\subfigure{
\includegraphics[scale=0.635]{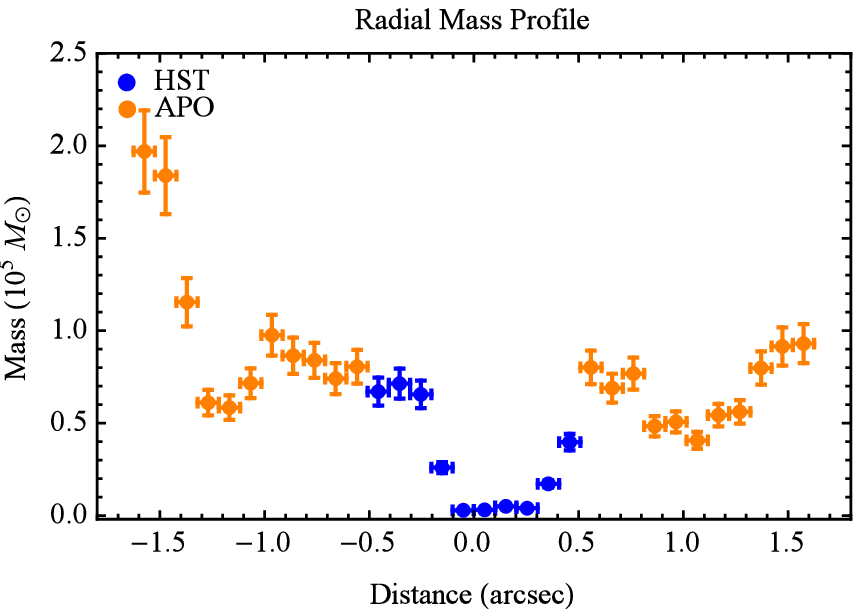}}
\subfigure{
\includegraphics[scale=0.635]{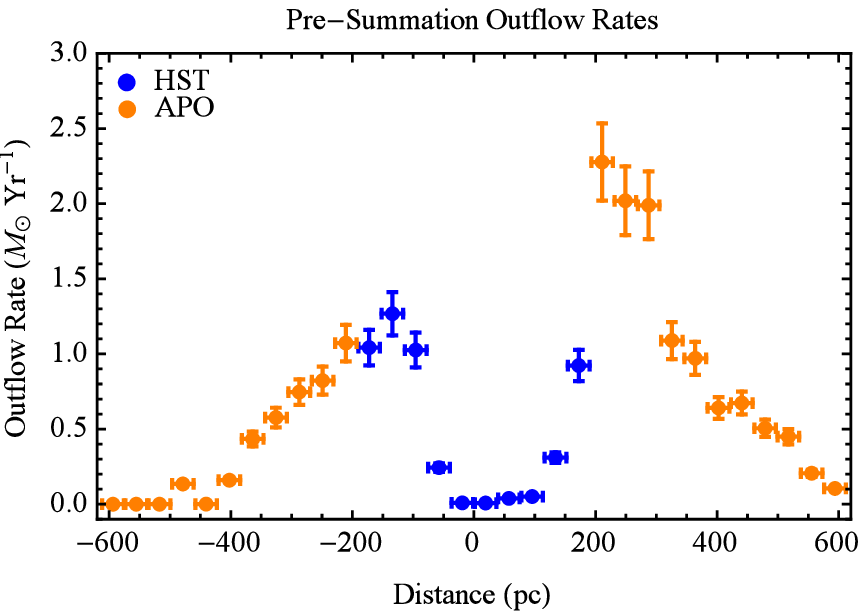}}
\vspace{-10pt}
\caption{Left: The derived scale factor at each location for converting [O~III] image fluxes to mass. The mean is indicated by the green dashed line. Middle: The derived mass profile in units of $10^5 M_{\odot}$ across the NLR. Right: The calculated mass outflow rates before azimuthal summation. The HST STIS and APO DIS data are in blue and orange, respectively. SE is to the left and NW is to the right.}
\label{scale}
\vspace{-5pt}
\end{figure*}

Our multi-component photoionization models are consistent with pure AGN ionization from the central source. As noted by others \citep{schlesinger2009}, we do not see any evidence for shock ionization in the outflow regions (see, e.g. \citealp{schlesinger2009, maksym2016, maksym2017, terao2016}, and references therein for discussions on shocks). Our composite models qualitatively agree with the IUE UV line ratios, indicating that we are likely encompassing a significant portion of the UV emitting gas in our models and resulting outflow rates.

\subsection{Physical Implications of the Models}

Using our models, we derived several physical quantities at each position that are given in Table 4. First, there is the surface area of the emitting clouds ($A = L_{\mathrm{H}\beta}/F_{\mathrm{H}\beta}$). Next, we confirmed that the physical thicknesses of the clouds ($N_{\mathrm{H}}/n_{\mathrm{H}}$) were smaller than our bin size to ensure they fit within the slit extraction (2 pixels). Finally, we calculated the summed depths of the clouds into the plane of the sky by dividing the cloud area by the projected slit width ($\sim$ 70 pc), to verify they were smaller than the scale height of the disk and ionizing bicone. It is important to note that each component may not be co-located within the extraction bin, as the emission is spread across $0\farcs2$ (4 pixels) in the spectral direction \citep{collins2009}.

\setlength{\tabcolsep}{0.159in}
\tabletypesize{\footnotesize}
\begin{deluxetable*}{l|c|c|c|c|c|c|c|c|c|c|}
\vspace{-26pt}
\tablenum{5}
\tablecaption{Predicted Cloudy Model Emission Line Ratios \vspace{-5pt}}
\tablehead{
\colhead{Line} & \colhead{$0\farcs46$} & \colhead{$0\farcs36$} & \colhead{$0\farcs25$} & \colhead{$0\farcs15$} & \colhead{$0\farcs05$} & \colhead{--$0\farcs05$} & \colhead{--$0\farcs15$} & \colhead{--$0\farcs25$} & \colhead{--$0\farcs36$} & \colhead{--$0\farcs46$} \vspace{-3pt}}
\startdata
\vspace{-0.5pt} He II $\lambda$3203	 & 	0.15	 & 	0.28	 & 	0.17	 & 	0.19	 & 	0.21	 & 	0.19	 & 	0.23	 & 	0.18	 & 	0.18	 & 	0.14	\\ \relax
\vspace{-0.5pt} [Ne V] $\lambda$3346	 & 	0.48	 & 	0.85	 & 	0.55	 & 	0.59	 & 	0.63	 & 	0.76	 & 	0.70	 & 	0.44	 & 	0.32	 & 	0.44	\\ \relax
\vspace{-0.5pt} [Ne V] $\lambda$3426	 & 	1.32	 & 	2.32	 & 	1.50	 & 	1.63	 & 	1.73	 & 	2.09	 & 	1.91	 & 	1.20	 & 	0.87	 & 	1.22	\\ \relax
\vspace{-0.5pt} [Fe VII] $\lambda$3586	 & 	0.22	 & 	0.36	 & 	0.30	 & 	0.43	 & 	0.40	 & 	0.31	 & 	0.23	 & 	0.11	 & 	0.29	 & 	0.17	\\ \relax
\vspace{-0.5pt} [O II] $\lambda$3727	 & 	2.20	 & 	1.48	 & 	2.24	 & 	0.84	 & 	0.88	 & 	0.52	 & 	0.66	 & 	0.99	 & 	1.13	 & 	2.20	\\ \relax
\vspace{-0.5pt} [Fe VII] $\lambda$3759	 & 	0.30	 & 	0.50	 & 	0.42	 & 	0.60	 & 	0.56	 & 	0.43	 & 	0.32	 & 	0.15	 & 	0.40	 & 	0.23	\\ \relax
\vspace{-0.5pt} [Ne III] $\lambda$3869	 & 	1.56	 & 	0.95	 & 	1.36	 & 	1.42	 & 	1.56	 & 	1.36	 & 	1.20	 & 	1.21	 & 	1.13	 & 	1.62	\\ \relax
\vspace{-0.5pt} He I	 $\lambda$3889 & 	0.35	 & 	0.49	 & 	0.31	 & 	0.23	 & 	0.33	 & 	0.38	 & 	0.33	 & 	0.35	 & 	0.29	 & 	0.36	\\ \relax
\vspace{-0.5pt} [Ne III] $\lambda$3969	 & 	0.63	 & 	0.45	 & 	0.57	 & 	0.59	 & 	0.63	 & 	0.58	 & 	0.53	 & 	0.53	 & 	0.50	 & 	0.65	\\ \relax
\vspace{-0.5pt} [S II] $\lambda$4074	 & 	0.10	 & 	0.05	 & 	0.08	 & 	0.09	 & 	0.11	 & 	0.09	 & 	0.08	 & 	0.09	 & 	0.09	 & 	0.09	\\ \relax
\vspace{-0.5pt} H$\delta$ $\lambda$4102	 & 	0.26	 & 	0.26	 & 	0.26	 & 	0.27	 & 	0.26	 & 	0.26	 & 	0.27	 & 	0.26	 & 	0.26	 & 	0.26	\\ \relax
\vspace{-0.5pt} H$\gamma$ $\lambda$4340	 & 	0.47	 & 	0.47	 & 	0.47	 & 	0.48	 & 	0.47	 & 	0.48	 & 	0.48	 & 	0.47	 & 	0.47	 & 	0.47	\\ \relax
\vspace{-0.5pt} [O III] $\lambda$4363	 & 	0.25	 & 	0.25	 & 	0.19	 & 	0.20	 & 	0.37	 & 	0.30	 & 	0.20	 & 	0.19	 & 	0.17	 & 	0.26	\\ \relax
\vspace{-0.5pt} He II $\lambda$4686	 & 	0.37	 & 	0.68	 & 	0.40	 & 	0.45	 & 	0.48	 & 	0.44	 & 	0.54	 & 	0.44	 & 	0.42	 & 	0.35	\\ \relax
\vspace{-0.5pt} [Ar IV] $\lambda$4711	 & 	0.06	 & 	0.07	 & 	0.04	 & 	0.04	 & 	0.01	 & 	0.03	 & 	0.05	 & 	0.05	 & 	0.04	 & 	0.07	\\ \relax
\vspace{-0.5pt} [Ar IV]	 $\lambda$4740 & 	0.06	 & 	0.07	 & 	0.05	 & 	0.06	 & 	0.07	 & 	0.10	 & 	0.09	 & 	0.08	 & 	0.05	 & 	0.06	\\ \relax
\vspace{-0.5pt} H$\beta$ $\lambda$4861	 & 	1.00	 & 	1.00	 & 	1.00	 & 	1.00	 & 	1.00	 & 	1.00	 & 	1.00	 & 	1.00	 & 	1.00	 & 	1.00	\\ \relax
\vspace{-0.5pt} [O III] $\lambda$4959	 & 	5.05	 & 	3.63	 & 	4.01	 & 	5.06	 & 	4.63	 & 	5.08	 & 	4.52	 & 	4.71	 & 	4.07	 & 	5.38	\\ \relax
\vspace{-0.5pt} [O III] $\lambda$5007	 & 	15.21	 & 	10.93	 & 	12.07	 & 	15.23	 & 	13.94	 & 	15.28	 & 	13.61	 & 	14.17	 & 	12.24	 & 	16.19	\\ \relax
\vspace{-0.5pt} [O I] $\lambda$6300	 & 	1.68	 & 	1.12	 & 	1.48	 & 	0.79	 & 	1.09	 & 	0.74	 & 	0.59	 & 	0.56	 & 	1.12	 & 	1.60	\\ \relax
\vspace{-0.5pt} [S III] $\lambda$	6312 & 	0.05	 & 	0.03	 & 	0.05	 & 	0.04	 & 	0.04	 & 	0.02	 & 	0.04	 & 	0.05	 & 	0.04	 & 	0.05	\\ \relax
\vspace{-0.5pt} [O I] $\lambda$6363	 & 	0.54	 & 	0.36	 & 	0.47	 & 	0.25	 & 	0.35	 & 	0.23	 & 	0.19	 & 	0.18	 & 	0.36	 & 	0.51	\\ \relax
\vspace{-0.5pt} [Fe X] $\lambda$6375	 & 	0.03	 & 	0.01	 & 	0.02	 & 	0.26	 & 	0.23	 & 	0.01	 & 	0.05	 & 	0.00	 & 	0.40	 & 	0.01	\\ \relax
\vspace{-0.5pt} [N II] $\lambda$6548	 & 	0.76	 & 	0.62	 & 	0.79	 & 	0.48	 & 	0.41	 & 	0.25	 & 	0.40	 & 	0.44	 & 	0.60	 & 	0.74	\\ \relax
\vspace{-0.5pt} H$\alpha$ $\lambda$6563	 & 	2.88	 & 	2.84	 & 	2.86	 & 	2.83	 & 	2.82	 & 	2.82	 & 	2.82	 & 	2.85	 & 	2.81	 & 	2.88	\\ \relax
\vspace{-0.5pt} [N II] $\lambda$6584	 & 	2.25	 & 	1.82	 & 	2.33	 & 	1.42	 & 	1.21	 & 	0.73	 & 	1.19	 & 	1.30	 & 	1.78	 & 	2.18	\\ \relax
\vspace{-0.5pt} [S II] $\lambda$6716	 & 	0.72	 & 	0.39	 & 	0.50	 & 	0.32	 & 	0.22	 & 	0.17	 & 	0.23	 & 	0.33	 & 	0.57	 & 	0.68	\\ \relax
\vspace{-0.5pt} [S II] $\lambda$6731	 & 	0.56	 & 	0.29	 & 	0.41	 & 	0.40	 & 	0.32	 & 	0.25	 & 	0.30	 & 	0.39	 & 	0.52	 & 	0.53
\enddata
\tablecomments{The predicted Cloudy emission line ratios for our final composite models with the fractional weightings of each component given in Table 4. Cloudy version 13.04 does not predict O III $\lambda$3133.}
\vspace{-1pt}
\end{deluxetable*}

\section{Calculations}

We calculate the mass at each point along the HST STIS slit using the extinction-corrected H$\beta$ and Cloudy model H$\beta$ fluxes. For our multi-component models, the mass in each photoionized component is calculated separately by dividing up the H$\beta$ luminosity in each component, and then they are summed.

\subsection{Ionized Mass in the Slit}

The ionized gas mass in each slit extraction ($M_{\mathrm{slit}}$) for a given H$\beta$ luminosity is given by \citep{peterson1997, crenshaw2015}
\begin{equation}
M_{\mathrm{slit}} = N_\mathrm{H} \mu m_p \left(\frac{L(\mathrm{H}\beta)}{F(\mathrm{H}\beta)_{m}}\right),
\end{equation}
\noindent
where $N_\mathrm{H}$ is the total model hydrogen column density (predominantly ionized, with only trace amounts of neutral and molecular phases), $\mu$ is the mean mass per proton ($\sim$ 1.36 for solar, $\sim$ 1.40 for our abundances), $m_p$ is the mass of a proton, $F(\mathrm{H}\beta)_{m}$ is the H$\beta$ model flux, and $L(\mathrm{H}\beta)$ is the luminosity of H$\beta$ calculated from the observed flux and distance. That is,
\begin{equation}
L(\mathrm{H}\beta) = 4 \pi D^2 F(\mathrm{H}\beta)_{i},
\end{equation}
\noindent
where $D$ is the distance to the galaxy and $F(\mathrm{H}\beta)_{i}$ is the intrinsic flux corrected for extinction at each point (\S3.3). Conceptually, Equation 6 determines the area of the emitting clouds through the ratio of the luminosity and flux, and then multiplies this by the column density, or projected number of particles per unit area, to yield the total number of particles. Multiplying this by the mean mass per particle gives the total ionized mass.

\subsection{Ionized Mass Profile from [O~III] Imaging}

These calculations yield a direct conversion between the luminosity of the H$\beta$ emission line and the ionized gas mass, specifically, the number of H$\beta$ photons emitted per unit mass at a given density. If high quality H$\beta$ emission line imaging of the NLR were available, then those fluxes and our model densities could be used to directly calculate the gas mass at all spatial locations. For Mrk 573, only high spatial resolution [O~III] imaging is available. The flux of H$\beta$ and [O~III] are related simply by the [O~III]/H$\beta$ ratio, and thus the mass per unit H$\beta$ luminosity is also proportional to the observed [O~III] flux within a scale factor. Specifically,
\begin{equation}
S = \left(\frac{M_{\mathrm{slit}} n_\mathrm{H}}{F_{\lambda 5007}} \right),
\end{equation}
\noindent
where $M_{\mathrm{slit}}$ is the ionized mass in the slit calculated from Equation 6, $n_\mathrm{H}$ is the fractional weighted mean hydrogen number density (cm$^{-3}$) for all components, and $F_{\lambda 5007}$ is the extinction-corrected [O~III] emission line flux from our spectra. We take an average value of the scale factors from each location (Figure \ref{scale}), which exhibit some scatter due to variations in the [O~III]/H$\beta$ ratios across the NLR. This scale factor allows us to derive masses from observed [O~III] image fluxes rather than H$\beta$ luminosities. The total ionized mass for a given image flux is then
\begin{equation}
M_{\mathrm{ion}} = S \left(\frac{F_\mathrm{[O~III]}}{n_\mathrm{H}} \right).
\end{equation}
\noindent
For this analysis, $F_\mathrm{[O~III]}$ is the flux in each image semi-annulus of width $\delta r$ (Figure \ref{imaging}) and n$_\mathrm{H}$ is the hydrogen number density. 

The calculated scale factors are shown in Figure \ref{scale}. The mean scale factor is S $= 1.25 \pm 0.14 \times 10^{21}$ $M_{\odot}$ cm$^{-1}$ erg$^{-1}$ s, and the 1$\sigma$ error corresponds to a fractional uncertainty of 11.3\%. For position $-0\farcs36$ the calculated scale factor was $>3\sigma$ from the mean, possibly due to an anomalous corrected H$\beta$ flux, and was replaced with a mean value. The scale factor uncertainty in \cite{crenshaw2015} was calculated using a standard error, while we have adopted the standard deviation. This can result in larger fractional errors for the mass outflow rates, but should yield a more realistic estimate of the uncertainty at any individual point given the [O~III]/H$\beta$ variations across the NLR.

Ideally our density law and resulting masses would be determined from detailed photoionization models at all locations along the NLR. However, at distances of $r > 0\farcs5$ from the nucleus, only the [O~III] and H$\alpha$ emission lines are strong enough to get reliable measurements in our high spatial resolution HST data. This is due to intrinsically lower fluxes further from the nucleus, in combination with the PA of the HST slit that does not follow the linear feature of bright emission line knots. 

To obtain masses for $r = 0\farcs5 - 1\farcs7$ from the nucleus, we derived a hybrid technique employing our scale factor and then derived the gas density at each distance from our power law fits to the [S~II] lines in our APO DIS data, as shown in Figure \ref{tempdensity}. The APO observations have lower spatial resolution, but the wider slit encompasses significantly more NLR emission. Our testing showed that at distances of $r \geq0\farcs5$ from the nucleus the ionization state of the gas drops enough such that the density derived from [S~II] is approximately equal to that of a multi-component model. In this way we were able to extend our mass outflow measurements from 175 pc to 600 pc.

\begin{figure*}
\vspace{-13pt}
\centering
\subfigure{
\includegraphics[scale=0.96]{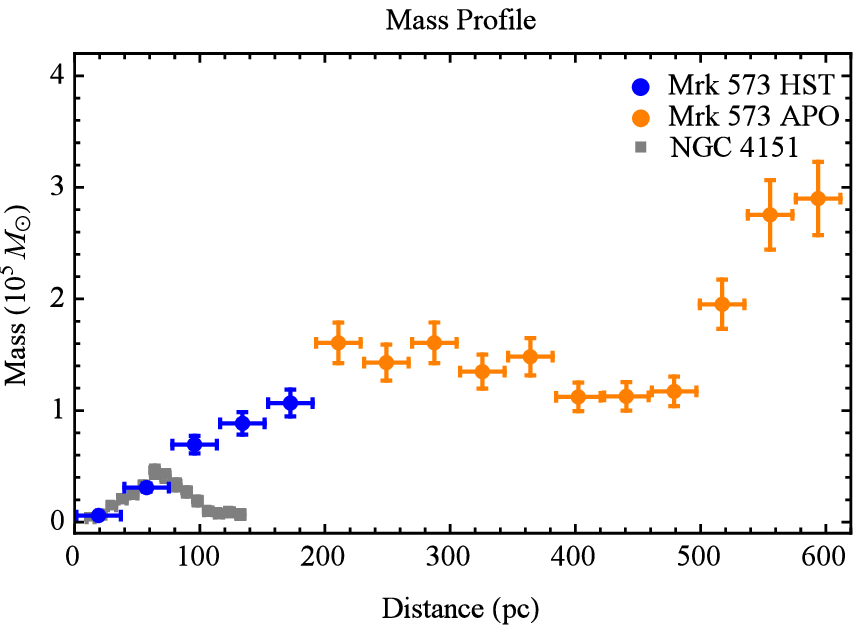}}
\vspace{-11pt}
\subfigure{
\includegraphics[scale=0.96]{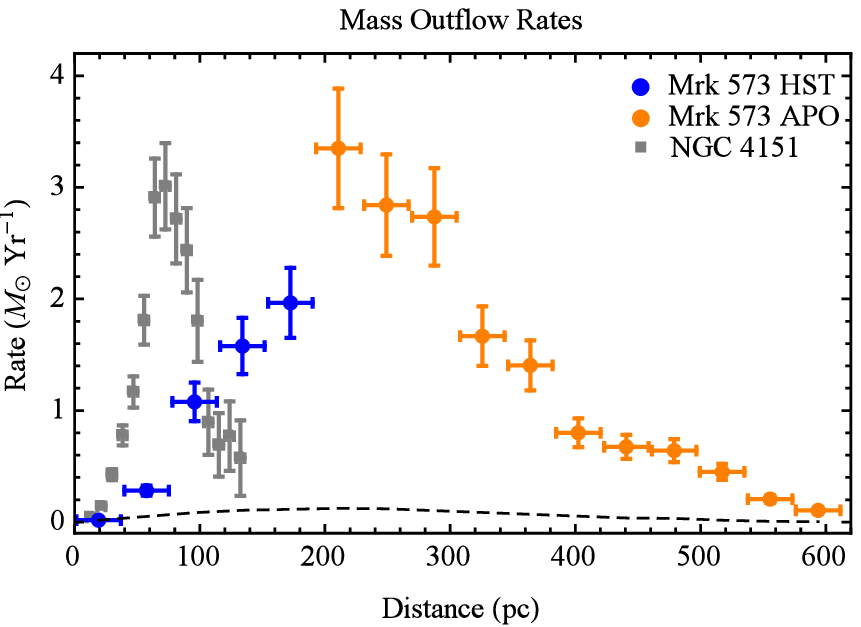}}
\subfigure{
\includegraphics[scale=0.96]{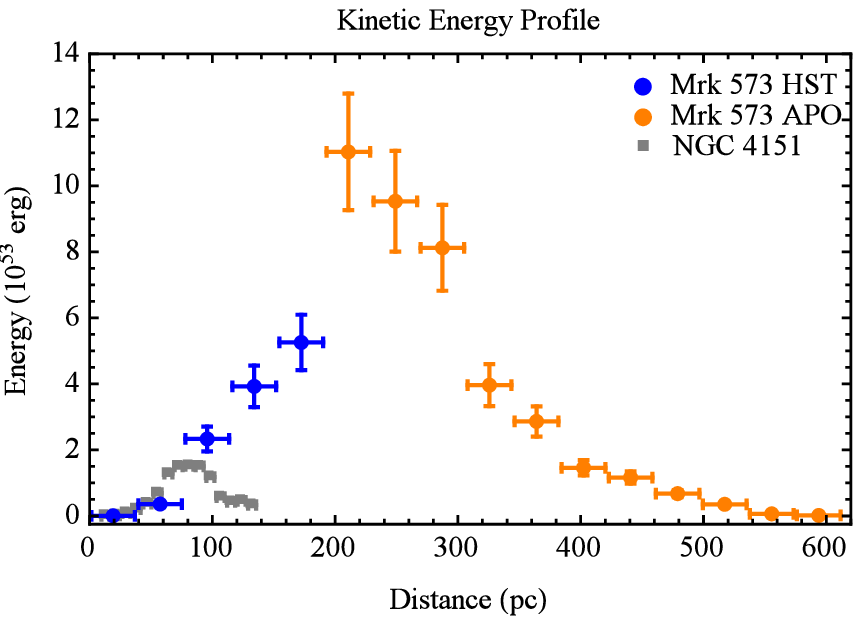}}
\vspace{-5pt}
\subfigure{
\includegraphics[scale=0.96]{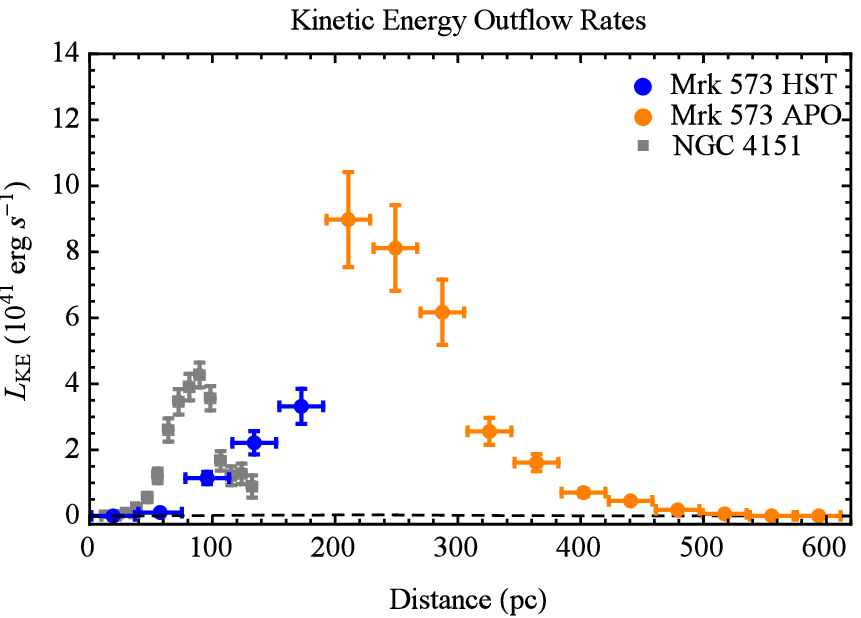}}
\subfigure{
\includegraphics[scale=0.96]{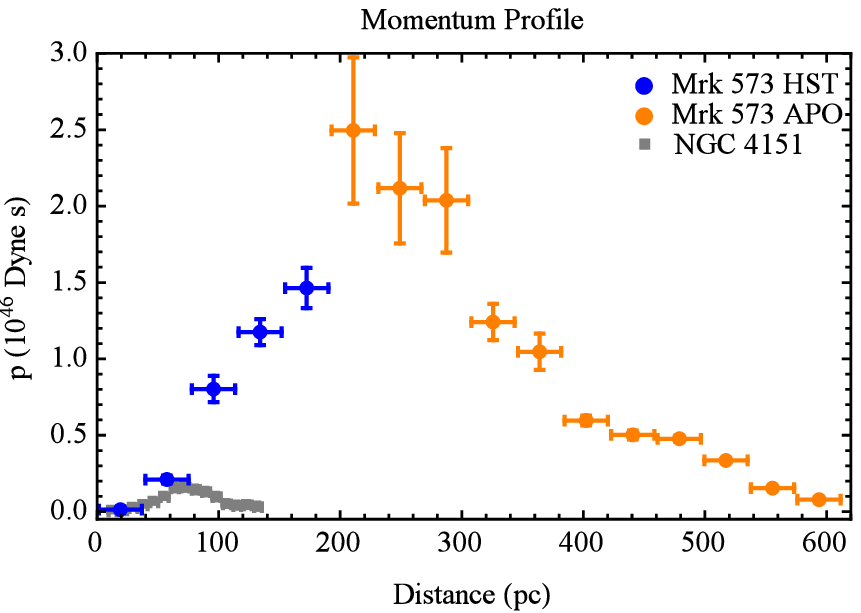}}
\subfigure{
\includegraphics[scale=0.96]{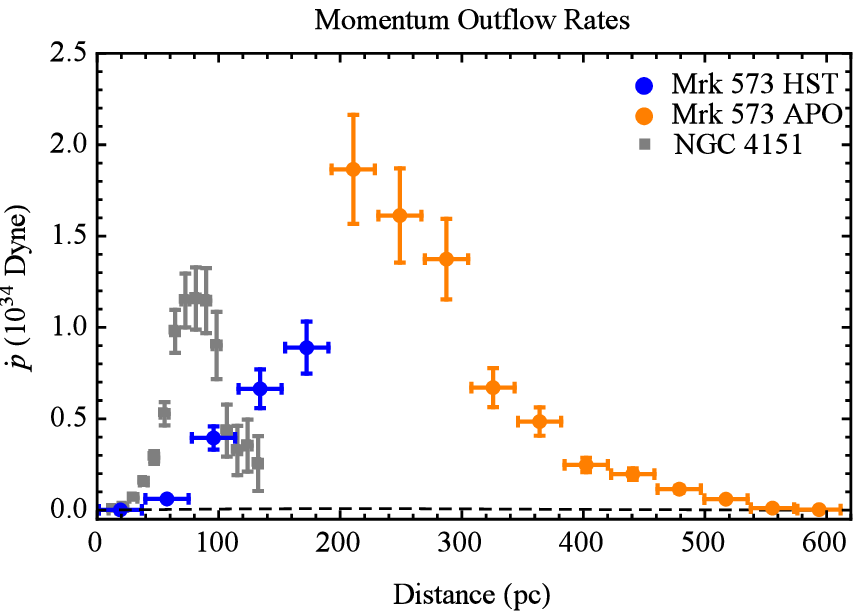}}
\vspace{-8pt}
\caption{From top left to bottom right are the azimuthally summed mass profiles, mass outflow rates, kinetic energy profiles, kinetic energy outflow rates, momentum profiles, and momentum outflow rates for Mrk 573 and NGC 4151 \citep{crenshaw2015}. The dashed lines represent the profiles that would result from the mass in the center bin ($M \approx 5.9 \times 10^3 M_{\odot}$) traveling through the velocity profile. The value at each distance is the quantity contained within that extraction bin of width $\delta r = 38.3$ pc.}
\label{results}
\end{figure*}

Using our scale factor and the densities from our photoionization models (for $r < 0\farcs5$) and [S~II] power law fits (for $r = 0\farcs5 - 1\farcs7$), we calculated the total mass in each image semi-annulus from Equation 9. The NLR mass profile is shown before (Figure \ref{scale}) and after (Figure \ref{results}) azimuthal summation. The bump in the mass profile between 500 and 600 pc is due to the partial inclusion of the bright arc of emission in the southwest. The total mass of ionized gas in the NLR for $r < 1\farcs7$ is $\sim 2.2 \times 10^6 M_{\odot}$ with $\sim$ 10\% of that contained in the HST spectral slit used to create our photoionization models and scale factor.

\subsection{Outflow Parameters}

Finally, we calculate the mass outflow rates ($\dot{M}_{\mathrm{out}}$) in units of $M_{\odot}$ yr$^{-1}$ at each position along the NLR using
\begin{equation}
\dot{M}_{\mathrm{out}} = \left(\frac{Mv}{\delta r} \right),
\end{equation}
\noindent
where $M$ is the semi-annular mass, $v$ is the deprojected velocity corrected for inclination and position angle on the sky (\S 3.2), and $\delta r$ is the deprojected width for each extraction. Deprojecting the distances results in a bin width that is 7.8\% larger than the observed value; thus each deprojected measurement spans $\delta r \approx$ 38.3 pc. 

In addition to mass outflow rates, a variety of energetic quantities can be determined, including kinetic energies, momenta, and their flow rates. These quantites yield information about the amount of AGN energy deposited into the NLR. The kinetic energy is given by
\begin{equation}
E = \frac{1}{2}{M}_{\mathrm{out}} v^2,
\end{equation}
The time derivative of this is the kinetic luminosity (also referred to as the energy injection or flow rate),
\begin{equation}
L_{KE} = \dot E = \frac{1}{2} \dot{M}_{\mathrm{out}} v^2,
\end{equation}
where we only include contributions from pure radial outflow (a $\sigma_v$ term is sometimes added to the energy budget to account for gas turbulence). Finally, the momenta ($p = M_{\mathrm{out}}v$) and momenta flow rates ($\dot p = \dot M_{\mathrm{out}}v$) are useful quantities that can be compared to the AGN bolometric luminosity, as well as the photon momentum ($L/c$), to quantify the efficiency of the NLR in converting radiation from the AGN into the radial motion of the outflows \citep{zubovas2012, costa2014}.

We obtain a single radial profile for each quantity by azimuthally summing the values derived for the SE and NW semi-annuli. The mass outflow rates prior to summation can be seen in Figure \ref{scale}, with the asymmetry due to the nature of the velocity laws and mass distributions.

\setlength{\tabcolsep}{0.145in}
\tabletypesize{\footnotesize}
\begin{deluxetable*}{c|c|c|c|c|c|c|c|}
\vspace{-26pt}
\tablenum{6}
\tablecaption{Radial Mass Outflow and Energetic Results\vspace{-6pt}}
\tablehead{
\colhead{Distance} & \colhead{Velocity} & \colhead{Mass} & \colhead{$\dot{M}$} & \colhead{Energy} & \colhead{$\dot{E}$} & \colhead{Momentum} & \colhead{$\dot{P}$}\\
\colhead{(pc)} & \colhead{(km s$^{-1}$)} & \colhead{(10$^5$ $M_{\odot}$)} & \colhead{($M_{\odot}$ yr$^{-1}$)} & \colhead{(10$^{53}$ erg)} & \colhead{(10$^{41}$ erg s$^{-1}$)} & \colhead{(10$^{46}$ dyne s)} & \colhead{(10$^{34}$ dyne)}\\
\colhead{(1)} & \colhead{(2)} & \colhead{(3)} & \colhead{(4)} & \colhead{(5)} & \colhead{(6)} & \colhead{(7)} & \colhead{(8)}
}
\startdata
19.2	 & 	106.7	 & 	0.06	$\pm$	0.01	 & 	0.02	$\pm$	0.00	 & 	0.01	$\pm$	0.00	 & 	0.00	$\pm$	0.00	 & 	0.01	$\pm$	0.00	 & 	0.00	$\pm$	0.00	\\
57.5	 & 	342.0	 & 	0.31	$\pm$	0.03	 & 	0.28	$\pm$	0.05	 & 	0.36	$\pm$	0.06	 & 	0.11	$\pm$	0.02	 & 	0.21	$\pm$	0.02	 & 	0.06	$\pm$	0.01	\\
95.8	 & 	580.7	 & 	0.69	$\pm$	0.08	 & 	1.08	$\pm$	0.17	 & 	2.33	$\pm$	0.37	 & 	1.15	$\pm$	0.18	 & 	0.80	$\pm$	0.09	 & 	0.39	$\pm$	0.06	\\
134.1	 & 	667.8	 & 	0.89	$\pm$	0.10	 & 	1.58	$\pm$	0.25	 & 	3.92	$\pm$	0.63	 & 	2.22	$\pm$	0.35	 & 	1.18	$\pm$	0.08	 & 	0.66	$\pm$	0.11	\\
172.4	 & 	689.8	 & 	1.07	$\pm$	0.12	 & 	1.96	$\pm$	0.31	 & 	5.25	$\pm$	0.84	 & 	3.32	$\pm$	0.53	 & 	1.46	$\pm$	0.13	 & 	0.89	$\pm$	0.14	\\
210.7	 & 	781.3	 & 	1.61	$\pm$	0.18	 & 	3.35	$\pm$	0.54	 & 	11.03	$\pm$	1.76	 & 	8.98	$\pm$	1.44	 & 	2.50	$\pm$	0.48	 & 	1.87	$\pm$	0.30	\\
249.1	 & 	744.7	 & 	1.43	$\pm$	0.16	 & 	2.84	$\pm$	0.45	 & 	9.53	$\pm$	1.52	 & 	8.11	$\pm$	1.30	 & 	2.12	$\pm$	0.36	 & 	1.61	$\pm$	0.26	\\
287.4	 & 	637.7	 & 	1.61	$\pm$	0.18	 & 	2.74	$\pm$	0.44	 & 	8.12	$\pm$	1.30	 & 	6.17	$\pm$	0.99	 & 	2.04	$\pm$	0.34	 & 	1.37	$\pm$	0.22	\\
325.7	 & 	463.0	 & 	1.35	$\pm$	0.15	 & 	1.67	$\pm$	0.27	 & 	3.96	$\pm$	0.63	 & 	2.56	$\pm$	0.41	 & 	1.24	$\pm$	0.12	 & 	0.67	$\pm$	0.11	\\
364.0	 & 	355.1	 & 	1.48	$\pm$	0.17	 & 	1.40	$\pm$	0.22	 & 	2.86	$\pm$	0.46	 & 	1.61	$\pm$	0.26	 & 	1.05	$\pm$	0.12	 & 	0.48	$\pm$	0.08	\\
402.3	 & 	267.3	 & 	1.12	$\pm$	0.13	 & 	0.80	$\pm$	0.13	 & 	1.46	$\pm$	0.23	 & 	0.71	$\pm$	0.11	 & 	0.60	$\pm$	0.03	 & 	0.25	$\pm$	0.04	\\
440.6	 & 	224.0	 & 	1.13	$\pm$	0.13	 & 	0.67	$\pm$	0.11	 & 	1.16	$\pm$	0.19	 & 	0.46	$\pm$	0.07	 & 	0.50	$\pm$	0.03	 & 	0.20	$\pm$	0.03	\\
478.9	 & 	204.8	 & 	1.17	$\pm$	0.13	 & 	0.64	$\pm$	0.10	 & 	0.68	$\pm$	0.11	 & 	0.18	$\pm$	0.03	 & 	0.48	$\pm$	0.02	 & 	0.11	$\pm$	0.02	\\
517.3	 & 	86.3	 & 	1.95	$\pm$	0.22	 & 	0.45	$\pm$	0.07	 & 	0.35	$\pm$	0.06	 & 	0.06	$\pm$	0.01	 & 	0.33	$\pm$	0.01	 & 	0.06	$\pm$	0.01	\\
555.6	 & 	28.0	 & 	2.75	$\pm$	0.31	 & 	0.21	$\pm$	0.03	 & 	0.06	$\pm$	0.01	 & 	0.00	$\pm$	0.00	 & 	0.15	$\pm$	0.00	 & 	0.01	$\pm$	0.00	\\
593.9	 & 	13.6	 & 	2.90	$\pm$	0.33	 & 	0.10	$\pm$	0.02	 & 	0.02	$\pm$	0.00	 & 	0.00	$\pm$	0.00	 & 	0.08	$\pm$	0.00	 & 	0.00	$\pm$	0.00
\enddata
\tablecomments{Numerical results for the mass and energetic quantities as a function of radial distance. Columns are (1) deprojected distance from the nucleus, (2) mass weighted mean velocity, (3) gas mass in units of 10$^{5}$ $M_{\odot}$, (4) mass outflow rates, (5) kinetic energies, (6) kinetic energy outflow rates, (7) momenta, and (8) momenta flow rates. These results, shown in Figure \ref{results}, are the sum of the individual radial profiles calculated for the SE and NW semi-annuli (see Figures \ref{imaging} and \ref{scale}). The value at each distance is the quantity contained within the annulus of width $\delta r = 38.3$ pc.}
\vspace{-3pt}
\end{deluxetable*}
\section{Results}

\subsection{Mass Outflow Rates \& Energetics}

In Figure \ref{results} and Table 6 we present our mass outflow rates and energetics as functions of distance from the nucleus for Mrk 573. We also show the results for NGC 4151 from \cite{crenshaw2015} for comparison. Data are available in machine-readable form by request to M.R. The outflow has a maximum radial extent of 600 pc from the nucleus and contains a total ionized gas mass of $M \approx 2.2 \times 10^6 M_{\odot}$. This mass is similar to other Seyfert galaxies such as Mrk 3 \citep{collins2009}. The total kinetic energy summed over all distances is $E \approx 5.1 \times 10^{54}$ erg.

The mass outflow rates rise to a peak value of $\dot M_{out} \approx$ 3.4 $\pm$ 0.5 $M_{\odot}$ yr$^{-1}$ at a distance of 210 pc from the nucleus and then steadily decrease to zero at $\sim$ 600 pc, which is the extent of our velocity law exhibiting outflow. The kinematics at further distances are consistent with rotation. The overall shapes of the profiles are a convolution of the velocity laws and mass profiles, exhibiting minor fluctuations on top of the overall increasing followed by decreasing trends. The dashed lines represent the mass outflow rates and energetics that would be observed if the amount of mass in the central bin ($M \approx 5.9 \times 10^ 3 M_{\odot}$) was allowed to propagate through the velocity profile. At 210 pc where the outflow peaks, this is $\sim$ 27 times smaller than the observed value, indicating that the outflow is not a steady state nuclear outflow, but that material is accelerated in-situ from its local location in the NLR.

The mass profiles for the SE and NW semi-annuli (Figure \ref{scale}) are asymmetric, such that their summed radial mass outflow rates (Table 6) are not represented by a simple average of the velocity laws, but by a mass weighted mean. The appropriate mean velocity profile is found by solving Equation 10 with the final mass and mass outflow rates (Table 6). The mean velocity profile does not reach the peak deprojected velocity of 1100 km s$^{-1}$, as the two halves of the velocity law peak at different radial distances (Figure \ref{kinematics}).

Compared to the AGN bolometric luminosity of Mrk 573, Log($L_{\mathrm{bol}}$) = 45.5 $\pm$ 0.6 erg s$^{-1}$ \citep{kraemer2009}, the peak kinetic luminosity reaches $<0.1$\% of $L_{\mathrm{bol}}$, but see also the discussion in \S8.2. The momentum flow rate can be compared to the photon momentum ($L/c$) of ionizing flux emanating from the AGN. The photon momentum from the bolometric luminosity is $\dot{p} \approx 1.05 \times 10^{35}$ Dyne, and the peak momentum flow rate is $\dot{p} \approx 2.5 \times 10^{34}$ Dyne. Thus the peak outflow momentum rate is $\sim$ 25\% of the AGN's photon momentum.

The outflow velocities trending to zero near the nucleus naturally leads to small outflow rates at small radial distances. In \cite{fischer2017} we found evidence of multiple high velocity (FWHM $\sim$ 1000 km s$^{-1}$) kinematic components near the nucleus using our high spatial resolution Gemini Near-Infrared Field Spectrograph (NIFS) observations. If the FWHM is taken to be a more representative signature of the outflow velocities near the nucleus, then the three innermost outflow rates would increase to $\sim$ 2.2, 3.6, and 4.8 $M_{\odot}$ yr$^{-1}$, respectively.

We also assumed that the NLR material is moving radially along the NLR major axis (PA = 128$\degr$), rather than the STIS slit PA. If this angle were used, the projection effects would be more significant ($\varphi = 12\degr$), with the peak deprojected velocities reaching $\sim$ 2700 km s$^{-1}$, and the mass outflow rates would increase by a factor of $\sim$ 2.48. From our modeling in \cite{fischer2017}, and the typical observed velocities in Seyferts, we consider this to be less probable and retain our conservative result.

Furthermore, we have neglected contributions to the mass outflow rates and energetics from ablation of gas off the spiral dust lanes at distances of 600--750 pc. Here the kinematics are generally consistent with rotation; however, the [O~III] and H$\alpha$ velocity centroids show a systematic separation $\sim$ 100 km s$^{-1}$ that is not seen at larger radial distances. This separation is most likely due to ablation of material off the faces of the ionized arcs in rotation, and we do not include it in our results.

Finally, our assumptions about outflows and the specific velocity and density laws may not be accurate for material outside of the nominal bicone, along the NLR minor axis. If we restrict our semi-annuli to azimuthal angles within the ionizing bicone, which has a large opening angle (\S 8.1.1), the mass and outflow rates decrease by $\sim$ 20\%.

In the ENLR ($r>1\farcs7$), the density drops very slowly and $n_e \approx$ 150--200 cm$^{-3}$. Adopting this density range, our scale factor, and extended [O~III] image fluxes, we find the ENLR mass is $\sim 6-7\times$10$^6$ $M_{\odot}$. Thus the mass of the NLR+ENLR is $\sim 8-9\times$10$^6$ $M_{\odot}$, indicating that $\sim$ 25\% of the ionized gas exhibits outflow kinematics.

\subsection{Comparison with NGC 4151}

The mass outflows in Mrk 573 are significantly more powerful than those in NGC 4151, as shown by the energetics in Figure \ref{results}. This is because the total ionized NLR mass participating in the outflow is $\sim$ 2.2$\times$10$^6$ $M_{\odot}$ in Mrk 573, a factor of seven greater than NGC 4151's $\sim$ 3$\times$10$^5$ $M_{\odot}$ \citep{crenshaw2015}. Another notable difference is the extent of the outflows, which reach $\sim$ 600 pc in Mrk 573, but only $\sim$ 140 pc for NGC 4151.

These results can be understood by comparing the physical properties of these two AGN. Mrk 573 has a SMBH mass of Log($M_{\mathrm{BH}}$) = 10$^{7.28}$ $M_{\odot}$ \citep{woo2002, bian2007}, a bolometric luminosity of $L_\mathrm{bol} \approx 10^{45.5}$ erg s$^{-1}$, and a corresponding accretion rate of $\dot m_{\mathrm{acc}} \approx$ 0.44 $M_{\odot}$ yr$^{-1}$ (assuming $\dot m_{\mathrm{acc}} = L/\eta c^2$ with $\eta = 0.1$, \citealp{peterson1997}). For NGC 4151 these values are Log($M_{\mathrm{BH}}$) = 10$^{7.66}$ $M_{\odot}$ \citep{bentz2006}, $L_\mathrm{bol} \approx 10^{43.9}$ erg s$^{-1}$, and $\dot m_{\mathrm{acc}} \approx$ 0.013 $M_{\odot}$ yr$^{-1}$. Thus despite the similar SMBH masses in these two objects their bolometric luminosities and corresponding accretion rates differ by $\sim$ 1.6 dex, yielding $L/L_{\mathrm{Edd}} \approx 0.75$ for Mrk 573 and $L/L_{\mathrm{Edd}} \approx 0.01$ for NGC 4151.

Mrk 573 is releasing significantly more energy into the NLR, allowing for higher velocity outflows containing more mass that are driven to larger distances. Interestingly, the masses, velocities, and extraction sizes conspire so both objects have peak outflow rates $\sim$ 3 $M_{\odot}$ yr$^{-1}$. For this reason, comparing the outflow energetics between objects may be more insightful.

\section{Discussion}

\subsection{Comparison with Global Outflow Rates}

We refer to single value mass outflow rates that are derived from mean conditions across the entire NLR as ``global'' outflow rates. There are two common techniques for obtaining global outflow rates. The first is to derive a geometric model, typically an ionized bicone, and fill it with material diluted by a filling factor to account for clumpiness. The second converts the observed luminosity of a hydrogen recombination line (e.g. H$\beta$, P$\beta$) to mass based on a mass-luminosity scaling relationship. We examine our results in the context of these techniques to explore systematics and uncertainties, and to compare with other AGN in a broader framework.

\subsubsection{Geometric Approach}

The geometric approach can take the form $\dot{M}_{out} = 2 m_p n_e v A f$, where $m_p$ is the proton mass, $n_e$ is the electron density, $v$ is the outflow velocity, $A$ is the area of the bicone, $f$ is a volume filling factor, and the factor of two accounts for two symmetric bicones (e.g. \citealp{mullersanchez2011}). The filling factor accounts for the clumpiness of the gas and the fact that it does not fill the entire volume of the ionization cone. This method has the advantage of yielding quick estimates once a geometric model is adopted, but variations in the filling factor from object to object and across the NLR can result in uncertainties $>$ 1 dex. This is compounded by assuming the outflow rates of each bicone are symmetric, which is not accurate for Mrk 573 (Figure \ref{scale}). When models are not available, this discrepancy might be reduced by estimating the filling factor observationally for individual objects, which can be done using the luminosity of $P\beta$ when spectra are available \citep{riffel2011b, mullersanchez2016}.

For comparing with this technique, we adopt a mean velocity (rather than a maximum) that is representative of the majority of the outflow. We also adopt the hydrogen number density as compared to the electron density, with the two related by $n_\mathrm{H} \approx 0.85 \times n_e$ (\S 3.4.4). We use our range of observed [S~II] densities, a biconical geometry with a half opening angle of $38\degr$, and radial extent of 600 pc to encompass the observed emission, and a range of NLR filling factors from the literature ($f \approx 0.001- 0.1$, \citealp{storchibergmann2010, mullersanchez2011, nevin2018}). We find $\dot{M}_{out} \approx 8-800$ $M_{\odot}$ yr$^{-1}$ for $n_e = 200$ cm$^{-3}$, and $\dot{M}_{out} \approx 65-6500$ $M_{\odot}$ yr$^{-1}$ for $n_e = 1500$ cm$^{-3}$. These values are significantly higher than those derived using our photoionization models. 

This discrepancy can be traced to the filling factors. From the volume of our biconical geometry intercepted by the slit, and the volumes of our model clouds, we calculate a mean filling factor of $f \approx 5.9\times 10^{-5}$. Using this filling factor, we find $\dot{M}_{out} \approx 0.5-4$ $M_{\odot}$ yr$^{-1}$ for $n_e = 200-1500$ cm$^{-3}$, which comfortably encompass our model derived outflow rates. It is important to note that our filling factor is arbitrarily low. If we adopted a geometry with material constrained to a disk, the filling factor would increase as the corresponding volume decreases, yielding the same mass outflow rates. For these reasons it is critical to calculate filling factors for individual objects.

Using the geometric technique to estimate mass outflow rates for the NLR of NGC 4151, \cite{storchibergmann2010} found a global outflow rate of $\sim 2.4$ $M_{\odot}$ yr$^{-1}$ using $f = 0.11$ (biconical) or $f = 0.025$ (spherical). Similarly, \cite{mullersanchez2011} found $\sim 9$ $M_{\odot}$ yr$^{-1}$ using $f = 0.001$. These values are in overall agreement with the peak value of $\sim 3.0$ $M_{\odot}$ yr$^{-1}$ from \cite{crenshaw2015}. This indicates that the two techniques can derive comparable mass outflow rates when calculated from physically motivated choices for the velocity, density, geometry, and filling factor of the system.

\subsubsection{Luminosity Approach}

The second technique that is closer to the methodology employed here is to convert an observed luminosity (e.g. H$\beta$, H$\alpha$, [O~III]) to mass using a simple relationship that assumes uniform NLR conditions and that scales with density. This type of relationship is the same as that given in Equation 6, with the mass typically determined using a scaling relationship based on a single emission line and density, in contrast to our multi-component models that account for material of different densities and ionization states at the same spatial location. Employing the techniques of \cite{nesvadba2006} and \cite{bae2017}, we calculate the NLR mass as $M = (9.73\times10^8 M_{\odot}) \times L_{\mathrm{H}\alpha, 43} \times n^{-1}_{e,100}$, where $L_{\mathrm{H}\alpha, 43}$ is the H$\alpha$ luminosity in units of 10$^{43}$ erg s$^{-1}$, and $n^{-1}_{e,100}$ is the electron density in units of 100 cm$^{-3}$. Here we used the [O~III] luminosity scaled by the average H$\alpha$/[O~III] ratio as a proxy for $L_{\mathrm{H}\alpha}$. From this we derive $\dot{M}_{out} \approx 5-35$ $M_{\odot}$ yr$^{-1}$ for $n_e = 1500-200$ cm$^{-3}$, the full range of our observed [S~II] densities. The corresponding NLR mass estimate is $\sim$ 1-7 times larger than that found from our models ($\sim$ 2.2$\times$10$^6$ $M_{\odot}$), highlighting the difference between employing a single density and multi-density gas phases at each location.

This method can have the advantage of deriving mass outflow rates with smaller systematic uncertainties than the geometric approach, but requires an accurate Balmer emission line luminosity or proxy obtained from spectroscopy or narrow-band imaging. As luminosity and density are physical indicators of the total gas mass \citep{peterson1997}, we strongly encourage the use of luminosity based methods for deriving total masses and outflow rates to avoid uncertain filling factors and geometries.

\subsubsection{Recommendations for Comparison}

The methodology used here that was modeled after \cite{crenshaw2015} has the advantage of deriving spatially resolved mass outflow rates with small uncertainties that are critical for probing AGN feedback on scales of tens of parsecs in the NLRs of nearby AGN. However, this requires high quality spectroscopy and imaging, or optical Integral Field Unit (IFU) spectroscopy and detailed, time-intensive photoionization modeling. For targets where these data are unavailable, both of the techniques above can provide estimates of the global mass outflow rate with larger uncertainties.

The derived mass outflow rates will depend strongly on the choices of velocity, density, and geometry of the system. Different conventions throughout the literature can result in estimated mass outflow rates spanning $\sim 3$ orders of magnitude for individual objects! (see, e.g \citealp{karouzos2016, bischetti2017, nevin2018, perna2017} for discussions).

When spatially resolved kinematics and density profiles are unfeasible, we would recommend using an average or flux weighted average velocity, as this will be more representative of the majority of the outflow than the peak velocity. In addition, densities should be determined from spectra whenever possible, as individual objects can vary significantly from typical NLR values, and the density profile as a function of radius can span more than a factor of 10 as shown here. [S~II] is typically a strong choice in optical spectra, with the caveat that it will generally yield a lower limit to the electron density, as [S~II] is a lower ionization line with a peak emissivity toward the neutral zone of a cloud such that additional material can be hiding in the neutral phase because it is not contributing electrons to the ionized gas. In addition, higher ionization gas will generally have a lower density that is not probed by [S~II]. Finally, the radial extent of the outflow must be determined precisely, which is only possible with spatially resolved spectroscopy.

The mass, kinetic energy, and momentum at each distance (Figure \ref{results}) may be added to obtain enclosed totals; however, the rates cannot. A continuous flow originating near the nucleus, shown by the dotted line in Figure \ref{results}, represents the minimum outflow rate. When in-situ acceleration is included,  gas is driven at all distances and the total mass reaching the outer boundary will be larger. This requires an integration over time, assuming the gas travels without being destroyed, and is not equivalent to summing the mass outflow rates. The average outflow rate across all bins, and the area under the outflow curve representing the total momentum, are invariant to the number of bins. For these reasons, comparing the total outflow energetics between objects is more straightforward.

For a direct comparison with global outflow rates, we consider larger spatial extractions such that each bin contains more mass and must travel a larger $\delta r$. Reducing this to a single point with $M \approx 2.2 \times 10^6 M_{\odot}$ and $\delta r = 600$ pc results in a global mass outflow rate of $\sim 2 M_{\odot}$ yr$^{-1}$ for a mean velocity of 550 km s$^{-1}$.

\subsection{Implications for Feedback}

The global kinetic luminosity ($M = 2.2 \times 10^6 M_{\odot}$, $\delta r = 600$ pc, $V = 550$ km s$^{-1}$) of the outflow in Mrk 573 is $L_{\mathrm{KE}}/L_{\mathrm{Bol}} \approx $ 0.4--0.8\% of the AGN bolometric luminosity ($L_\mathrm{bol} \approx 10^{45.5}$ erg s$^{-1}$, \citealp{kraemer2009}). This is in the range of 0.5\%--5\% used in some models of efficient feedback \citep{dimatteo2005, hopkins2010} and is similar to the values reported for NGC 4151 \citep{storchibergmann2010, mullersanchez2011, crenshaw2015}. However, it is important to note that local AGN already have well-established bulges, so comparing to models of effective feedback for higher redshift AGN requires further investigation.

Our results indicate that most of the outflow is accelerated in-situ within the NLR and does not originate from near the nucleus. This is seen in that the peak mass outflow rate is much greater than what would result from the amount of mass near the nucleus following the radial velocity law, as shown by the dashed line in Figure \ref{results}. The only nuclear outflow scenario that could produce the observed result would be if the nuclear outflow rate decreased and increased in a fashion exactly matching the shape and travel time of a cloud along the velocity profile, requiring inordinate fine tuning.

Our modeling is consistent with the conclusions of \cite{fischer2017} that the NLR outflows are radiatively driven. Recently, \cite{mou2017} also explored the possibility that these outflows are circumnuclear clouds accelerated by an accretion disk wind. Their numerical simulations of the NLR outflows in NGC 4151 match the mass outflow rates and kinetic luminosities, with some discrepancy in the velocity turnover at larger radii. The high temperatures of the model winds indicate they may be difficult to detect observationally. The comparison of these types of models to radiative driving for the more powerful outflows in Mrk 573 and other AGN should yield valuable physical insight into the launching mechanisms responsible for NLR outflows.

A variety of metrics are employed in the literature to determine if outflows deliver ``effective'' feedback to their host galaxies by impacting star formation \citep{leung2017}. This can include triggering star formation through positive feedback \citep{silk2013, mahoro2017}, quenching star formation through negative feedback \citep{wylezalek2016}, or more complex interactions \citep{zubovas2017}. These comparisons require accurate star formation rates (SFRs), which are typically estimated through H$\alpha$ luminosities. Determining SFRs for AGN is difficult due to the contamination by AGN ionized gas \citep{imanishi2011}. Successful techniques include using estimates in wavebands where the AGN is weakly emitting, utilizing high spatial resolution optical IFU data to separate the emission \citep{davies2016}, and the emission from hydrocarbons that are excited by star forming regions \citep{shipley2013, shipley2016}. A detailed investigation of the SFR is beyond the scope of this work, and future observations with the {\it James Webb Space Telescope} (JWST) will enable spatially resolved distribution studies of star-forming excited hydrocarbons \citep{kirkpatrick2017}.

\subsection{Missing Mass: X-ray \& Molecular Outflows}

These results account for the optical and some UV emission line gas; however, AGN outflows are also seen in more highly ionized UV/X-ray gas that is not accounted for in our mass outflow rates. For NGC 4151, \cite{wang2011} found a mass outflow rate of  $\dot{M}_{\mathrm{out}} \approx 2.1$ $M_{\odot}$ yr$^{-1}$ for the X-ray emitting gas, while \cite{crenshaw2012} derived a mass outflow rate of $\dot{M}_{\mathrm{out}} \approx 0.3-0.7$ $M_{\odot}$ yr$^{-1}$ for the UV/X-ray absorbers, indicating two additional important outflow components. The presence of an ultrafast outflow (UFO) traveling at 0.1c with a much smaller mass outflow rate ($\dot{M}_{\mathrm{out}} \approx 0.003-0.04$ $M_{\odot}$ yr$^{-1}$) than the NLR outflows, but comparable kinetic luminosity ($\dot{E}\sim10^{42} - 10^{43}$ erg s$^{-1}$), underscores the need to study outflows across all spatial scales and energy regimes \citep{schurch2003, kraemer2005, kraemer2006, piro2005, tombesi2010, tombesi2011, tombesi2012}.

Studies of Mrk 573 in the X-rays \citep{ferland1986, awaki1991} with {\it Chandra} and {\it XXM-Newton} \citep{guainazzi2005, paggi2012, reynaldi2012} have found very highly ionized gas including the Fe XXV K$\alpha$ line. Photoionization modeling by \cite{bianchi2010} and \cite{gonzalezmartin2010} found that two high ionization components were needed to describe the X-ray gas. It is interesting to note that the ionization parameters of our highest ionization components begin to approach the conditions of the components modeled in the X-rays, suggesting a natural continuum of physical conditions in the NLR, as mentioned by \cite{gonzalezmartin2010}. If this X-ray material is outflowing, it may contribute significantly to feedback.

At much lower temperatures, AGN driven outflows of molecular hydrogen (H$_2$) have been observed \citep{sturm2011, feruglio2015, janssen2016, rupke2017}. There are several H$_2$ lines present in near infrared spectra of Mrk 573 \citep{veilleux2007, fischer2017} that probe warmer molecular gas, and we presented a kinematic map showing signatures of outflow in \cite{fischer2017}. In future work we will address the contribution of warm molecular outflows to the overall feedback and energetics of Mrk 573. While the total mass in the warm molecular gas phase is significantly smaller than the optical emission line gas, it may represent the warm skin of the more massive cold molecular reservoirs that form stars. Probing the cold H$_2$ in detail will require radio observations with observatories such as the Atacama Large Millimeter Array (ALMA).

\section{Conclusions}

We used long slit spectroscopy, [O~III] imaging, and Cloudy photoionization models to determine the mass outflow rates and energetics as functions of distance from the nucleus in the Seyfert 2 galaxy Mrk 573. This is the second spatially resolved outflow rate for an AGN, and the first for a type 2. Our conclusions are as follows:
\begin{enumerate}
\item The outflow contains $M \approx 2.2 \times 10^6 M_{\odot}$ of ionized gas, with a total kinetic energy of $E \approx 5.1 \times 10^{54}$ erg. This is significantly more ionized gas and energy than the NLR outflow in the lower luminosity Seyfert 1 galaxy NGC 4151 \citep{crenshaw2015}.

\item We find that the outflows extend to $\sim$ 600 pc, reaching a peak mass outflow rate of $\dot M_{out} \approx$ 3.4 $\pm$ 0.5 $M_{\odot}$ yr$^{-1}$ at a distance of 210 pc from the SMBH. Our spatially resolved measurements are consistent with in-situ acceleration of the circumnuclear gas.

\item The global kinetic luminosity of the outflow is $\mathrm{L_{KE}/L_{Bol}} \approx 0.4-0.8$\% of the AGN bolometric luminosity. This is similar to values used in feedback models and those reported for NGC 4151.

\item Methods for determining global outflow rates are subject to larger uncertainties if photoionization models are not used, and luminosity-based methods are preferred over geometric, as they invoke physical tracers of the gas mass. All techniques require accurate gas densities, mean velocities, and system geometry. Spatially resolved outflow rates cannot be added to obtain global estimates.

\item Our results account for the UV/optical emission line gas, and multi-wavelength studies of individual AGN are needed to understand the importance of additional outflow components including hot X-ray and cold molecular gas phases.
\end{enumerate}

In the future we will apply these techniques to additional Seyferts to uncover correlations between the NLR mass outflows and properties of the host AGN, and compare with estimated SFRs to quantify how effective NLR outflows are at delivering feedback in nearby AGN.

\section*{Acknowledgements}
The authors thank the anonymous referee for their helpful comments that improved the clarity of this paper. M.R. would like to thank Justin Cantrell and Sushant Mahajan for cluster computing assistance, Kathryn Lester and Wesely Peters for helpful comments, Joseph Weingartner for discussions on dust grains, and Gary Ferland, Peter van Hoof, and Eric Pellegrini for Cloudy support.

M.R. gratefully acknowledges support from the National Science Foundation through the Graduate Research Fellowship Program (GRFP). This material is based upon work supported by the National Science Foundation Graduate Research Fellowship Program under Grant No. DGE-1550139. Any opinions, findings, and conclusions or recommendations expressed in this material are those of the author(s) and do not necessarily reflect the views of the National Science Foundation.

Support for this work was provided by NASA through grant number HST-AR-14290.001-A from the Space Telescope Science Institute, which is operated by AURA, Inc., under NASA contract NAS 5-26555. Basic astrophysics research at the Naval Research Laboratory is supported by 6.1 base money. T.C.F. was supported by an appointment to the NASA Postdoctoral Program at the NASA Goddard Space Flight Center, administered by the Universities Space Research Association under contract with NASA.

This paper used the photoionization code Cloudy, which can be obtained from \url{http://www.nublado.org} and the Atomic Line List available at \url{http://www.pa.uky.edu/~peter/atomic/}. This research has made use of the NASA/IPAC Extragalactic Database (NED), which is operated by the Jet Propulsion Laboratory, California Institute of Technology, under contract with the National Aeronautics and Space Administration. This research has made use of NASA's Astrophysics Data System.

\facility{{\it Facilities: }HST (STIS, WFPC2), ARC (DIS)}

\end{document}